\documentclass[twocolumn]{aastex631}

\usepackage{comment}
\usepackage{acronym}
\usepackage{stackengine}
\usepackage{tikz}
\usetikzlibrary{shapes, arrows}
\usepackage{graphicx}
\usepackage{newtxmath}
\usepackage{amsmath,bm}

\DeclareMathOperator*{\argmin}{arg\,min}
\usepackage{amstext}
\usepackage{amssymb}
\usepackage{bbm}
\usepackage{makecell}
\usepackage{threeparttablex, tablefootnote}

\shorttitle{Direct Exoplanet Detection with ConStruct}
\shortauthors{Wolf et al.}

\graphicspath{{./}{figures/}}

\newcommand{\SNR}{\ac{SNR}}

\newcommand{\HCI}{\ac{HCI}}
\newcommand{\PSF}{\ac{PSF} }
\newcommand{\AO}{\ac{AO} }
\newcommand{\IWA}{\ac{IWA}}
\newcommand{\ADI}{\ac{ADI}}

\newcommand{\RDI}{\ac{RDI}}
\newcommand{\LOCI}{\ac{LOCI}}
\newcommand{\KLIP}{\ac{KLIP}}

\newcommand{\ROI}{\ac{ROI}}
\newcommand{\ReLU}{\ac{ReLU}}

\newcommand{\RMS}{\ac{RMS}}
\newcommand{\PCA}{\ac{PCA}}
\newcommand{\KOA}{\ac{KOA}}
\begin{document}

\newacro{PSF}[PSF]{Point Spread Function}
\newacro{RV}[RV]{Radial Velocity}
\newacro{IWA}[IWAs]{inner-working-angles}
\newacro{AO}[AO]{adaptive optics}
\newacro{HCI}[HCI]{high contrast imaging}
\newacro{LOCI}[LOCI]{Locally Optimized Combination of Images}
\newacro{KLIP}[KLIP]{Karhunen-Lo\'eve Image Projection}
\newacro{ADI}[ADI]{Angular Differential Imaging}
\newacro{SDI}[SDI]{Spectral Differential Imaging}
\newacro{RDI}[RDI]{Reference Differential Imaging}
\newacro{SNR}[SNR]{signal-to-noise ratio}
\newacro{PCA}[PCA]{Principal Component Analysis}
\newacro{IFU}[IFU]{Integral Field Unit}
\newacro{LLSG}[LLSG]{Locally Low-rank Sparse Gaussian Analysis}
\newacro{ROI}[ROI]{region of interest}
\newacro{ReLU}[ReLU]{rectified linear activation function}
\newacro{GPU}[GPU]{Graphical Processing Unit}
\newacro{NN}[NN]{neural network}
\newacro{RMS}[RMS]{root mean square}
\newacro{KOA}[KOA]{Keck Observatory Archive}

\title{Direct Exoplanet Detection Using Deep Convolutional Image Reconstruction (\texttt{ConStruct}):\\ A New Algorithm for Post-Processing High-Contrast Images} 

\author[0000-0002-1406-8829]{Trevor N. Wolf}
\affiliation{Department of Aerospace Engineering and Engineering Mechanics, The University of Texas at Austin,\\
2617 Wichita St., Austin, TX 78712, USA}
\affiliation{Department of Astronomy, The University of Texas at Austin,\\
2515 Speedway Blvd., Austin, TX 78712, USA}
\affiliation{Center for Planetary Systems Habitability, The University of Texas at Austin,\\
2305 Speedway Stop C1160, Austin, TX 78712, USA}

\author[0000-0003-3480-6320]{Brandon A. Jones}
\affiliation{Department of Aerospace Engineering and Engineering Mechanics, The University of Texas at Austin,\\
2617 Wichita St., Austin, TX 78712, USA}
\affiliation{Center for Planetary Systems Habitability, The University of Texas at Austin,\\
2305 Speedway Stop C1160, Austin, TX 78712, USA}

\author[0000-0003-2649-2288]{Brendan P. Bowler}
\affiliation{Department of Astronomy, The University of Texas at Austin,\\
2515 Speedway Blvd., Austin, TX 78712, USA}

\begin{abstract}
We present a novel machine-learning approach for detecting faint point sources in high-contrast adaptive optics imaging datasets. The most widely used algorithms for primary subtraction aim to decouple bright stellar speckle noise from planetary signatures by subtracting an approximation of the temporally evolving stellar noise from each frame in an imaging sequence. Our approach aims to improve the stellar noise approximation and increase the planet detection sensitivity by leveraging deep learning in a novel direct imaging post-processing algorithm. We show that a convolutional autoencoder neural network, trained on an extensive reference library of real imaging sequences, accurately reconstructs the stellar speckle noise at the location of a potential planet signal. This tool is used in a post-processing algorithm we call Direct Exoplanet Detection with Convolutional Image Reconstruction, or \texttt{ConStruct}. The reliability and sensitivity of \texttt{ConStruct} are assessed using real Keck/NIRC2 angular differential imaging datasets. Of the 30 unique point sources we examine, \texttt{ConStruct} yields a higher S/N than traditional PCA-based processing for 67$\%$ of the cases and improves the relative contrast by up to a factor of 2.6. This work demonstrates the value and potential of deep learning to take advantage of a diverse reference library of point spread function realizations to improve direct imaging post-processing. \texttt{ConStruct} and its future improvements may be particularly useful as tools for post-processing high-contrast images from the James Webb Space Telescope and extreme adaptive optics instruments, both for the current generation and those being designed for the upcoming 30 meter-class telescopes.

\end{abstract}


\section{Introduction} \label{sec:intro}

A complete statistical census of exoplanet demographics is needed to test and guide planet formation and evolutionary models \citep[e.g.,][]{Burrows2001, Alibert2005, Gaudi2021}. Planets detected with indirect methods, particularly using radial velocities or transits \citep{Seager2008, Lovis2010}, comprise the bulk of known discoveries. These methods are effective for finding companions at close separations to their host stars, but are less sensitive at wider orbital distances. Over the past two decades, \HCI\hspace{0.1ex} has emerged as an effective tool to study long-period planets by probing the architectures of planetary systems from the outside in, while also enabling spectroscopic characterization of their atmospheres \citep{Bowler2016, Baron2019, Nielsen2019, Vigan2021}.

Dedicated hardware, including high-order \AO \citep{Guyon2005} and coronagraphy \citep{Guyon2006, Oppenheimer2009}, are necessary to reach the planet$/$star contrasts required to detect faint substellar companions with direct imaging. Advanced post-processing algorithms play a crucial role in pushing the sensitivity of imaging surveys to smaller \IWA\hspace{0.1ex} to maximize the scientific yield of these instruments. Central to this is accurately modeling and removing correlated quasi-static speckle noise in the imaging data. For ground-based instruments, residual atmospheric wavefront errors uncorrected with \AO and instrumental aberrations produce speckle noise with correlation lengths that range between seconds and hours \citep{Hinkley2007, Martinez2012}. Efficient post-processing is especially necessary at small \IWA\hspace{0.3ex} where speckles are often brighter and exhibit similar spatial characteristics to planetary signatures \citep[e.g.,][]{Fitzgerald2005}.

Post-processing strategies are tied to the observation approach. With \ADI\hspace{0.3ex}, instruments are configured to observe in pupil-tracking mode so that on-sky sources rotate deterministically around the optical axis of the instrument, while slowly evolving speckle noise realizations remain fixed in the image plane \citep{Liu_2004, Marois2005}. Post-processing algorithms such as \LOCI\hspace{0.3ex} \citep{Lafreniere2007} and \KLIP\hspace{0.3ex} \citep{Soummer2012} leverage the spatial de-coupling between speckle noise and planetary signals with \ADI\hspace{0.3ex} to estimate a best-fit reconstruction of the speckle noise in each frame, conditioned on other frames in the sequence. The reconstructed images are subsequently subtracted from each respective frame, and the residual images are de-rotated and averaged to recover real sources.

Several variations and extensions of LOCI and KLIP have been developed to improve these algorithms in their original forms. For instance, at very close separations, the sky rotation is sometimes insufficient to decouple planetary signals from speckle noise, causing the reconstructed frame to partially fit any real source that might be present. To prevent excessive over-subtraction, \citet{Pueyo2012} introduced a damped version of \LOCI\hspace{0.3ex} which regularizes the least-squares reconstruction. \RDI\hspace{0.3ex} is another strategy to address over-subtraction by making use of a library of reference images \citep{Lafreniere2009, Xie2022, Sanghi2022}, or images of a reference star sampled concurrently with the target \citep{Wahhaj2021}, to create the linear reconstruction of each science frame. However, the \RDI\hspace{0.3ex} image basis is not necessarily linearly independent to the expected signal profile of an on-sky object, so some self-subtraction can still occur.

To improve the linear speckle reconstruction methods like those used in both \ADI\hspace{-0.1ex} and \RDI, we present a machine learning-based method for speckle noise reconstruction we call Direct Exoplanet Detection with Convolutional Image Reconstruction, or \texttt{ConStruct}. Our approach explicitly partitions image patches\footnote{We define image patches as small sub-frames sampled from individual \ADI\hspace{-0.1ex} frames. In this work, the size of each patch is much smaller than the original ADI frame.} sampled in \ADI\hspace{-0.1ex} frames into hypothesized planet absent$/$present sectors for post-processing. It uses the spatial speckle correlations across neighboring pixels in the planet absent sector to predict a speckle noise reconstruction, independent of potential signals, in the planet-present sector. For this, an autoencoder neural network is trained in a self-supervised learning architecture with thousands of real examples, which can be applied to science targets without further training. 

By leveraging a library of archival \HCI\hspace{0.3ex} data, \texttt{ConStruct} is similar to \RDI, but it can encode information from thousands of training examples without manually selecting a reference library of images. Our approach is motivated by algorithms developed in the machine learning community for filling corrupted or missing regions in images with deep learning, commonly known as image inpainting. \cite{Elharrouss2019} provides a review of the literature related to image inpainting.

Machine learning approaches for post-processing direct imaging data have received growing interest in recent years. A supervised learning framework for detecting faint point sources is introduced in \citet{Gonzalez2018} which uses both random forests and neural networks to classify likely candidate point sources. Similarly, in \citet{Flasseur_2023} the authors use convolutional neural networks for both detection and characterization of post-processed images, generated with the \texttt{PACO} algorithm \citep{Flasseur_2018, Flasseur_2020}.   \citet{Yip2019} apply generative adversarial networks to create synthetic coronagraphic image realizations. The synthetic images then train a convolutional neural network to classify regions that contain potential bright planets in single science frames. \citet{Gebhard2022} built a regularized linear model of speckle noise based on causal predictors to fill in speckle noise in a region of interest. Our approach shares some similarities; however, we utilize a highly nonlinear model to complete the speckle prediction with solely local speckle correlation.

This paper is organized as follows. In Section \ref{sec:problem_formulation} we explain \texttt{ConStruct}'s principal and how it is used for detecting planetary companions. We also compare the operating mechanisms of \texttt{ConStruct} to those used in linear speckle reconstruction approaches like \KLIP. Section \ref{sec:inpainting_for_exoplanet_detection} details the autoencoder neural network used in \texttt{ConStruct} as well as an additional linear correction to leverage the temporal correlations between frames in \ADI\hspace{0.3ex} sequences. In Section \ref{sec:algorithm_performance_and_tuning} we discuss how \texttt{ConStruct} is tuned to maximize the \SNR\hspace{0.3ex} of substellar sources in these data. We apply \texttt{ConStruct} to 30 unique point sources in data sets from the W.M. Keck Observatory's NIRC2 imaging camera in Section \ref{sec:results}, and compare the results with a PCA reduction approach.

\section{Framework of Construct} \label{sec:problem_formulation}
\subsection{Speckle Noise Reconstruction}
\begin{figure}[t!]
\epsscale{1.2}
\plotone{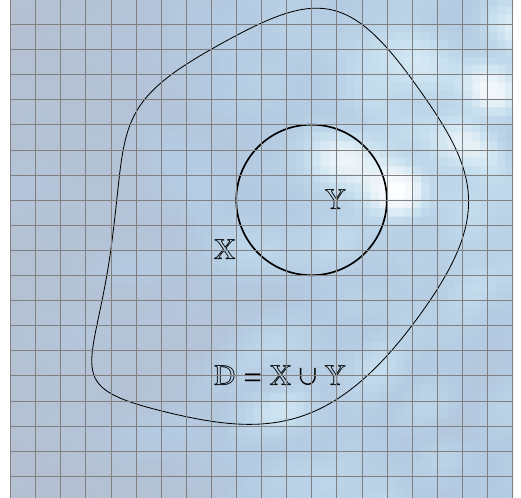}
\caption{The partitioning of regions $\mathbb{X}$, $\mathbb{Y}$, and $\mathbb{D}$ in an \ADI\hspace{-0.1ex} frame. Each region can in principle be any shape and size. In forming the partitions, the region $\mathbb{Y}$ is hypothesized to contain a point source, and $\mathbb{X}$ contains only speckle noise. \label{fig:tikz_diagram}}
\end{figure}

In this section, we first introduce the framework behind \texttt{ConStruct} and place it in the context of existing approaches for processing high-contrast \ADI\hspace{-0.1ex} sequences. A general region in an \ADI\hspace{-0.1ex} frame can be partitioned into sub-regions $\mathbb{X}$ and $\mathbb{Y}$ as illustrated in Figure \ref{fig:tikz_diagram}. The sub-region $\mathbb{X}$ is assumed to include only pixels from speckle noise, and $\mathbb{Y}$ contains speckle noise spatially correlated with $\mathbb{X}$ and a potential signal from an on-sky object. The pixel-wise intensities for the $t^{\mathrm{th}}$ frame contained in regions $\mathbb{X}$, $\mathbb{Y}$, and their union, $\mathbb{D}$, are,
%


\begin{align}
    \label{eqn:general_hypth_test}
    \bm I^t_{\mathbb{X}} &= \bm s^t_{\mathbb{X}} + \bm \eta\\
    \bm I^t_{\mathbb{Y}} &= \bm s^t_{\mathbb{Y}} + \mathbbm{1}_{\mathcal{H}_1}(\gamma) \bm h_{\mathbb{Y}} +  \bm \eta \mbox{ \hspace{1ex} where, \hspace{1ex}} \mathbbm{1}_{\mathcal{H}_1}(\gamma) \coloneqq 
    \begin{cases}
        1 \mbox{ if } \gamma \in \mathcal{H}_1\\ 
        0 \mbox{ if }  \gamma \in \mathcal{H}_0
    \end{cases} \nonumber\\
    \bm I^t_{\mathbb{D}} &= [\bm I^{t\top}_{\mathbb{X}}, \bm I^{t\top}_{\mathbb{Y}}]^\top \mbox{.}\nonumber 
\end{align}
Here, $\bm s_{\mathbb{X}}^t$ is the intensity of the speckle noise contained in region $\mathbb{X}$ for frame $t$, and $\bm s_{\mathbb{Y}}^t$ is the intensity of the speckle noise contained in region $\mathbb{Y}$ for frame $t$. The pixel intensities contained in $\mathbb{Y}$ are modeled as an additive combination of the speckle noise, $\bm s_{\mathbb{Y}}^t$, and a potential point source, $\bm h_{\mathbb{Y}}$. The term $\bm \eta$ includes any irreducible uncertainty in the data. Here, we denote $\gamma$ to be the sample statistic which is a function of the excess intensity in the region $\mathbb{Y}$. In this formulation, if the intensity from speckle noise is known perfectly, then detecting an on-sky point source is limited only by $\bm \eta$. To that end, post-processing algorithms seek an unbiased minimum error estimate, $\hat{\bm s}^t_{\mathbb{Y}}$. \PCA\hspace{-0.1ex}-based reconstruction approaches like \KLIP\hspace{0.3ex} estimate the weighting coefficients $\{\hat{\bm w}\}_{t = 1}^T$ of an optimal low-rank image reconstruction basis matrix $\hat{V}_q$ which together generate this estimate. Without explicitly partitioning the regions $\mathbb{X}$ and $\mathbb{Y}$, minimizing 
\begin{equation}
     \hat{V}_q, \{\hat{\bm w}^t\}_{t = 1}^T = \argmin_{V_q,  \{\bm w\}_{t = 1}^T} \sum_{t = 1}^T || \bm {I}^{t}_{\mathbb{D}} - V_q\bm w^{t} ||^2_2
     \label{eqn:general_ls}
\end{equation}
produces a minimum mean squared error reconstruction of the region $\mathbb{D}$ for each of the $t^{\mathrm{th}}$ frames. \citet{Hastie2001} show that Equation \ref{eqn:general_ls} is equivalent to
\begin{align}
    \hat{\bm w}^t &= V_q^T \bm I_{\mathbb{D}}^t\\
    \hat{V}_q &= \min_{V_q} \sum_{t = 1}^T || \bm I^{t}_{\mathbb{D}} - V_q V_q^T \bm I^{t}_{\mathbb{D}} ||^2_2\mbox{.} \label{eqn:PCA_cost}
\end{align}
The solution to Equation \eqref{eqn:PCA_cost} is found via singular value decomposition of the data matrix $X$, where,
\begin{align}
    \tilde{X} &= [\bm I_{\mathbb{D}}^1,\ldots,\bm I_{\mathbb{D}}^T]\\
    \tilde{X} &= UD\hat{V}^{\top}\mbox{.}
\end{align}
The matrix $U$ is an orthonormal matrix and the columns contain the left-singular vectors of the decomposition. The matrix $D$ is a diagonal matrix containing the singular values of the matrix $X$. The columns of $\hat{V}$ are the right singular vectors, or principal component vectors, which are ordered in the directions of maximum variance. $\hat{V}_q$ is obtained by truncating the columns of $\hat{V}$ up to a desired integer $q$. In \PCA--based reduction approaches, the regions $\mathbb{D}$ are sometimes partitioned into concentric annuli, sectors, or full \ADI\hspace{0.1ex} frames. Generating the speckle estimate for each region $\mathbb{D}$ consists of two symmetric operations:
\begin{enumerate}
    \item The intensity vector is mapped onto a low-rank \PCA\hspace{-0.1ex} feature space with $\hat{\bm w}^t = \hat{V}_q^\top \bm I_{\mathbb{D}}^t$.
    \item The \PCA\hspace{-0.1ex} features are mapped back into the original pixel space to produce an estimate of the speckle noise in region $\mathbb{D}$ with $\bm \hat{\bm s}_{\mathbb{D}}^t = \hat{V}_q \hat{ \bm w}^t$.
\end{enumerate}

In this work, we generalize $V_q$ and $V_q^T$ to be two nonlinear functions $\bm f(\cdot)$ and $\bm g(\cdot)$ parameterized by a neural network. To alleviate partial fitting to the potential source, we explicitly exclude the region $\mathbb{Y}$ in building the reconstruction. Minimizing the cost function
  
\begin{align}
\centering
     \min_{\bm f(\cdot)\mbox{, } \bm g(\cdot)} \Big (\sum_{i = 1}^N | \bm I^{i}_{\mathbb{D}} - \bm g(\bm f(\bm I^{i}_{\mathbb{X}})) |^p \Big )^{\frac{1}{p}} \label{eqn:inpainting_cost}
\end{align}
results in the optimal nonlinear reconstruction functions $\widehat{\bm f}(\cdot)$ and $\widehat{\bm g}(\cdot)$. Here, $p$ is a positive scalar, and $N$ is the number of training samples in a reference library, each indexed by an integer $i$. We detail the architecture and implementation of this algorithm in Section \ref{sec:inpainting_for_exoplanet_detection}. Note that in this formulation, the speckle noise prediction relies solely on spatial correlations exhibited between the predictor $\mathbb{X}$ and the region of interest, $\mathbb{Y}$.


\subsection{Recovering Off-Axis Point Sources}

During \ADI\hspace{0.3ex} observations, a telescope is placed in pupil tracking mode causing on-sky point sources to rotate deterministically around the optical axis of the instrument \citep{Marois2005}. Accordingly, pixel intensities in a region $\mathbb{Y}_*$ label a set of locations, $\{\mathbb{Y}_t\}_{t = 1}^T$, across the \ADI\hspace{0.3ex} sequence which will all contain a point source if it is contained in $\mathbb{Y}_*$. Note that we use ``$*$" to denote the first frame of the ADI sequence. Re-arranging Equation \ref{eqn:general_hypth_test}, the maximum likelihood estimate of a potential point source across the set of time-dependent locations, associated to $\mathbb{Y}_*$ is,

\begin{equation}
    \hat{ \bm h}_{\mathbb{Y}_*} = \mathbb{E}_t[\bm I_{\mathbb{Y}_t}^t - \bm \tilde{s}_{\mathbb{Y}_t}^t - \bm \epsilon]\mbox{.} \label{eqn:max_likelihood_h_estimate}
\end{equation}
Here, $\tilde{s}_{\mathbb{Y}_t}^t$ is the predicted speckle noise intensity for $\mathbb{Y}_t$ in the $t^{\mathrm{th}}$ frame, which is generated through our prediction function $\bm g \circ \bm f( \bm I_{\mathbb{X}_t}^t)$. The operator $\mathbb{E}_t(\cdot)$ denotes the statistical expectation taken with respect to time.

Constraining a hypothesized point source requires systematically comparing the estimated signal to its surrounding resolution elements. This is formalized in a hypothesis test where,
\begin{align}
    \mathcal{H}_0: \gamma &= 0 \mbox{,}\\
    \mathcal{H}_1: \gamma &> 0 \mbox{,}\nonumber
\end{align}
with $\hat{\gamma} = \| \hat{\bm h}_{\mathbb{Y}_*} \|_{\infty}$, where the operator $\|\cdot\|_{\infty}$ takes the maximum value of the expected source intensity vector. This is calculated in practice by placing a circular aperture around the potential source and retrieving the maximum value within the aperture's extent. We assume the test statistic $\hat{\gamma}$ is a Gaussian distributed random variable so that $\hat{\gamma} \sim \mathcal{N}(\gamma, \bar{\sigma})$, where $ \bar{\sigma}$ is the empirical \RMS\hspace{0.3ex} of the residual intensity in the surrounding elements. Note that we assume that residual pixel-wise intensities are independent and identically distributed (i.i.d). The optimal decision in the sense of minimizing the false negatives, subject to a fixed probability of false positives, is given by the log-likelihood ratio test,
\begin{equation}
    \Lambda(\mathcal{H}_1, \mathcal{H}_0) = \log \frac{p(\hat{\gamma} | \gamma_1, \bar{\sigma})}{p(\hat{\gamma} | 0, \bar{\sigma})}  \lessgtr \lambda\mbox{.}\label{eqn:GLRT}
\end{equation}
Because we are assuming Gaussian statistics, Equation \ref{eqn:GLRT} reduces to,
\begin{equation}
    \hat{\gamma} \stackunder{\stackon{$\gtrless$}{\tiny $\mathcal{H}_1$}}{\tiny $\mathcal{H}_0$} \tilde{\lambda} \bar{\sigma}\mbox{.}\label{eqn:GLRT_reduced}
\end{equation}
The scaling variable $\tilde{\lambda}$ is used to gate the hypothesis test as a multiplicative factor of the residual speckle \RMS. Due to spatial correlations in the residual speckle intensity, the samples are not i.i.d. A more statistically sound treatment of the residual speckle noise statistics is considered in \citet{Mawet2014}.

\section{Autoencoder Neural Network for Image Reconstruction} \label{sec:inpainting_for_exoplanet_detection}
\subsection{Reconstruction Function Architecture}

\begin{figure*}[htb!]
\epsscale{1.2}
\plotone{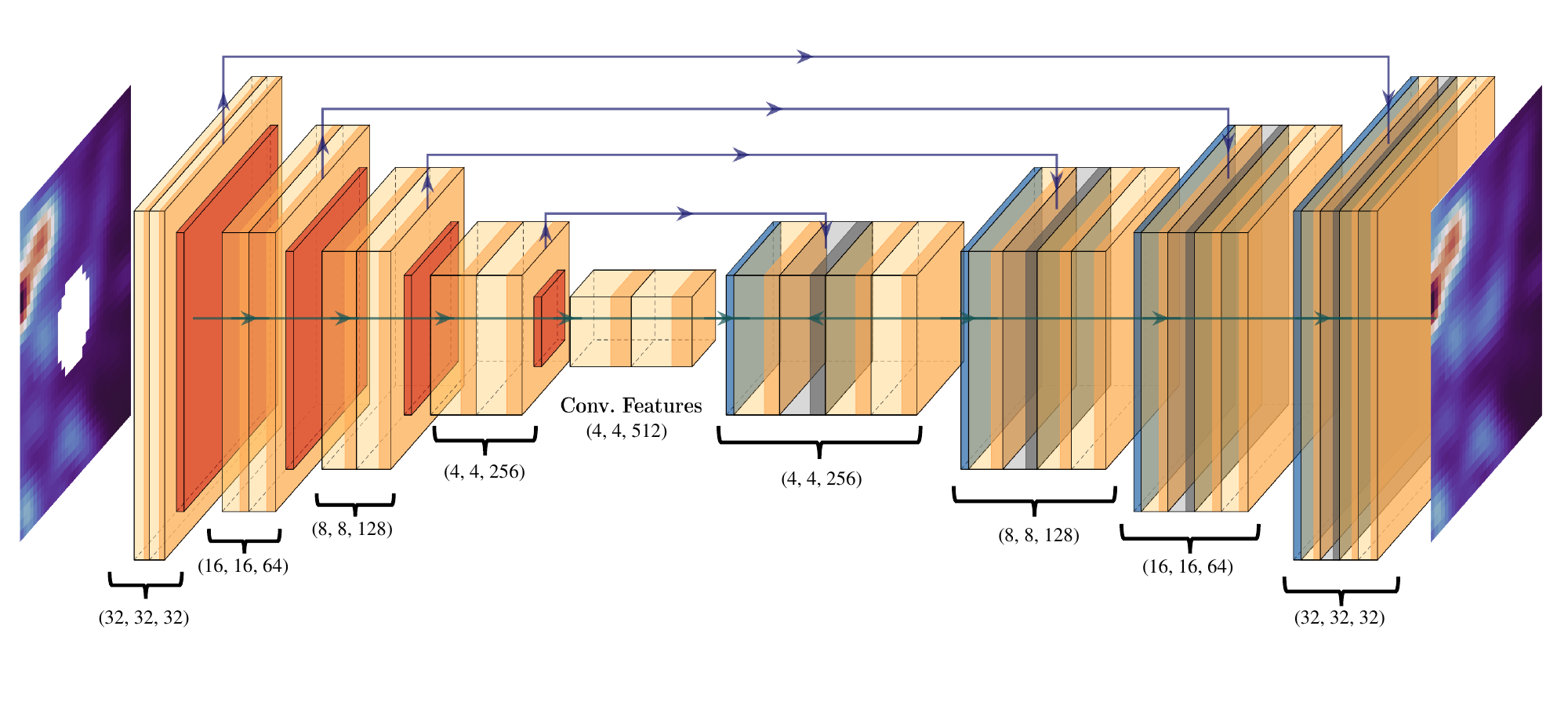}
\caption{Graphical illustration of the autoencoder architecture used in \texttt{ConStruct}. From left to right, an image region of interest is fed into the network, with the central region $\mathbb{Y}$ masked. In this example, the autoencoder is constructed such that it is compatible with a $32\times32$ pixel image patch. The encoder and decoder portions of the network act sequentially to first transform the masked image into low-dimensional feature space, then back into the original input space. By training the network with many examples, the network learns to accurately predict the masked central region. Here, we include the sizes for the output from each encoder and decoder block. \label{fig:NN_architecture}}
\end{figure*}
%

\begin{figure*}[htb!]
\epsscale{1.2}
\plotone{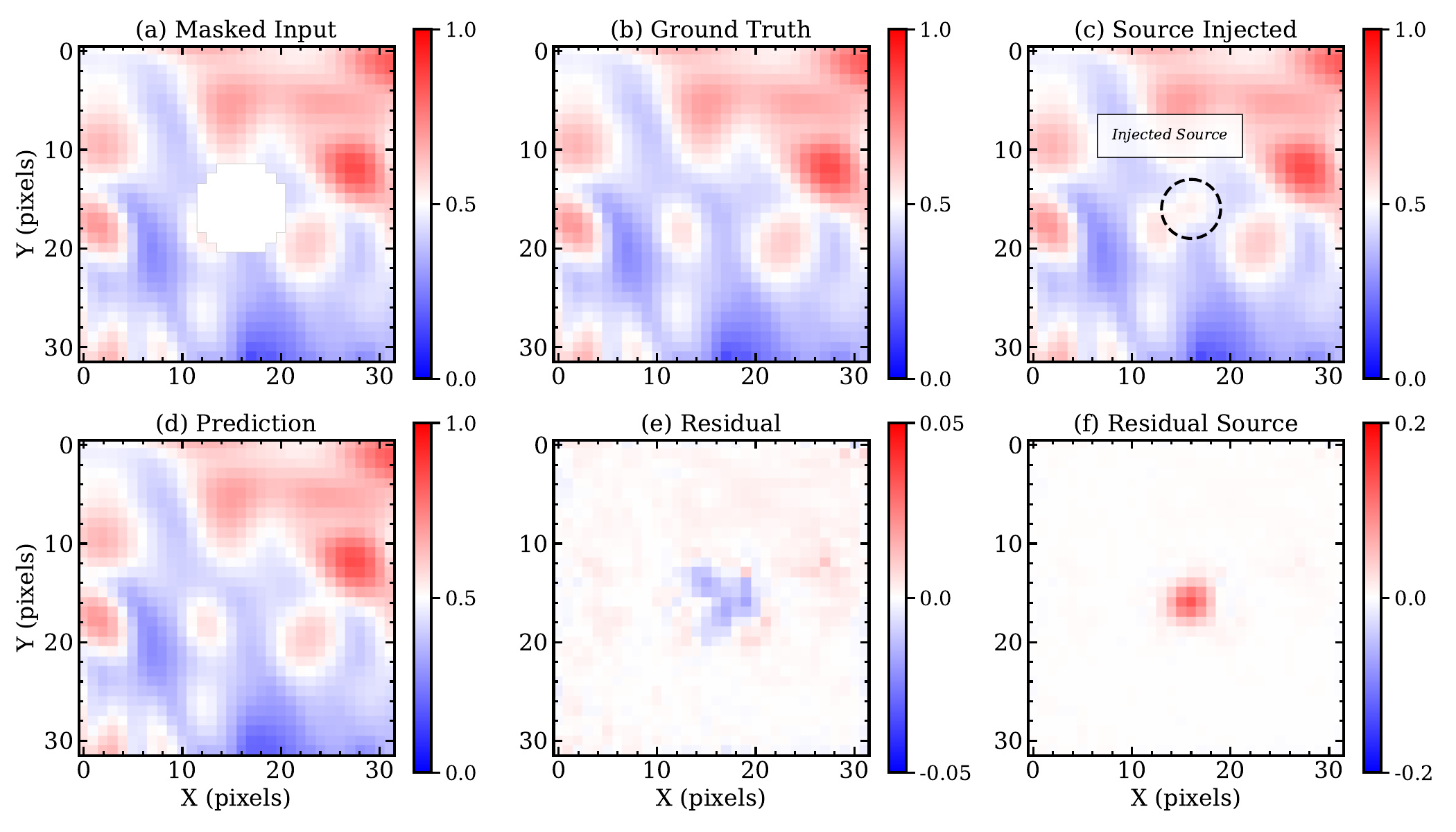}
\caption{Autoencoder prediction of the speckle noise intensity in a masked region of an ADI sequence. The image patch is extracted from a frame in Sequence 9 in Table \ref{tab:testing_data}. It is linearly scaled to values between zero and one before being fed into the autoencoder. Panels (a) and (b) are the masked autoencoder input and the ground-truth \ROI\hspace{0.3ex}, respectively. Panel (c) is the ground truth, with an injected Gaussian signal with a peak scaled intensity of 0.2. Panel (d) is the prediction formed by passing the masked input into our network. Panels (e) and (f) are the residual images produced by subtracting the predicted image region from the ground truth without and with the injected source, respectively. \texttt{ConStruct} can accurately reconstruct the ground truth over the mask, and recover the injected source. \label{fig:example_side_by_side}}
\end{figure*}

In this section, we detail how an autoencoder neural network serves as a surrogate for the functions $\bm f(\cdot)$ and $\bm g(\cdot)$ given in Equation \ref{eqn:inpainting_cost}. An autoencoder is a feed-forward neural network that reconstructs a corrupted input as its output. An encoder and decoder network portion act symmetrically to first transform a high-dimensional input into a low-dimensional feature space, and subsequently back into the original input space. \texttt{ConStruct} uses a fully-convolutional version of an autoencoder, similar to the U-net network \citep{Ronneberger2015}. Convolutional neural networks excel in image-based tasks, so this architecture choice is attractive for our application. We provide a graphical illustration of our network architecture in Figure \ref{fig:NN_architecture} and show the output sizes from each block in the network. 

The encoder portion of the network consists of a sequence of four convolutional blocks. Each block contains two convolutional layers, followed by a max-pooling layer. The convolutional layers create a set of feature maps by translating a $3\times3$ convolutional filter over the output of the previous layer and applying a \ReLU\hspace{0.3ex}. Unique filters correspond to each feature map. Through training, the network learns a set of convolutional filters that enable it to accurately predict a corrupted, or in our case, masked region of interest. The max-pooling layer applies the maximum operator over a $2\times2$ filter translated across the feature map. This operation downsizes each dimension of the feature map by one-half.

The decoder portion uses a set of low-dimensional convolutional features produced by the encoder to reconstruct a prediction of the speckle noise in the masked region of interest. The decoder contains four up-convolutional blocks. Similar to the encoder, each block contains two sequential convolutional layers, but these are followed by a deconvolutional layer that transforms many lower-dimensional features into one of higher dimension. The encoder features congruent to each decoder block are concatenated to the deconvolved layer, and the new set is passed into the next block. This technique, sometimes referred to as residual connections or skip connections, helps to alleviate the issue of vanishing gradients common in deep neural network architectures. The final layer in our architecture consists of a single convolutional layer with a sigmoid activation function which reconstructs a scaled speckle prediction in the masked region, along with a reconstruction of the original image region fed into the network\footnote{To facilitate stability in predicting the speckle region, all images used by \texttt{ConStruct} are scaled into the range of $[0, 1]$. The output from the autoencoder is re-scaled to produce the final speckle prediction.}. In Appendix \ref{sec:Appendix_D}, we include the layer-wise parameters for the autoencoder used by \texttt{ConStruct}.

Figure \ref{fig:example_side_by_side} shows an example of using the trained autoencoder to predict speckle noise in an \ADI\hspace{-0.1ex} image patch. In this example, an image patch is extracted from a frame contained in an ADI sequence of the HR 8799 system, corresponding to Sequence 9 in Table \ref{tab:testing_data}. A synthetic 2-D Gaussian source is injected into the center of the image patch. The central region is then masked before being fed into the network. \texttt{ConStruct} accurately predicts the speckle in the missing masked region, and this prediction is subtracted from the original image patch to reveal the synthetic source. In Appendix \ref{sec:Appendix_E}, we included two additional prediction examples extracted from spatially disparate realizations of the the speckle field.


\subsection{Data Selection}

A collection of representative data is needed to train our autoencoder. We select a library of archival \ADI\hspace{0.3ex} sequences observed with the Keck NIRC2 near-infrared camera for this task. The data are downloaded from the \KOA\hspace{-0.1ex}, along with accompanying calibration frames. In total, 92 unique \ADI\hspace{-0.1ex} sequences are used for training our algorithm. 

\KOA\hspace{0.3ex} contains a large number of \ADI\hspace{-0.1ex} sequences that we can potentially include for training. We downselect our sample to only include $K_{\text{s}}$-, $K_{\text{p}}$-, $CH_4 S$-, $Y$-, and $H$-band filters. This choice has repercussions on which datasets we can deploy our trained algorithm: because the spatial frequency of speckle noise is wavelength dependent \citep{Sparks2002}, we expect our algorithm to perform best when used on \ADI\hspace{0.3ex} sequences taken with the same filters. Additionally, only data using the 400, 600, and 800 mas-diameter Lyot coronagraphs are used. The sequences were searched on the archive primarily using a list of stars with directly imaged planets and low-mass brown dwarf companions contained in \citet{Bowler2016}. Most training sequences have known point sources, but we assume that the number density of the point sources is sufficiently small across these data sets so as to not significantly bias the speckle realizations. Each sequence is visually inspected to determine its quality for training. Sequences containing saturated pixels or degraded Strehl ratios were excluded. The data used for training are detailed in Appendix \ref{sec:Appendix_A}.

\subsection{Processing the Training Dataset}\label{sec:training_processing}

Each of the ADI sequences used for training our network are pre-processed using standard procedures for HCI. Using dome flat calibration frames, we generate a bad-pixel mask for each sequence with the {\fontfamily{lmtt}\selectfont ccdproc Astropy} module \citep{Craig2017}. The mask is applied to each calibration and science frame, and bad pixels are removed with nearest-neighbor interpolation. Dark subtraction and flat fielding are carried out for each frame in the science sequence. Registration is performed by centering each frame at the pixel coordinate closest central stellar \PSF\hspace{-1.0ex}, visible inside the occulting coronagraph. 

\begin{figure}[t!]
\epsscale{1.2}
\plotone{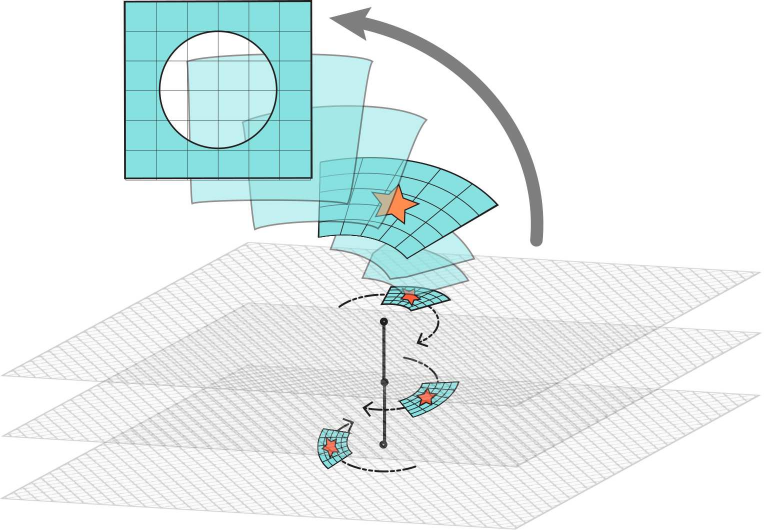}
\caption{Graphical illustration of how sector regions are extracted and processed by \texttt{ConStruct} on \ADI\hspace{0.3ex} sequences. Each plane represents a frame in the \ADI\hspace{-0.1ex} sequence. Sectors are extracted from each frame, and geometrically transformed into square arrays for processing. The center of each array is masked before being fed into the autoencoder.  \label{fig:extraction_diagram}}
\end{figure}

In generating training data, we draw an ADI sequence from our library by sampling a distribution who's probability of selection is proportional to the number of frames in the respective sequence. Patch coordinates in the sequence are then uniformly sampled in azimuth, radial separation, and frame number. These define the central coordinate of a sector which is projected into a square array for training. Figure \ref{fig:extraction_diagram} illustrates the projection operation. An angular and radial width defines the dimension of each sector. Depending on the radial coordinate, we choose the angular width such that the square projection approximately maintains the number of pixels in the sector. The projection uses cubic spline interpolation. We find the information degradation resulting from this transformation is sufficiently small to not affect the performance of \texttt{ConStruct}.

The pixel values in each projected square array are linearly scaled between zero and one. A circular mask of zeros is applied to each of these scaled samples, corresponding to the region $\mathbb{Y}$. The original array, $\bm y$, and the masked sample, $\bm x$, together form a training sample $(\bm x, \bm y)$. We use 30,000 samples to train the network. The complete set is partitioned into a $90/10\%$ train-validation split. The prediction accuracy of the neural network is measured after each training epoch by testing it on the validation set. This is useful for imposing a stopping criterion during training: when the loss in the validation set begins to increase (i.e., the prediction accuracy decreases), the algorithm is over-fitting the training set, and training should terminate.

\subsection{Autoencoder Implementation and Training}

We implement our autoencoder network in \texttt{Keras} \citep{Chollet2015}, which is an application program interface for the \texttt{TensorFlow} machine learning library \citep{Abadi2015}. In training, the component $\bm x$ of each sample is first propagated forward through the network, and the network's prediction is subtracted from the ground truth, $\bm y$, to form a residual array. Repeatedly backpropagating the residual errors adjusts the autoencoder parameters such that the prediction accuracy in the masked region increases. We use a 32-sample mini-batch gradient descent with the \textit{Adam} optimizer \citep{Kingma2015} and a Huber loss function given as, 

\begin{equation}
    \mathcal{L}_\epsilon = \sum_{i = 1}^{N_{\text{batch}}} \mathcal{L}_\epsilon^i(\boldsymbol{x}^i, \boldsymbol{y}^i)  
\end{equation}
where,
\begin{equation}
    \mathcal{L}_{\epsilon}^i(\boldsymbol{x}^i, \boldsymbol{y}^i) =  \begin{cases}
        \frac{1}{2}\|\boldsymbol{y}^i - \boldsymbol{g} \circ \boldsymbol{f}(\boldsymbol{x}^i)\|_2 \hspace{1ex} \mbox{if}\\ \hspace{5ex} \|\boldsymbol{y}^i - \boldsymbol{g} \circ \boldsymbol{f}(\boldsymbol{x}^i)\|_1 \leq \epsilon \mbox{,}\\ 
        \epsilon \|\boldsymbol{y}^i - \boldsymbol{g} \circ \boldsymbol{f}(\boldsymbol{x}^i)\|_1  - \frac{1}{2}\epsilon^2 \hspace{1ex} \mbox{otherwise.}
    \end{cases} \nonumber\\
\end{equation}

The Huber loss uses the minimum squared error when deviations are small (i.e., within some defined deviation $\epsilon$), and minimum absolute error when deviations are large. This reduces the importance placed on outlier samples in the training process. 30 training epochs are used, and we observe that the errors in the validation set consistently decrease with all epochs. For our application, more training epochs can likely be used, but with only marginal improvement in the prediction accuracy.

\subsection{Correcting the Autoencoder Prediction Residuals}

We leverage the temporal correlation of speckle noise across \ADI\hspace{-0.1ex} frames to further correct the residual errors of the autoencoder predictions with an L2 regularized linear regression model, otherwise known as ridge regression. In our formulation, we use the output from the autoencoder as the predictor variables for the model. For a fixed spatial region $\mathbb{D}$ in the $t^{\text{th}}$ frame partitioned into sub-regions $\mathbb{X}$ and $\mathbb{Y}$, the speckle prediction from the autoencoder is $\tilde{\bm s}^t_{\mathbb{D}} = \hat{\bm g} \circ \hat{\bm f}(\bm I^t_\mathbb{X})$. The prediction residual in the region $\mathbb{Y}$ is an additive combination of the residual speckle noise, and a potential point source, 
\begin{align}
    \bm \zeta_{\mathbb{Y}}^t &= \bm I_{\mathbb{Y}}^t -  \tilde{\bm s}^t_{\mathbb{Y}}\\\nonumber
    &= \mathbbm{1}_{\mathcal{H}_1}(\gamma) \bm h_\mathbb{Y} + \bm \eta_\mathbb{Y}^t\mbox{.}
\end{align}

\begin{figure}[htb!]
\centering
\epsscale{1.2}
\plotone{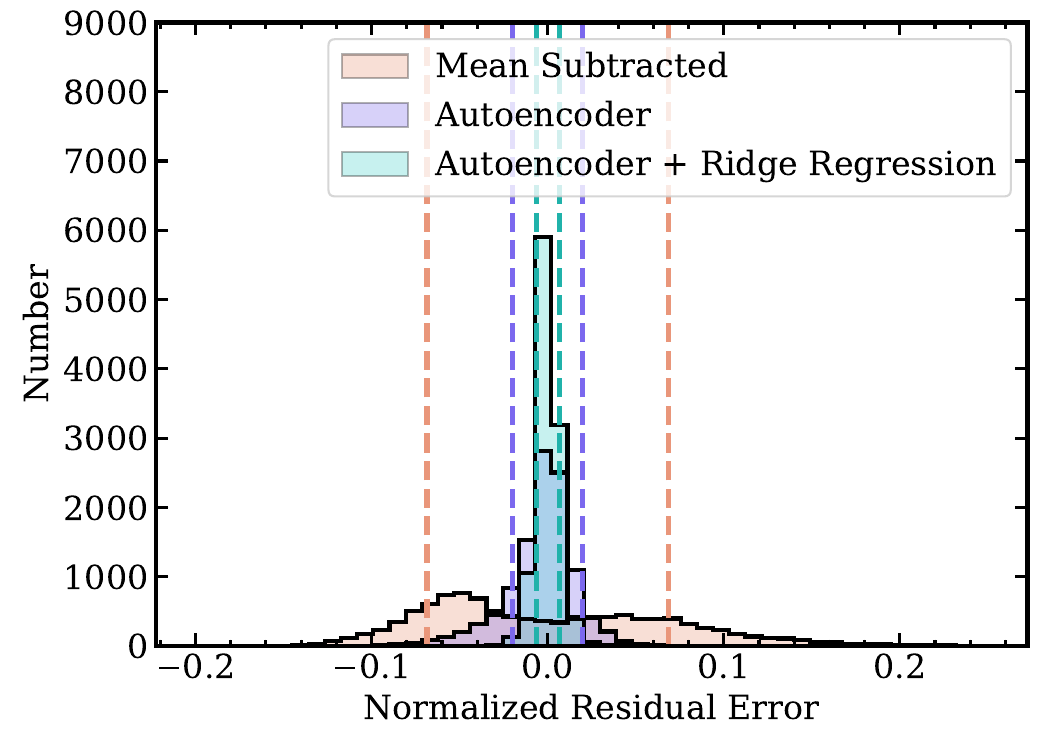}
\caption{Pixel-wise speckle prediction residuals in a fixed spatial patch, $\mathbb{Y}$, accumulated over all frames in an \ADI\hspace{-0.1ex} sequence. Each of the samples is first linearly scaled between zero and one. The orange distributions are residuals after subtracting the time average pixel-wise intensity from each frame, the purple distributions are the autoencoder prediction residuals, and the green distributions are the prediction residuals augmented with an additional linear regression. The vertical dashed lines indicate the $\pm 1 \sigma$ bounds on the residuals. The autoencoder prediction and regularized least-squares regression in \texttt{ConStruct} progressively improves the prediction accuracy, yielding the most accurate speckle prediction, shown in the green distribution. \label{fig:residuals_histogram}}
\end{figure}

\begin{figure}[htb!]
\centering
\epsscale{1.2}
\plotone{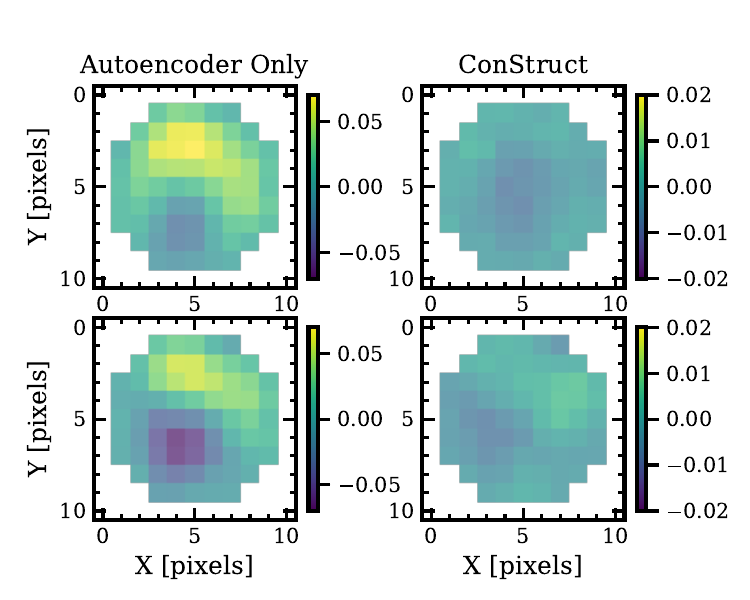}
\caption{Two examples of the residual speckle patches in the region of interest $\mathbb{Y}$ produced (left) with the autoencoder only and (right) with the autoencoder augmented with the regularized least-squares regression step. Both of the samples are extracted from the same spatial location, but at different frames in the ADI sequence. Each of the samples is first linearly scaled between zero and one. The autoencoder prediction removes most of the speckle noise in the prediction region. The bias in the autoencoder prediction is removed with the linear regression step.\label{fig:autoencoder_vs_ridge}}
\end{figure}

For each pixel in the region $\mathbb{Y}$, we solve a regularized linear regression to correct the residuals. Let $\boldsymbol{z}_j = [{}^j\zeta_\mathbb{Y}^1, {}^j\zeta_\mathbb{Y}^2, \cdots,  {}^j\zeta_\mathbb{Y}^T]^\top$ be the response variable in the regression model, where ${}^j\zeta_\mathbb{Y}^t$ is the residual intensity for the pixel indexed by the pre-superscript $j$ in the $t^{\text{th}}$ frame. Each regression model finds the coefficients $\boldsymbol{\beta}_j$ that minimize the loss function
\begin{equation}
    \mathcal{L}_j = (\boldsymbol{z}_j - \tilde{X}\boldsymbol{\beta}_j)^\top(\boldsymbol{z}_j - \tilde{X}\boldsymbol{\beta}_j) + \alpha \boldsymbol{\beta}_j^{\top}\boldsymbol{\beta}_j\mbox{,}\label{eqn:ridge_loss}
\end{equation} 
which has the analytic solution,
\begin{equation}
    \hat{\boldsymbol{\beta}}_j = (\tilde{X}^\top\tilde{X} + \alpha I_{T \times T})^{-1}X^\top \boldsymbol{z}_j\mbox{.}
\end{equation}
Here, $I_{T\times T}$ is the $T$-dimensional identity matrix. $\alpha$ is a positive scalar which acts as a tuning variable in \texttt{ConStruct}. Higher values of $\alpha$ increase the bias of the regression model, but help reduce overfitting to potential companion point sources.  $\tilde{X}$ is the design matrix, which are the horizontally stacked predictions produced by the autoencoder, 

\begin{equation}
    \tilde{X} = 
    \begin{bmatrix}
    \tilde{\boldsymbol{s}}_\mathbb{D}^{1\top}\\
    \tilde{\boldsymbol{s}}_\mathbb{D}^{2\top}\\
    \vdots\\
    \tilde{\boldsymbol{s}}_\mathbb{D}^{T\top}\\
    \end{bmatrix}\mbox{.}
\end{equation}
This is repeated for all pixels contained in region $\mathbb{Y}$. The output of the regression step in \texttt{ConStruct} are the corrected residual intensities
\begin{equation}
    \delta Z =
    \begin{bmatrix}
        \delta \boldsymbol{z}_1^\top\\
        \delta \boldsymbol{z}_2^\top\\
        \vdots\\
        \delta \boldsymbol{z}_J^\top \mbox{,}
    \end{bmatrix} = 
    \begin{bmatrix}
        {}^1 \delta \zeta^1 & \cdots & {}^1 \delta \zeta^T\\
        \vdots & \ddots & \vdots\\
        {}^J \delta \zeta^1 & \cdots & {}^J \delta \zeta^T
    \end{bmatrix}
\end{equation}
where,
\begin{equation}
    \delta \boldsymbol{z}_j = \boldsymbol{z}_j -  \tilde{X} \hat{\boldsymbol{\beta}}_j
\end{equation}
Figure \ref{fig:residuals_histogram} illustrates how the autoencoder prediction and least-squares regression step sequentially reduces the residual speckle noise. We also include two examples showing the speckle residuals in the region $\mathbb{Y}$ from the autoencoder without and with the least-squares regression in Figure \ref{fig:autoencoder_vs_ridge}. The estimated intensity of the potential companion point source in the region $\mathbb{Y}$ in each frame $t$ is then,
\begin{equation}
    \bm \hat{h}^t_{\mathbb{Y}} = \text{col}_t( \delta Z)\mbox{,}
\end{equation}
where $\text{col}_t(\cdot)$ denotes the $t^{\text{th}}$ column of the matrix.

\subsection{Using \texttt{ConStruct} with ADI sequences}

\texttt{ConStruct} uses a flexible architecture that provides sub-pixel processing capabilities with \ADI\hspace{0.3ex} sequences. Based on a user-defined resolution, and lower and upper radial processing bounds, radial and azimuthal coordinates, $(r_*, \theta_*)$, are defined in the image plane that satisfy the resolution requirement. Here, we use the subscript $``*"$ to denote that these coordinates are located in the first frame of the ADI sequence and define the plane of the final flux map, $\mathcal{F}$, and S/N map, $\mathcal{S}$. Each coordinate $(r_*, \theta_*)$ labels a set of spatial patches $\{\mathbb{Y}_t\}_{t = 1}^T$ which are centered at the set of coordinates $\{(r_*, \theta_* + \phi_t)\}_{t = 1}^T$, where $\phi_t$ is the field rotation observed up to and including frame $t$. Post-processing is performed serially over radial resolution bins, and for each bin, the following steps are implemented:

\begin{enumerate}
    \item Each azimuthal and time coordinate defines the center of a sector area in the radial bin, and the sector is extracted and mapped to a square array, as illustrated in Figure \ref{fig:extraction_diagram}. This step is parallelized for all elements $(\theta_*, t)$ in the radial bin spanning the ADI sequence.
    
    \item The extracted square arrays are each independently scaled (i.e., a unique scaling function is fit to each spatial patch) between zero and one, and the central pixel values corresponding $\mathbb{Y}$ are masked with a circular template of zeros.
    
    \item The masked arrays are fed as a batch into the autoencoder network that allows for GPU parallelization. The autoencoder reconstructs the missing regions in each patch using solely the spatial correlations with the surrounding speckle noise.
    
    \item The predictions produced by the autoencoder are re-scaled by applying the inverse of the scaling operator fit to each patch in Step \#2.
    
    \item The re-scaled predictions are subtracted from the true image patches, creating a set of residual images. The residual images corresponding to a fixed azimuthal coordinate but spanning all frames, $\{\zeta^t_{\mathbb{Y}_*}\}_{t = 1}^T$, are further corrected with ridge regression described above. This is repeated for all azimuthal elements $\theta_*$ in the radial bin.
    
    \item The corrected residuals are formed into sets that may each contain a point source, based on a label azimuthal coordinate, and the parallactic rotation of the \ADI\hspace{0.3ex} sequence. Each of these sets are denoted as $\{\hat{\bm h}^t_{\mathbb{Y}_t}\}_{t = 1}^{T}$. This is repeated for all azimuthal coordinates $\theta_*$ in the radial bin.
    
    \item The images in the set $\{\hat{\bm h}^t_{\mathbb{Y}_t}\}_{t = 1}^{T}$ are averaged, and the center pixel value of the averaged array is used for the intensity of the coordinate $(r_*, \theta_*)$ in the final flux map, $\mathcal{F}$. This is repeated for all azimuthal coordinates $\theta_*$ in the radial bin.
    
\end{enumerate}

The intensity values of $\mathcal{F}$ contained within a radial window of $\pm 3$ pixels of $r_*$ are used to calculate the emperical RMS of each radial bin. The S/N values of $\mathcal{S}$ are then
\begin{equation}
    \mathcal{S}(r_*, \theta_*) = \frac{\mathcal{F}(r_*, \theta_*)}{\bar{\sigma}_{r_*}}\mbox{,}
\end{equation}
where $\bar{\sigma}$ denotes the emperical RMS.

%
%

\startlongtable
\begin{deluxetable*}{ccccccccccc}
\movetableright = 0.0in
\tabletypesize{\footnotesize}
\tablewidth{0pt}

\tablecaption{Keck/NIRC2 \ADI\hspace{-0.1ex} sequences used for the performance analysis of \texttt{ConStruct}. \label{tab:testing_data}}

\tablehead{
\colhead{Sequence \#} & \colhead{Companion} & \colhead{Date (UT)} & \colhead{\# of Frames} & \colhead{Int. Time (s)} & \colhead{Filter} & \colhead{Corona. (mas)} & \colhead{PI} & \colhead{S/N \texttt{ConStruct}} & \colhead{S/N PCA}}
\startdata 
1& HD 4747 B&	2015-01-09&	39&	2340& $K_\text{s}$& 600& Knutson& 10.9 & 15.3\\
2& HD 19467 B&	2011-08-30&	50&	1250&	$K_\text{p}$& 300& Crepp& 31.2 & 83.9\\
3& HD 19467 B&	2012-08-25&	53&	1325&	$K_\text{p}$& 300& Johnson& 11.4 & 14.6\\
4& HD 114174 B&	2011-02-22&	36&	720&	$K_\text{p}$& 300& Crepp& 25.6 & 19.3\\
5& HD 114174 B&	2012-02-02&	91&	1820&	$K_\text{p}$& 300& Knutson& 37.7& 24.6\\
6& HR 7672 B&	2011-05-15&	30&	450&	$K_\text{p}$& 300& Carpenter& 12.6& 18.7\\
7& Kappa And b&	2013-05-30&	38&	1520&	$K_\text{p}$& 400& Carpenter& 34.4 & 78.8\\
8& Kappa And b&	2018-01-30&	15&	450&	$K_\text{s}$& 600& Bowler& 11.1 & 6.7\\
9& HR 8799 b &	2010-07-13&	70&	1400&	$K_\text{s}$& 400& Barman& 61.6 & 57.4\\
9& HR 8799 c&--&--&--&--&--&--&48.0 & 36.5\\
9& HR 8799 d&--&--&--&--&--&--&15.0 &9.3\\
9& HR 8799 e&--&--&--&--&--&--&2.3 & 2.7\\
%
10& HR 8799 b& 2011-07-21&	148& 2960&	$K_\text{s}$& 400& Macintosh& 98.8 & 77.9\\
10& HR 8799 c&--&--&--&--&--&--& 66.7 & 51.4\\
10& HR 8799 d&--&--&--&--&--&--& 30.2 & 18.7\\
10& HR 8799 e&--&--&--&--&--&--& 5.7 & 3.0\\
%
11& HR 8799 b& 2012-10-26&	98&	1960&	$K_\text{s}$& 400& Cooray & 81.8 & 54.3\\
11& HR 8799 c&--&--&--&--&--&--& 52.8 & 25.5\\
11& HR 8799 d&--&--&--&--&--&--& 21.5 & 11.5\\
11& HR 8799 e&--&--&--&--&--&--& 5.0 & 4.1\\
%
12& HR 8799 b&	2012-07-22&	133& 3325&	$K_\text{s}$& 400& Macintosh& 67.2 & 60.5\\
12& HR 8799 c&--&--&--&--&--&--& 51.7 & 30.1\\
12& HR 8799 d&--&--&--&--&--&--& 25.0 & 9.2\\
12& HR 8799 e&--&--&--&--&--&--& 5.0 & 3.7\\
13& Gl 758 B&	2013-07-03&	28&	840&	$K_\text{s}$& 600& Crepp& 4.2& 10.2\\
14& Kappa And b&	2013-06-22&	22&	550&	$K_\text{p}$& 400& Hinkley& 26.9 & 60.4\\
15& HD 4747 B&	2012-08-25&	159& 3816&	$K_\text{p}$& 300& Johnson& 19.5& 17.8\\
16& HD 4747 B&	2015-01-09&	39&	2340&	$K_\text{s}$& 600& Knutson& 7.7& 14.1\\
17& HD 8375 B&	2010-10-13&	60&	1086&	$H$& 300&	      Crepp& 32.2& 25.1\\
18& HD 114174 B&	2012-05-29&	61&	1220&	$K_\text{p}$& 300& Crepp& 13.8& 15.8
\enddata
\begin{tablenotes}[flushright, para]\footnotesize
    \textbf{Sequence References.} (1, 16) \cite{Crepp_2016}, (2, 3) \cite{Crepp_2014}, (4, 5, 18) \cite{Crepp_2013}, (6) \cite{Crepp_2012}, (9, 10, 11, 12) \cite{Konopacky2016}, (13) \cite{Bowler_2018}, (17) \cite{Crepp_2013a}.
\end{tablenotes}
\end{deluxetable*}

%
%

%
%

\section{Algorithm Performance and Tuning} \label{sec:algorithm_performance_and_tuning}

\begin{figure*}[hbt!]
  \centering
  \centering \begin{tabular}[b]{@{}p{0.42\textwidth}@{}}
    \includegraphics[width=1.0\linewidth]{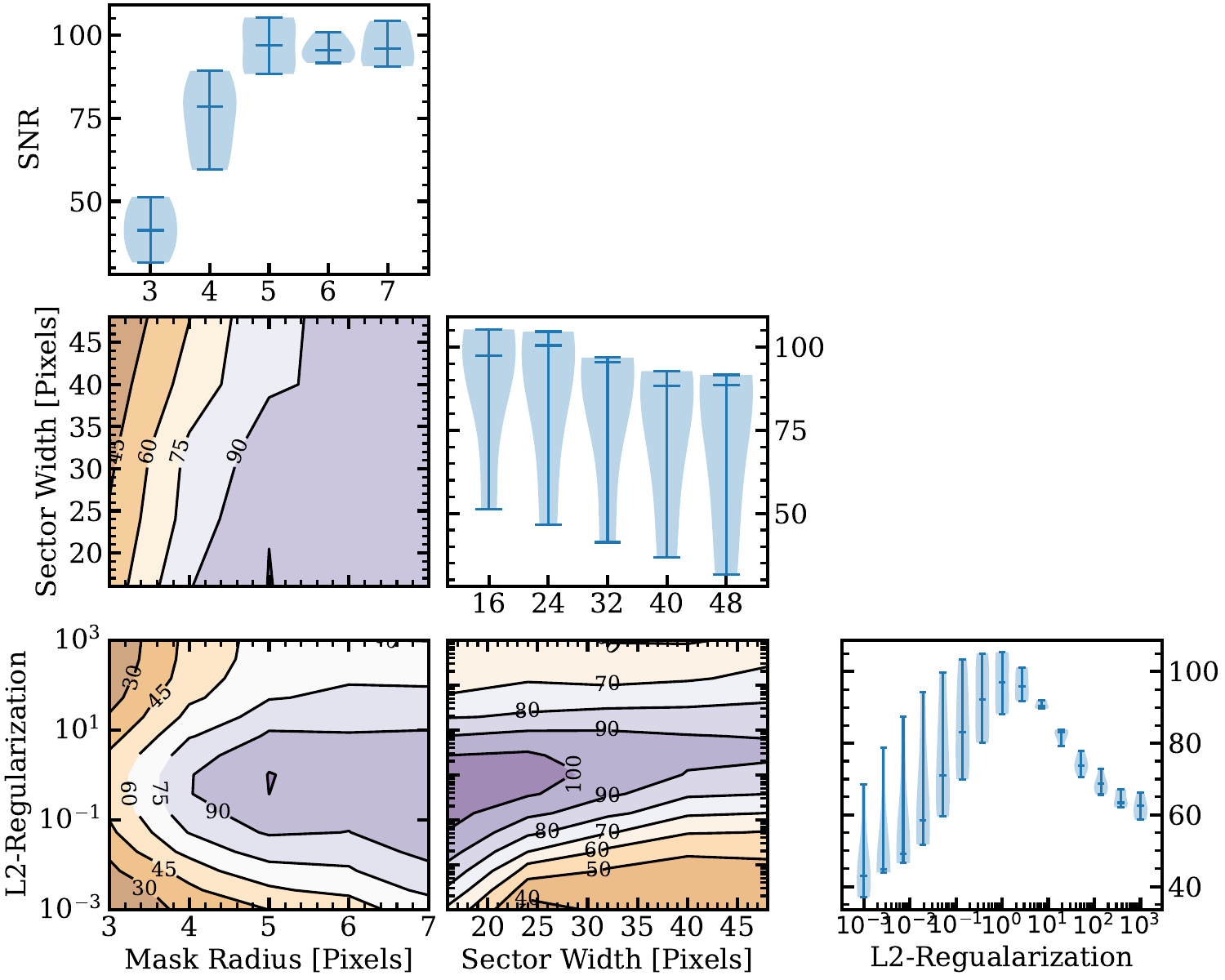} \\
    \centering \small (a) HR 8799 b
  \end{tabular}%
  \quad
  \begin{tabular}[b]{@{}p{0.42\textwidth}@{}}
    \includegraphics[width=1.0\linewidth]{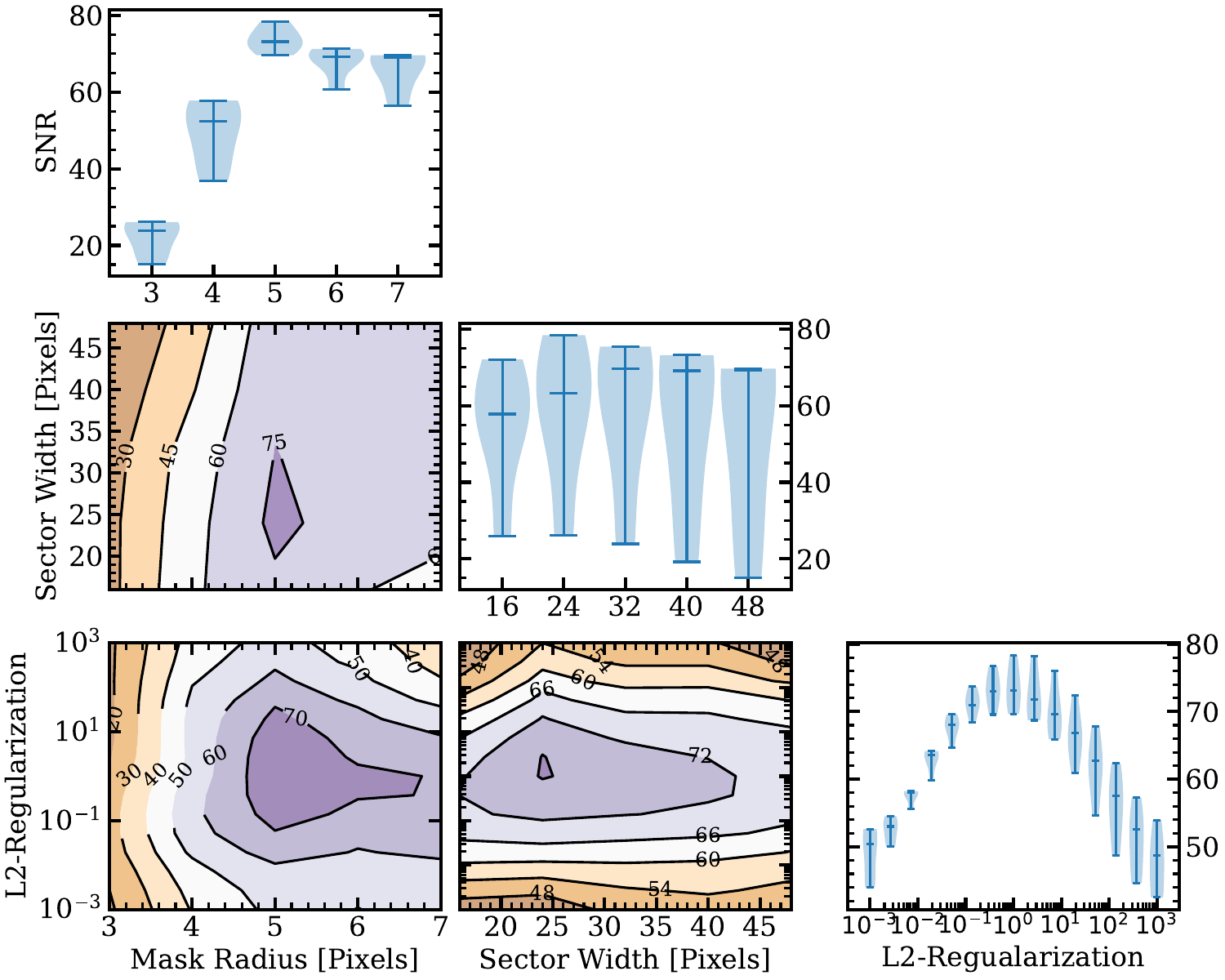} \\
    \centering \small (b) HR 8799 c
  \end{tabular}
  \begin{tabular}[b]{@{}p{0.42\textwidth}@{}}
    \vspace{2mm}\hspace{5mm}
    \includegraphics[width=1.0\linewidth]{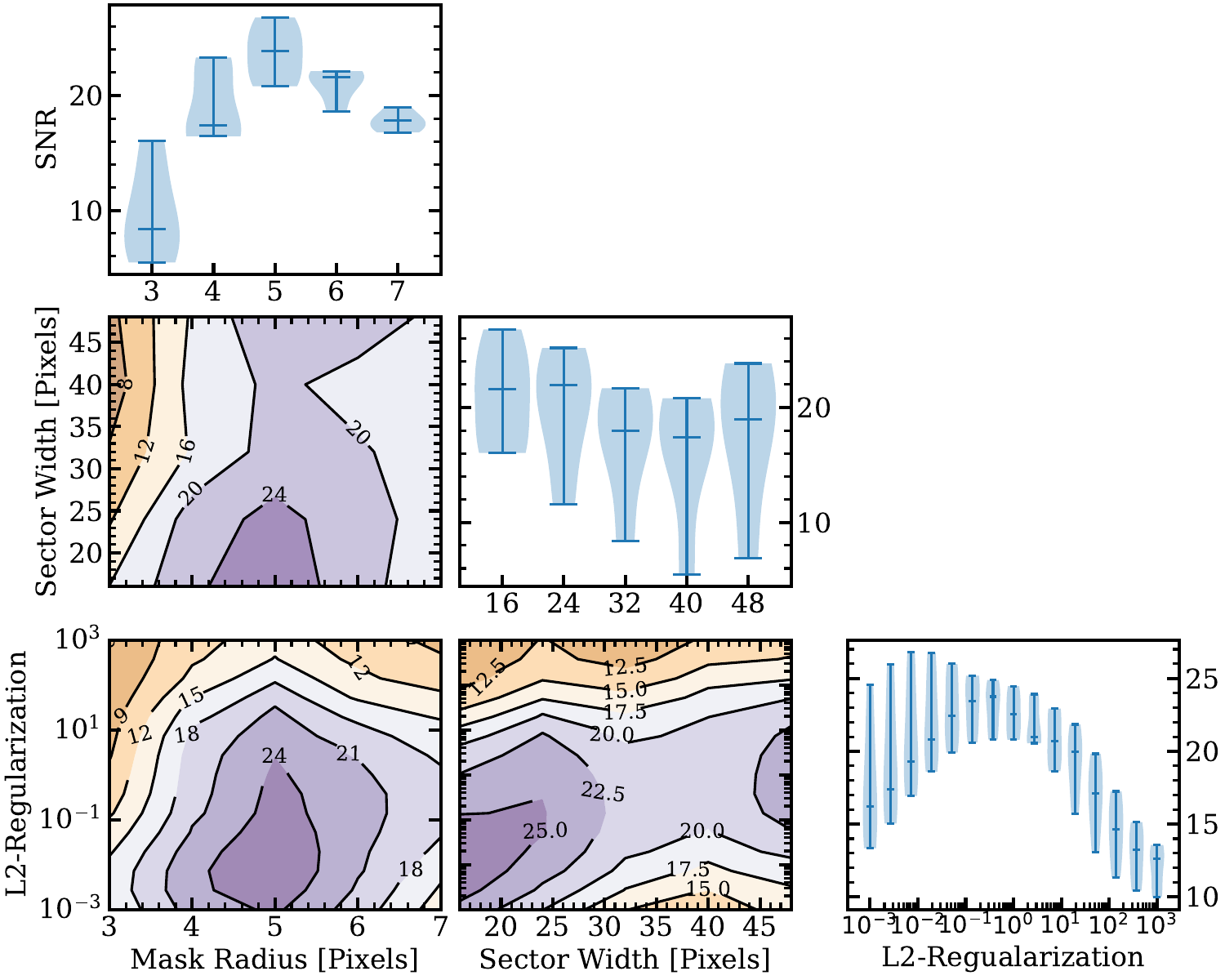} \\
    \centering \small \hspace{12mm} (c) HR 8799 d
  \end{tabular}%
  \quad
  \begin{tabular}[b]{@{}p{0.42\textwidth}@{}}
    \hspace{6mm}\includegraphics[width=1.0\linewidth]{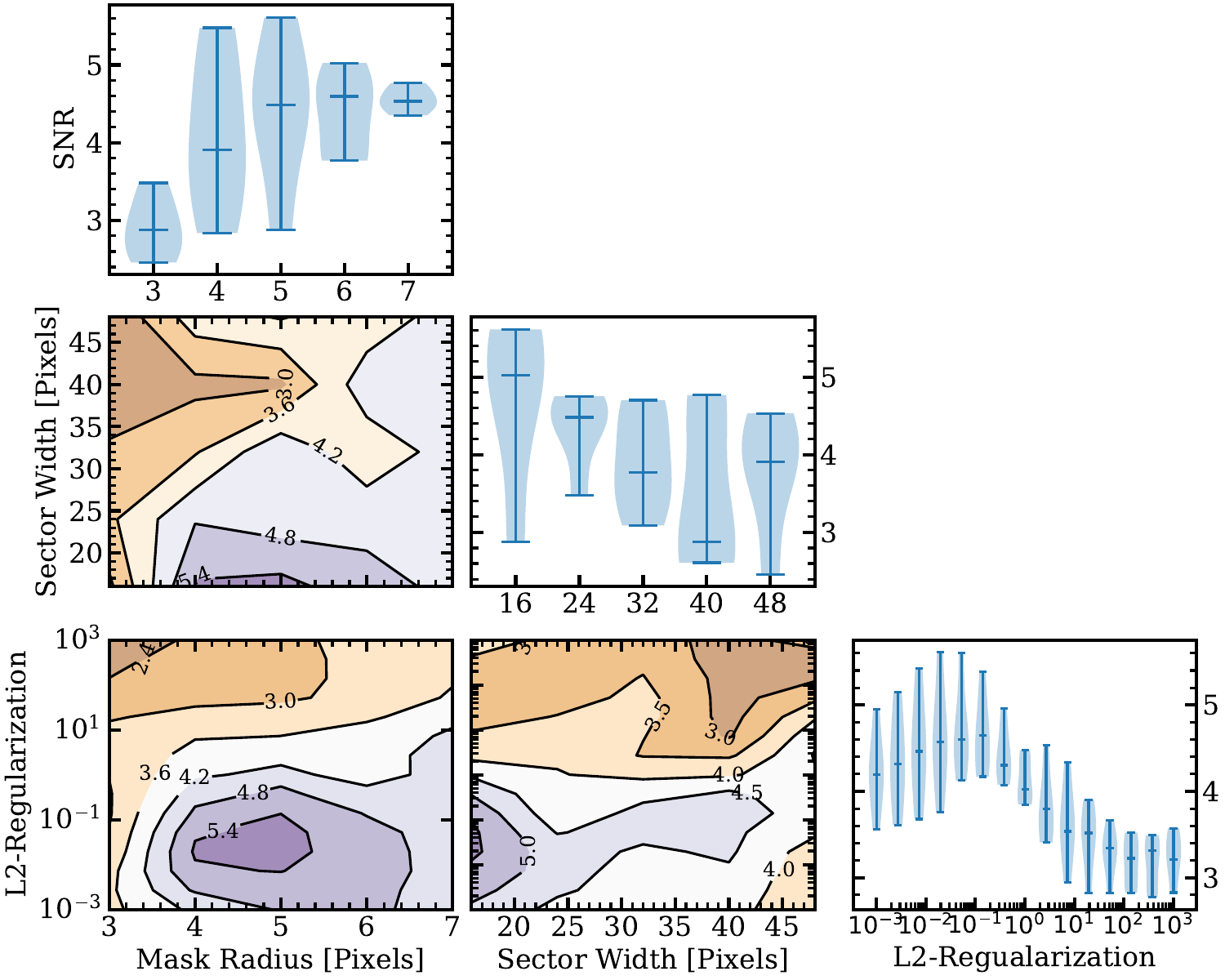} \\
    \centering \small \hspace{12mm} (d) HR 8799 e
  \end{tabular}
  \caption{Marginalized representations for our parametric parameter search for the $2011-07-21$ epoch of HR 8799. Each contour plot shows the S/N values projected onto two of the variables. The projection operation takes the maximum S/N along the axis of the marginalized variable. The violin plots show the spread in the S/N in each variable bin. Panels (a), (b), (c), and (d) correspond to HR 8799 b, c, d, and e, respectively.}
\label{fig:corner_plot_figs_20}
\end{figure*}

In this section, we optimize \texttt{ConStruct} for detecting substellar companions in \ADI\hspace{-0.1ex} sequences from NIRC2. We tune \texttt{ConStruct} over three selected algorithm parameters. NIRC2 \ADI\hspace{-0.1ex} sequences of the HR 8799 system are used to tune our algorithm on the four known planetary companions in these data sets. This procedure allows us to select a particular parameter set for applying \texttt{ConStruct} on other \ADI\hspace{-0.1ex} sequences. The goal is not to determine a universally optimal set of parameters, but rather to establish a reasonable set of parameter values that can then be applied to other data sets.

\subsection{Selecting Design Parameters for \texttt{ConStruct}}

We tune three design parameters in \texttt{ConStruct}: 1) the radius of the masked region, $\mathbb{Y}$, 2) the size of the sector used for the autoencoder prediction, $\mathbb{D}$, and 3) the regularization parameter, $\alpha$, in the linear correction model. Three HR 8799 \ADI\hspace{0.3ex} sequences from the Keck NIRC2 instrument serve as a test-bed for adjusting these design parameters. The objective is to find parameters that recover all four known planets in this sequence and maximize their S/N. These sequences were originally published in \citet{Konopacky2016}, and the details of the observations are summarized in Table \ref{tab:testing_data} in Sequences 9, 10, and 11.

We run \texttt{ConStruct} over a combinatorial grid of the design parameters at the locations of the four planets for each data set. The regularization is iterated uniformly in log-space between $10^{-3}$ and $10^{3}$, with 15 increments. The mask radius is tested at values of  4, 5, 6, and 7 pixels. The radial and azimuthal widths of the image sector patches are adjusted from 16, 24, 32, and 40 pixels. In total, this produces 240 unique parameter combinations. We determine the S/N of each companion for design parameters, which creates a 3D grid with each axis corresponding to a design variable. Figure \ref{fig:corner_plot_figs_20}, show slices of the 3-D S/N grid of each planet in Sequence 10 in Table \ref{tab:testing_data}. 

We perform the same analysis for all three HR 8799 data sets. These data have similar observation parameters (i.e., number of frames, field rotation, filter type), and we include the results for the other two sequences in Appendix \ref{sec:Appendix_C}.  From this study, we determine parameters that work well but are not necessarily optimal, for all planets across the three datasets in our sample. In deploying \texttt{ConStruct} on other datasets, we choose a sector size of 24 pixels, a mask radius of 5 pixels, and an L2 ridge regression regularization of 0.1.

\begin{figure}
  \centering
  \begin{tabular}[b]{@{}p{0.45\textwidth}@{}}
    \centering \hspace{-1.2cm} \includegraphics[clip, width=.9\linewidth]{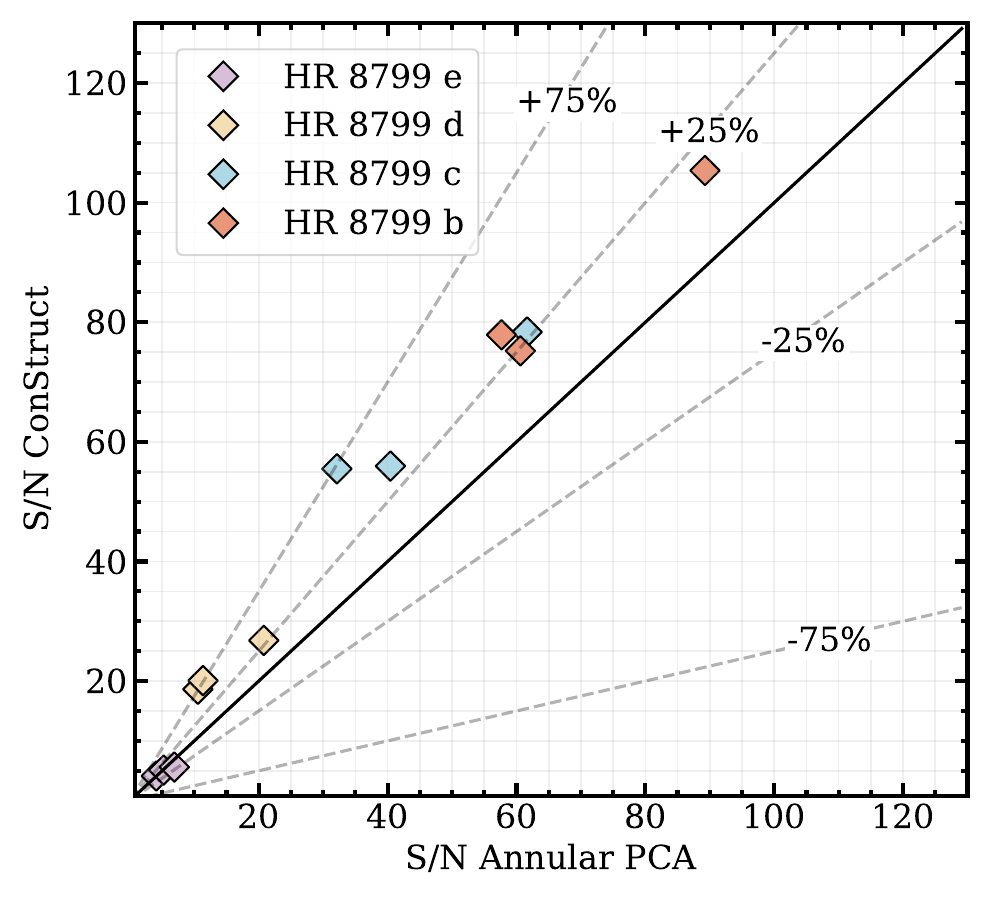} \\
    \centering\small (a) Maximum obtainable S/N for \texttt{ConStruct} and annular PCA on the tuning sample. 
  \end{tabular}%
  \quad
  \begin{tabular}[b]{@{}p{0.45\textwidth}@{}}
    \centering\includegraphics[clip, width=1.05\linewidth]{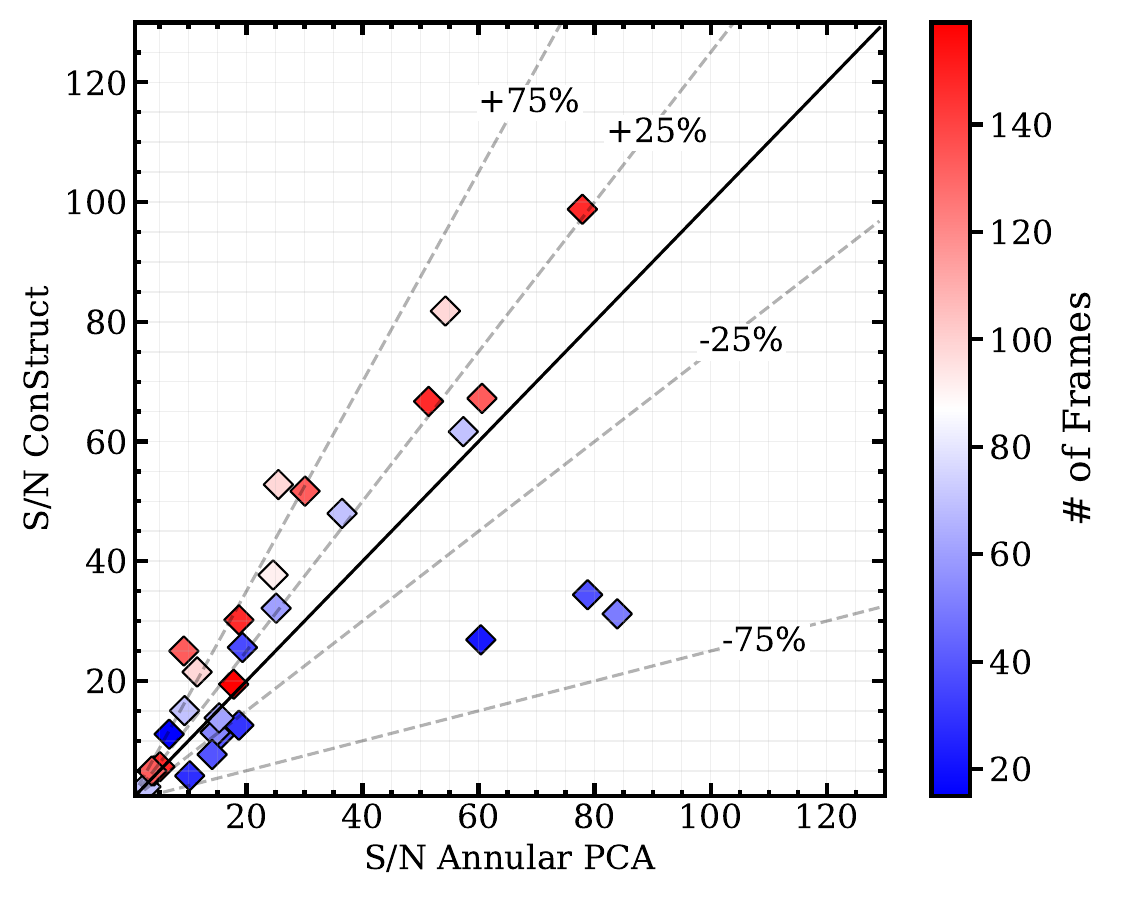} \\
    \centering\small (b) Sample S/N for \texttt{ConStruct} and annular PCA.
  \end{tabular}
  \vspace{2mm}
  \caption{Comparative analysis between \texttt{ConStruct} and annular \PCA\hspace{-0.1ex}. Each point source's S/N produced by \texttt{ConStruct} is plotted against that obtained with \PCA\hspace{-0.1ex}. The left panel shows the maximum obtainable S/N by tuning both \texttt{ConStruct} and PCA on each point source. In the right panel, we fix the design parameters for both \texttt{ConStruct} and \PCA\hspace{-0.1ex} to reduce 18 ADI sequences, which together contain 30 unique point sources. For this set of tuning parameters, we find the number of frames in the image sequence is an indication of the relative performance of \texttt{ConStruct}, with more frames (typically above 100) yielding improved results compared to the standard PCA-based approach.}
  \label{fig:max_and_sample_snr_comparisons}
\end{figure}

A standard annular PCA reduction serves to benchmark \texttt{ConStruct}'s performance on the HR 8799 \ADI\hspace{-0.1ex} sequences in Table \ref{tab:testing_data}. For this, a similar parameter search is performed where the annular width and number of \PCA\hspace{-0.1ex} components are iterated to find those that recover each of the planets at a maximal S/N. We iterate the annular width between 8 and 30 pixels in increments of 3 pixels, for a total of 7 different annular widths. The number of PCA\hspace{-0.1ex} components is iterated between 5 and the maximum allowable number, which is equivalent to the number of frames in the \ADI\hspace{-0.1ex} sequence, in increments of 4 components. In Appendix \ref{sec:Appendix_C2} we show the performance of our PCA-based processing approach over the grid of parameters for three HR 8799 datasets contained in Table \ref{tab:testing_data}. Figure \ref{fig:max_and_sample_snr_comparisons} compares these two approaches for all 12 sources in our tuning sample. For this, \texttt{ConStruct} produces a higher maximum S/N for planets b, c, and d than PCA over all datasets and produces comparable performance with the PCA-based reduction approach for planet e.

%
%
\section{Performance on NIRC2 ADI Data} \label{sec:results}

\subsection{Sample S/N Comparison with PCA and \texttt{ConStruct}}

We applied \texttt{ConStruct} to additional datasets obtained from \KOA\hspace{-0.1ex} and compared the performance with a standard annnular \PCA\hspace{-0.1ex} reduction approach. The 18 ADI sequences chosen contain known faint companions and are included in Table \ref{tab:testing_data}. In total, the sample includes 30 unique point sources. For all datasets, \texttt{ConStruct} uses the design parameters identified in Section \ref{sec:algorithm_performance_and_tuning}. To provide a fair comparison between \texttt{ConStruct} and annular PCA, we also fix a set of design variables for our PCA reduction approach over this testing dataset. We identified the suitable design variables to be 17 pixels for the annulus width while using the maximum allowable number of PCA components (i.e., the number of frames in the ADI sequence).
\begin{figure*}[hbt!]
  \centering
  \begin{tabular}[b]{@{}p{1.0\textwidth}@{}}
    \centering\includegraphics[width=0.95\linewidth]{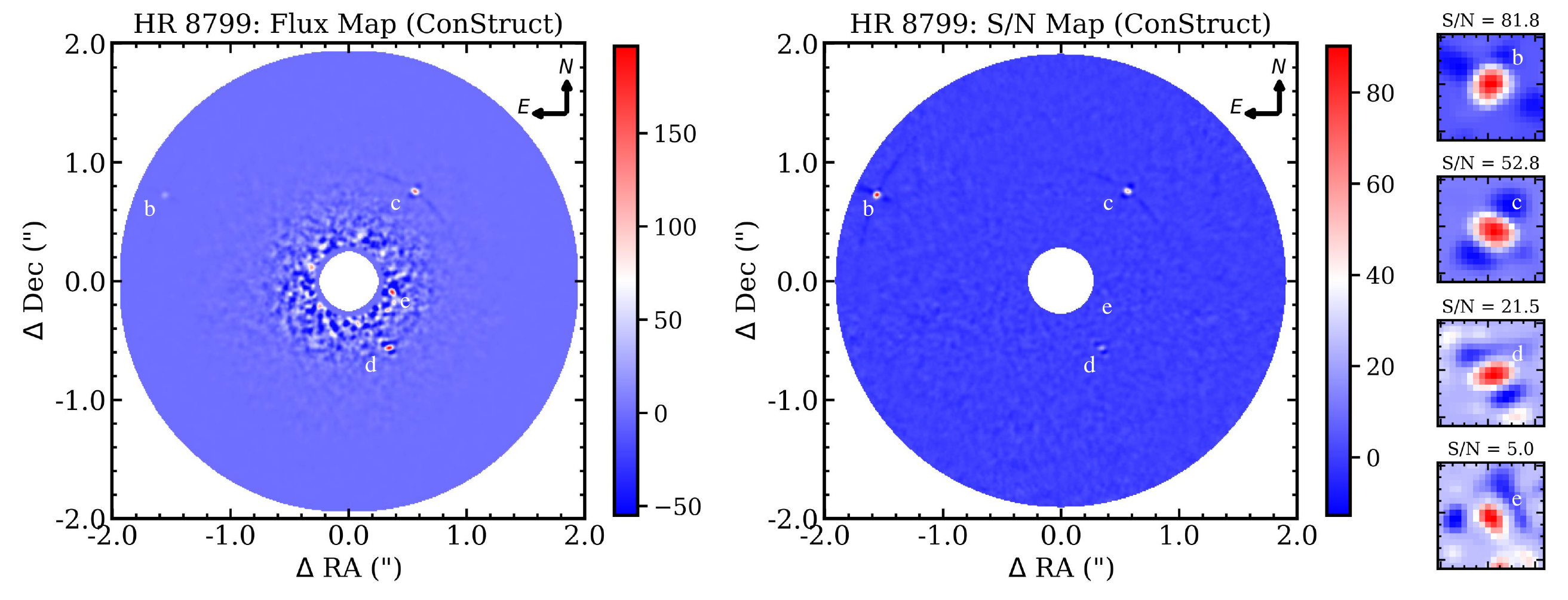} \\
    \centering\small \hspace{-12mm} (a) \texttt{ConStruct}
  \end{tabular}%
  \quad
  \begin{tabular}[b]{@{}p{1.0\textwidth}@{}}
    \centering\includegraphics[width=0.95\linewidth]{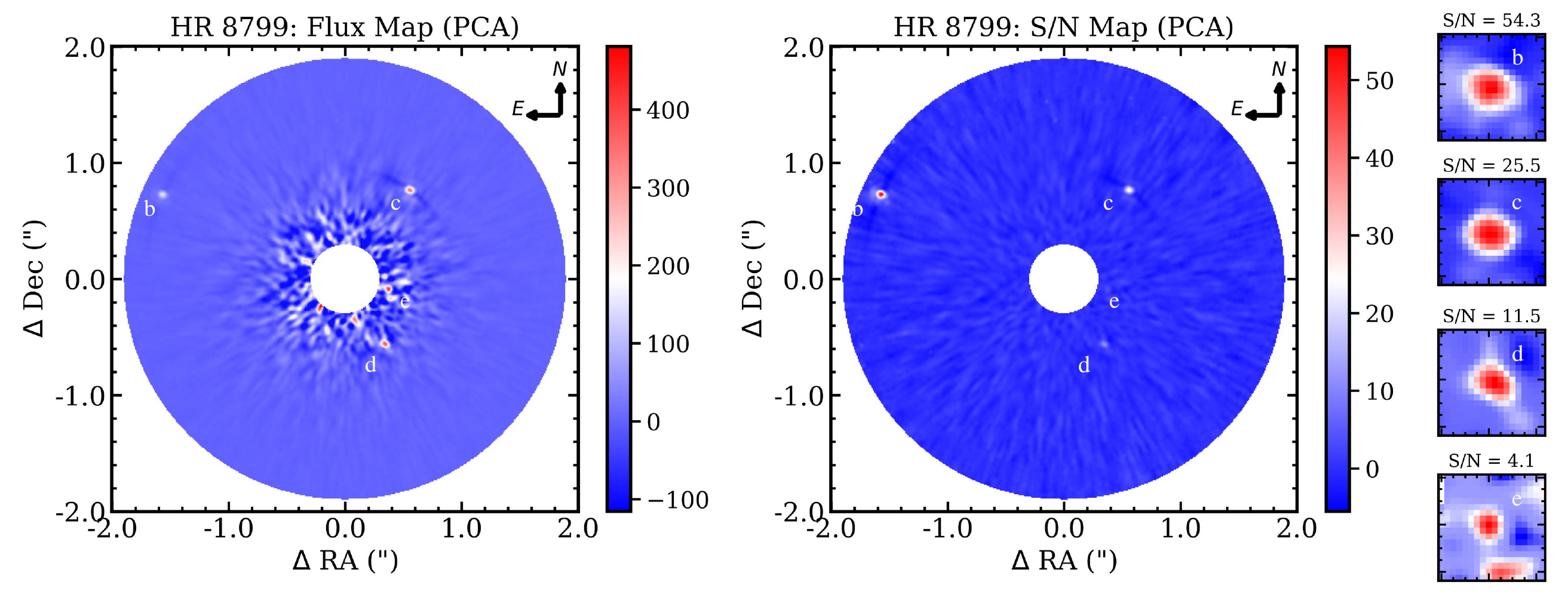} \\
    \centering\small \hspace{-10mm} (b) Annular PCA
  \end{tabular}
  \caption{Full-frame reductions with ConStruct and \PCA\hspace{-0.1ex} of HR 8799 \citep{Marois_2008, Marois_2010} for Sequence 11 in Table \ref{tab:testing_data}. The top two panels correspond to (left) the flux map and (right) the S/N map produced by \texttt{ConStruct}. The bottom panels are that produced with PCA. In the side panels, we show a cut-out view of each planetary point source, and the source's S/N produced by each algorithm.}
  \label{fig:reductions_19}
\end{figure*}

\begin{figure*}[hbt!]
  \centering
  \begin{tabular}[b]{@{}p{1.0\textwidth}@{}}
    \centering\includegraphics[width=.95\linewidth]{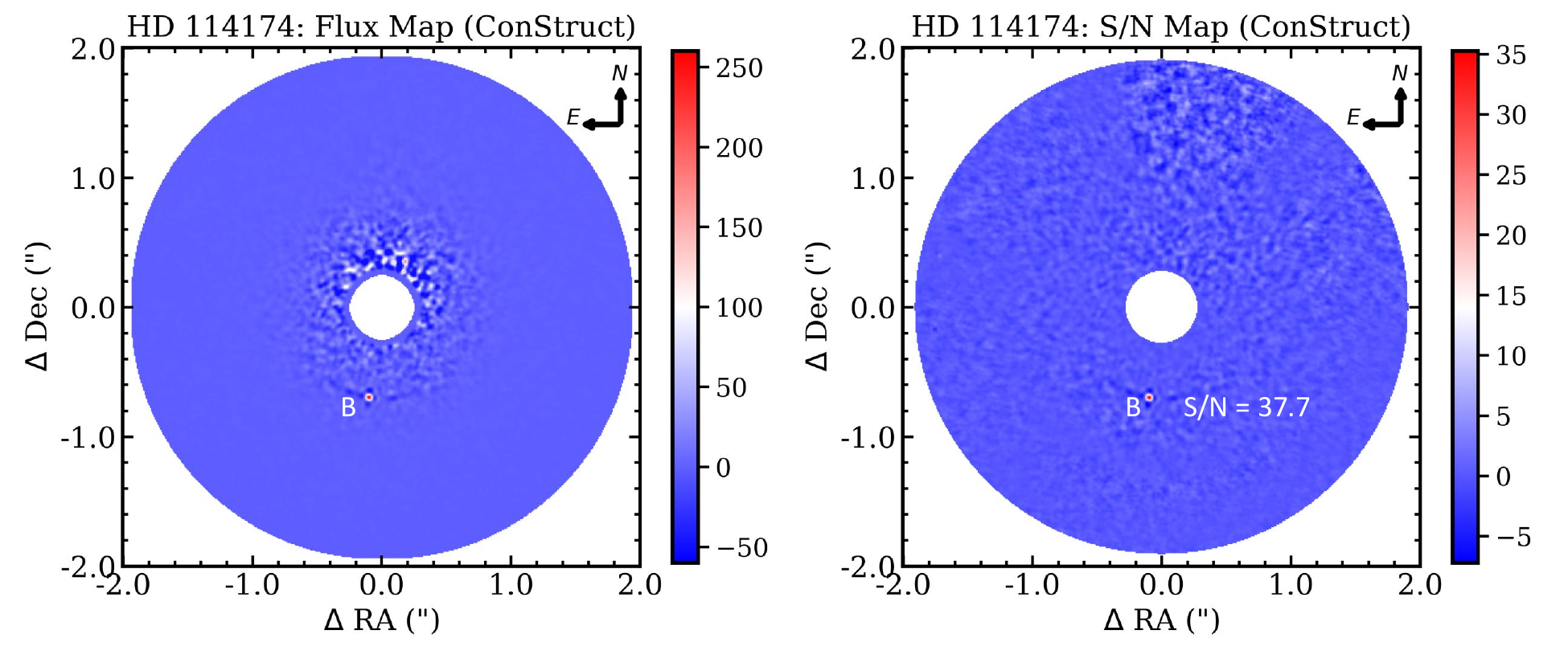} \\
    \centering\small (a) \texttt{ConStruct}
  \end{tabular}%
  \quad
  \begin{tabular}[b]{@{}p{1.0\textwidth}@{}}
    \centering\includegraphics[width=.95\linewidth]{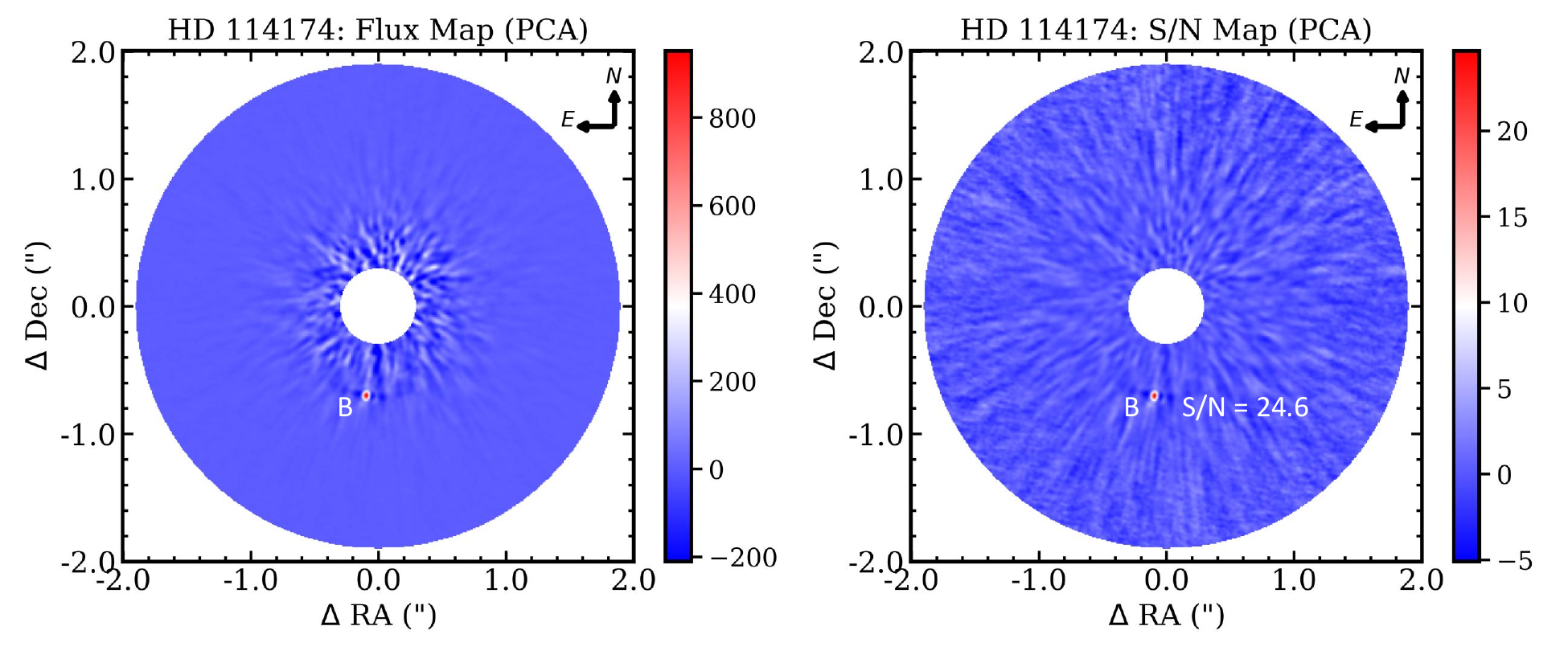} \\
    \centering\small (b) Annular PCA
  \end{tabular}
  \caption{Full-frame reductions with \texttt{ConStruct} and \PCA\hspace{-0.1ex} of HD 114174 \citep{Crepp_2013} for Sequence 5 in Table \ref{tab:testing_data}. The top two panels correspond to (left) the flux map and (right) the S/N map produced by \texttt{ConStruct}. The bottom panels are that produced with PCA.}
  \label{fig:reductions_13}
\end{figure*}

\begin{figure*}
  \centering
  \begin{tabular}[b]{@{}p{0.42\textwidth}@{}}
    \centering \hspace{-2.0cm}\includegraphics[width=1.0\linewidth]{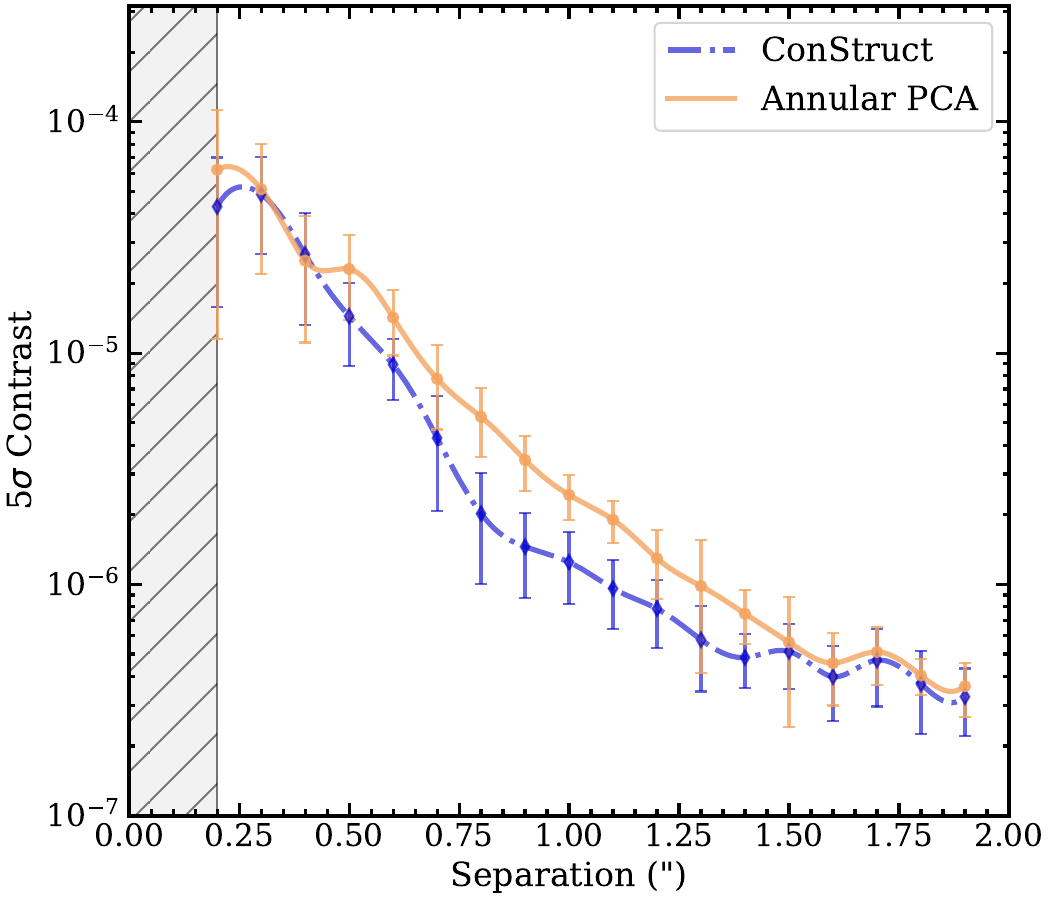} \\
    \hspace{-1.5cm}\centering\small (a) Sequence 11
  \end{tabular}%
  \quad
  \begin{tabular}[b]{@{}p{0.42\textwidth}@{}}
    \centering \hspace{-1cm}\includegraphics[width=1.0\linewidth]{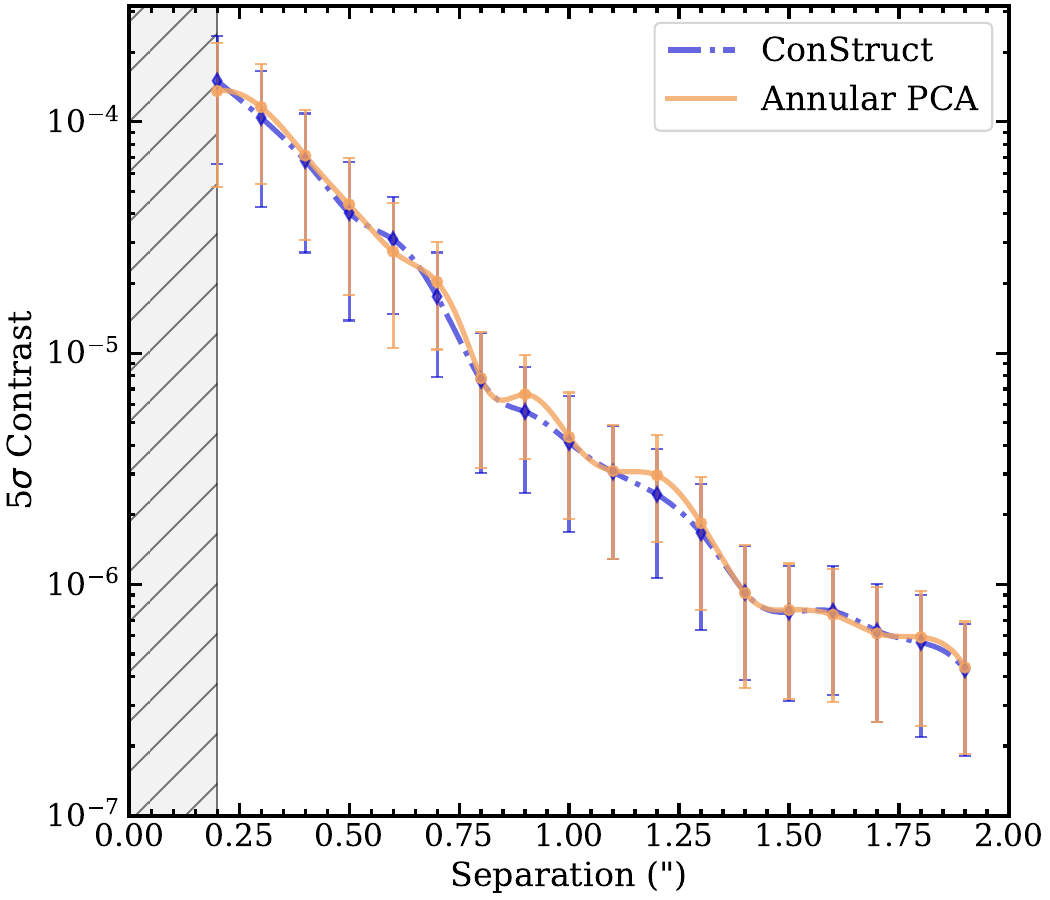} \\
    \hspace{-0.5cm}\centering\small (b) Sequence 12
  \end{tabular}
  \vspace{2mm}
  \caption{$5\sigma$ contrast curves with associated $1\sigma$ confidence bounds for (left) Sequence 11 and (right) Sequence 12 in Table \ref{tab:testing_data}. The blue and orange contrast curves correspond to \texttt{ConStruct} and annular PCA, respectively. For Sequence 11, \texttt{ConStruct} shows an improvement over PCA for most separation. Sequence 12, however, presents nearly identical 5$\sigma$ contrasts for both algorithms.}
\label{fig:contrast_curves}
\end{figure*}

For all the sources in our sample, we determine the recovered S/N produced from \texttt{ConStruct} and PCA\hspace{-0.1ex} with these fixed design parameters. Figure \ref{fig:max_and_sample_snr_comparisons} shows a one-to-one comparison of the two algorithms for all 30 sources in our sample. \texttt{ConStruct} recovers 20 out of the 30 point sources at a higher S/N than \PCA\hspace{-0.1ex}. Additionally, we find that the number of frames in the \ADI\hspace{-0.1ex} sequence is correlated with \texttt{ConStruct}'s performance, where more frames give a higher S/N for \texttt{ConStruct} than PCA. The L2 regularization coefficient, $\alpha$, chosen in tuning \texttt{ConStruct} is likely the primary driver for this trend\footnote{For sequences with few frames, choosing a non-zero regularization prevents over-fitting. In this study, \texttt{ConStruct} is tuned on sequences with many frames, so the algorithm sees better results when deployed on sequences with similar characteristics.}.

\subsection{Full Frame Reductions}

Figures \ref{fig:reductions_19} and \ref{fig:reductions_13} highlight two representative full-frame reductions for HR 8799 \citep{Marois_2008, Marois_2010} and HD 114174 \citep{Crepp_2013} using \texttt{ConStruct} and PCA. The top and bottom panels correspond to the reductions produced by \texttt{ConStruct} and PCA, respectively. For each, the flux map (left) and S/N map (right) of the reduced data sets are shown. Each S/N map is produced by dividing the flux values in each pixel-width annulus by the empirical RMS of the residuals taken over a buffered annular region that extends $\pm 3$ pixels radially. In Figure \ref{fig:reductions_19}, both \texttt{ConStruct} and \PCA\hspace{-0.1ex} recover all four known planets in the HR 8799 system, however \texttt{ConStruct} produces a higher S/N. In Figure \ref{fig:reductions_13}, both \texttt{ConStruct} and \PCA\hspace{-0.1ex} recover the white dwarf companion HD 114174 B with an S/N of 37.7 and 24.6, respectively. Interestingly, individual speckle noise realizations produced by \PCA\hspace{-0.1ex} appear more elongated radially whereas those produced by \texttt{ConStruct} have much less spatial covariance in the image plane. This effect may be attributed to the inclusion of data set diversity in training \texttt{ConStruct} over many ADI sequences.

\subsection{$5\sigma$ Contrast Curves}

We generate contrast curves to rigorously compare the performance of \texttt{ConStruct} with \PCA\hspace{-0.1ex} for two HR 8799 ADI sequences contained in Table \ref{tab:testing_data}. Synthetic sources are injected into each frame across varying radial separations. A source template is created by fitting a 2-D Gaussian profile to the central stellar PSF partially visible in the occulting disk of the coronagraph in NIRC2 images. Sources are injected over a grid of azimuthal coordinates and contrasts commensurate with the RMS of the residual speckle noise in a reduced image corresponding to the particular separation. For each frame, the grid is rotated in the negative parallactic rotation direction so that on-sky sources in the \ADI\hspace{-0.1ex} sequence do not add coherently during the derotation and coaddition step. We perform a least-squares fit between the intensity of the injected source and the recovered S/N for each radial separation. We use the resulting fit to find the intensity of the source that is recovered at a $5\sigma$ confidence level, and divide it by the flux of the central stellar PSF to obtain the $5\sigma$ contrast. The stellar flux is found using frames of the unsaturated PSF taken immediately before or after the ADI sequence. The flux values in the unsaturated PSF are scaled linearly so that the exposure is equivalent to the ADI frames. We also determine an associated $1\sigma$ confidence interval for the contrast value by determining the sample standard deviation of the contrast value over the grid of azimuthally spaced source locations.

In Figure \ref{fig:contrast_curves}, we show $5\sigma$ contrast curves for two HR 8799 datasets, corresponding to Sequences 11 and 12 in Table \ref{tab:testing_data}. For Sequence 11, \texttt{ConStruct} shows improvement over \PCA\hspace{-0.1ex} for nearly all separations. At its best, \texttt{ConStruct} improves the contrast by a factor of 2.6 at $0.8"$. Sequence 12, on the other hand, is nearly identical to \PCA. The contrast curves share similar behavior to the results of the S/N analysis when comparing \texttt{ConStruct} and PCA. The properties of each dataset are the likely driver of these results, however it is difficult to directly attribute any one factor.

\section{Conclusions} \label{sec:discussion}
%
%
In this work, we introduced \texttt{ConStruct}, a deep learning-based algorithm for identifying faint point sources in ADI sequences. This algorithm uses an autoencoder neural network in a self-supervised training architecture to embed information from thousands of ADI\hspace{0.3ex} image examples. The trained network is employed in a source present/absent processing framework to predict speckle noise, and recover faint signals in \ADI\hspace{0.3ex} sequences. We further augment the autoencoder neural network in \texttt{ConStruct} with a regularized least squares regression step to leverage temporal correlations in speckle noise across \ADI\hspace{-0.1ex} frames. \texttt{ConStruct} is tuned using three datasets of the HR 8799 system taken with Keck/NIRC2 to identify suitable design parameters for a detailed assessment of its performance on other NIRC2 datasets. We demonstrated \texttt{ConStruct} with a sample of 30 point sources in 18 NIRC2 ADI sequences taken from KOA and compared the performance with a standard PCA-based reduction approach. For the examples analyzed, \texttt{ConStruct} improves the S/N of detections by up to ${\sim}75 \%$ and contrast up to a factor of 2.6.

There are several potential directions to improve \texttt{ConStruct} and make it more generally applicable for other high-contrast imaging datasets. This includes testing
alternative data types for training and deploying \texttt{ConStruct}, including different filters and additional instruments besides NIRC2. It is also possible to assess a broader variety of learning architectures that may improve the accuracy of \texttt{ConStruct}. In \texttt{ConStruct}, we employ a regularized least-squares regression to use the temporal correlations in speckle noise across \ADI\hspace{-0.1ex} frames. This step can potentially be implemented directly in the neural network with the addition of a Long Short-Term Memory (LSTM) block -- a standard method in machine learning for predicting features with temporal correlations.

Additionally, there is growing interest in the high-contrast imaging community for processing procedures that operate under an inverse problem framework \citep{Cantalloube_2015, Flasseur_2018, Flasseur_2020}. \texttt{ConStruct} may be adapted for these methods by representing speckle noise predictions as a continuous probability distribution, which can be achieved with a variational autoencoder. This work only considers single spectral band \ADI\hspace{-0.1ex} sequences, but \texttt{ConStruct} can, in principle, be readily extended to multi-wavelength integral field units that take advantage of \ac{SDI} such as the Gemini Planet Imager \citep{Macintosh_2006}, the Spectro-Polarimetric High-contrast Exoplanet REsearch (SPHERE) instrument \citep{Beuzit_2019}, the Subaru Coronagraphic Extreme Adaptive Optics (SCExAO) instrument \citep{Jovanovic_2015}, and MagAO-X \citep{Males_2018, Males_2020} by using higher dimensional convolutional layers. Finally, \texttt{ConStruct} and its future improvements present a promising approach for reducing data collected from the James Webb Space Telescope and future extreme adaptive optics ground-based facilities.

\section*{Acknowledgments}
This research has made use of the Keck Observatory Archive (KOA), which is operated by the W. M. Keck Observatory and the NASA Exoplanet Science Institute (NExScI), under contract with the National Aeronautics and Space Administration. B.P.B. acknowledges support from the National Science Foundation grant AST-1909209, NASA Exoplanet Research Program grant 20-XRP20$\_$2-0119, and the Alfred P. Sloan Foundation.



\appendix
\section{Autoencoder Architecture}\label{sec:Appendix_D}
Here we show the parameters for the autoencoder neural network used in \texttt{ConStruct}. In total, the autoencoder consists of 32 unique layers. In Table \ref{tab:autoencoder_model}, we show the parameters for each layer in the network, which include the layer type and the size of the layer. 

\startlongtable
\begin{deluxetable}{ccccc}

\tabletypesize{\footnotesize}
\tablewidth{0pt}

 \tablecaption{Autoencoder network parameters for \texttt{ConStruct}. \label{tab:autoencoder_model}}

 \tablehead{\colhead{Layer $\#$} & \colhead{Type} & \colhead{Tensor Size}}

\startdata 
0&	Input&	(32, 32, 1)\\
1&	Conv.&	(32, 32, 32)\\
2&	Conv.&	(32, 32, 32)\\
3&	Max. Pool&	(16, 16, 32)\\
4&	Conv.&	(16, 16, 64)\\
5&	Conv.&	(16, 16, 64)\\
6&	Max. Pool&	(8, 8, 64)\\
7&	Conv.&	(8, 8, 128)\\
8&	Conv.&	(8, 8, 128)\\
9&	Max. Pool&	(4, 4, 128)\\
10&	Conv.&	(4, 4, 256)\\
11&	Conv.& 	(4, 4, 256)\\
12&	Max. Pool&	(2, 2, 256)\\
13&	Conv.& 	(2, 2, 512)\\
14&	Conv.& 	(2, 2, 512)\\
15&	Up Conv.&	(4, 4, 256)\\
16&	Concatenate& (4, 4, 512)\\
17&	Conv.&	(4, 4, 256)\\
18&	Conv.& 	(4, 4, 256)\\
19&	Up Conv.&	(8, 8, 128)\\
20&	Concatenate&	(8, 8, 256)\\
21&	Conv.&	(8, 8, 128)\\
22&	Conv.&	(8, 8, 128)\\
23&	Up Conv.&	(16, 16, 64)\\
24&	Concatenate& (16, 16, 128)\\
25&	Conv.& 	(16, 16, 64)\\
26&	Conv.&	(16, 16, 64)\\
27&	Up Conv.&	(32, 32, 32)\\
28&	Concatenate&	(32, 32, 64)\\
29&	Conv.&	(32, 32, 32)\\
30&	Conv.& 	(32, 32, 32)\\
31&	Output&	(32, 32, 1)
\enddata


\end{deluxetable}

\clearpage
\section{Additional Autoencoder Predictions of Speckle Noise}\label{sec:Appendix_E}
Here we show two additional examples demonstrating how the autoencoder neural network can predict the speckle noise in an image patch extracted from an \ADI\hspace{-0.1ex} sequence. In this example, the patches are extracted from a frame contained in an ADI seqeunce of the HR 8799 system, corresponding to Sequence 9 in Table \ref{tab:testing_data}. A synthetic 2-D Gaussian source is injected into the center of the the image patch. The central region is then masked before being fed into the network. 

\begin{figure}[b!]
  \centering
  \begin{tabular}[b]{@{}p{1.0\textwidth}@{}}
    \centering\includegraphics[width=0.75\linewidth]{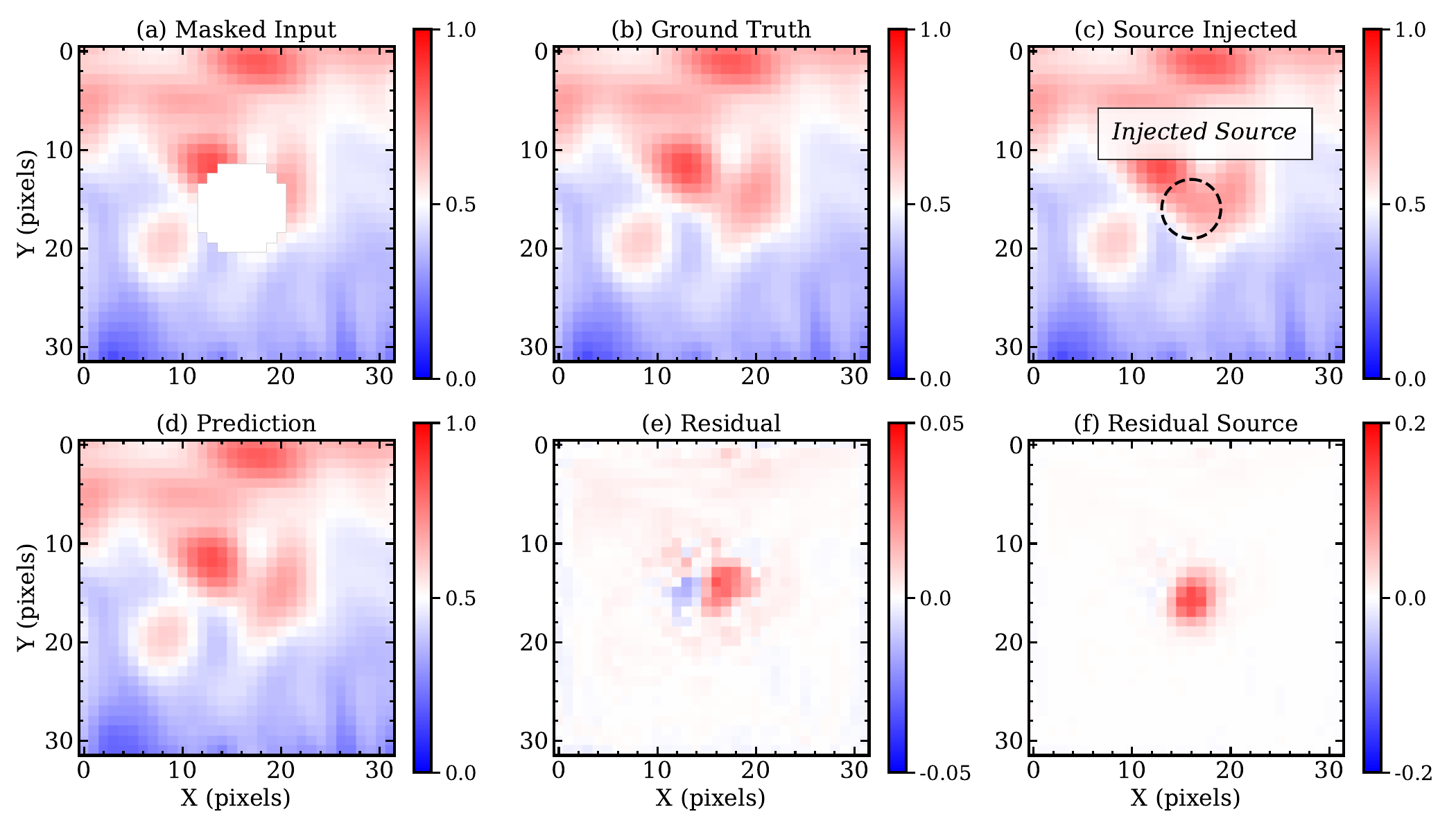} \\
    \centering\small (a)
  \end{tabular}%
  \quad
  \begin{tabular}[b]{@{}p{1.0\textwidth}@{}}
    \centering\includegraphics[width=0.75\linewidth]{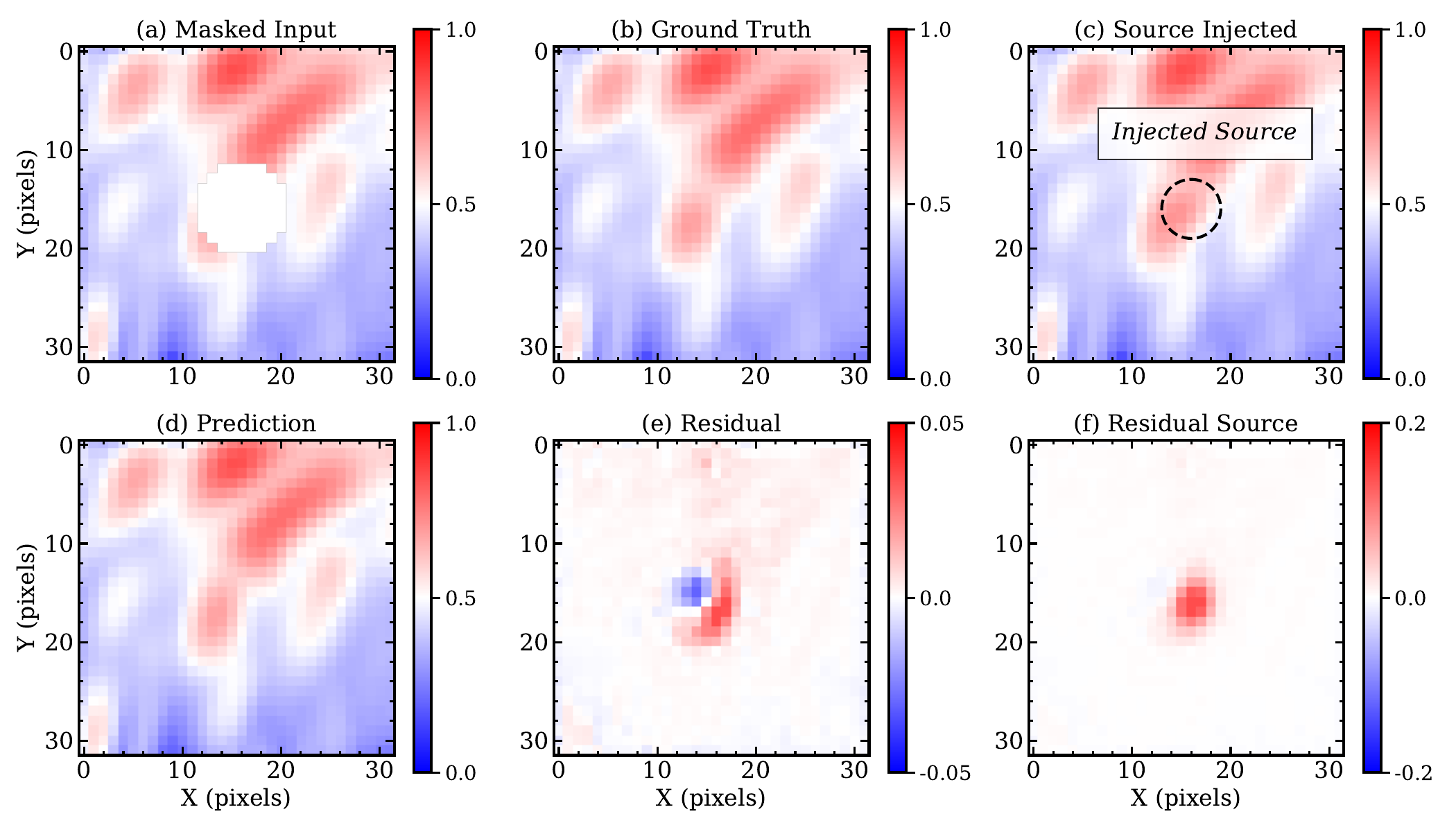} \\
    \centering\small (b)
  \end{tabular}
  \caption{Two autoencoder prediction examples for the speckle noise contained in a masked patch extracted from an ADI frame. Each of the patches are linearly scaled between zero and one. In each of the examples provided, panels (a) and (b) are the masked autoencoder input and the ground-truth region of interest (ROI), respectively. Panel (c) is the ground truth, with an injected Gaussian signal with a peak scaled intensity of 0.2. Panel (d) is the prediction formed by passing the masked input into our network. Panels (e) and (f) are the residual images produces by subtracting the predicted image regions from the ground through without and with the injected source, respectively.}
  \label{fig:figures}
\end{figure}

\clearpage
\section{Training Data}\label{sec:Appendix_A}
In Table \ref{tab:training_data}, we include the ADI sequences used for training \texttt{ConStruct}. These data are obtained from KOA. In total, 92 unique ADI sequences are used from various targets, many with known companions.

\startlongtable
\begin{deluxetable}{cccccccc}

\tabletypesize{\footnotesize}
\tablewidth{0pt}

 \tablecaption{ \texttt{ConStruct} training \ADI\hspace{0.3ex} sequences from Keck/NIRC2. \label{tab:training_data}}

 \tablehead{
 \colhead{Target} & \colhead{Date (UT)} & \colhead{\# of Frames} &\colhead{Int. Time (s)} & \colhead{Filter}& \colhead{Program PI}
 }

\startdata 
GJ 504&	2002-06-15&	20&	1200&	$H$&	Charbonneau\\
HN Peg&	2002-06-15&	20&	1200&	$H$&	Charbonneau\\
HR 6070&	2003-03-13&	15&	435.6&	$H$&	Liu\\
51 Eri&	2003-03-13&	15&	435.6&	$H$&	Liu\\
GI 355&	2003-03-14&	15&	435.6&	$H$&	Liu\\
V889 Her&	2003-06-18&	10&	300&	$H$&	Liu\\
49 Seti&	2003-11-11&	10&	600&	$H$&	Liu\\
HD 25457&	2003-11-11&	10&	435.6&	$H$&	Liu\\
HD 22049&	2003-11-11&	10&	435.6&	$H$&	Liu\\
HD 22049&	2003-12-07&	9&	540&	$H$&	de Pater\\
HR 4796&	2003-12-07&	21&	210&	$H$&	Kalas\\
HR 4796&	2003-12-07&	11&	330&	$K_\text{p}$&	Kalas\\
HIP 102409&	2004-05-30&	48&	2400&	$H$&	dePater\\
HIP 6276&	2004-09-08&	15&	450&	$H$&	Liu\\
HD 147513&	2005-04-18&	14&	840&	$H$&	Kalas\\
V889 Her&	2005-07-15&	10&	290.4&	$H$&	Liu\\
HD 191089&	2005-10-21&	15&	300&	$H$&	Kalas\\
HIP 17395&	2008-11-04&	95&	1900&	$K_\text{s}$&	Macintosh\\
HD 92945&	2008-12-03&	40&	1600&	$H$&	Kalas\\
HD 107146&	2008-12-17&	55&	880&	$K_\text{s}$&	Jayawardhana\\
HD 161868&	2009-04-13&	35&	525&	$K_\text{p}$&	Law\\
HD 984&	2009-07-31&	60&	1200&	$K_\text{p}$&	Macintosh\\
HIP 1134&	2009-08-07&	90&	1800&	$K_\text{p}$&	Macintosh\\
HIP 7345&	2009-08-07&	107&	2140&	$K_\text{p}$&	Macintosh\\
HIP 91043&	2009-08-07&	88&	1760&	$K_\text{p}$&	Macintosh\\
HIP 95793&	2009-11-01&	60&	2400&	$K_\text{s}$&	Macintosh\\
HIP 21547&	2009-11-02&	33&	1980&	$K_\text{s}$&	Macintosh\\
HD 61005&	2009-11-24&	205&	1025&	$H$&	Kalas\\
HD 61005&	2009-12-05&	58&	1740&	$K_\text{p}$&	Kalas\\
HD 131835&	2010-04-03&	13&	390&	$H$&	Fitzgerald\\
HD 131835&	2010-04-03&	92&	2208&	$H$&	Fitzgerald\\
HD 131835&	2010-04-03&	109&	3270&	$H$&	Fitzgerald\\
GL 758&	2010-05-02&	102&	3060&	$CH_4 S$&	Biller\\
HIP 95793&	2010-07-10&	45&	1350&	$K_\text{s}$&	Macintosh\\
HIP 76063&	2010-07-11&	40&	1200&	$K_\text{s}$ &	Macintosh\\
HIP 95347&	2010-07-11&	56&	1680&	$K_\text{s}$&	Macintosh\\
HIP 72197&	2010-07-12&	45&	1350&	$K_\text{s}$&	Macintosh\\
HD 10008&	2010-09-26&	70&	1680&	$K_\text{p}$&	Hinkley\\
HD 8375&	2010-10-13&	60&	1086&	$H$&	Crepp\\
HIP 53954&	2011-02-06&	100&	145.2&	$K$&	Hinkley\\
HIP 53954&	2011-02-06&	20&	290.4&	$K$&	Hinkley\\
HIP 24528&	2011-02-06&	60&	2400&	$K_\text{p}$ &	Hinkley\\
HIP 25453&	2011-02-06&	60&	960&	$K_\text{p}$ &	Hinkley\\
HD 114174&	2011-02-22&	37&	740&	$K_\text{p}$ &	Crepp\\
HR 7672&	2011-05-15&	15&	30&	    $K_\text{p}$ &	Carpenter\\
HR 7672&	2011-05-15&	30&	450&	$K_\text{p}$ &	Carpenter\\
HD 206860&	2011-07-17&	60&	304.9&	$H$ &	Morales\\
HD 206860&	2011-07-17&	41&	208.4&	$K_\text{p}$ &	Morales\\
HD 19467&	2011-08-30&	50&	1250&	$K_\text{p}$ &	Crepp\\
HD 61005&	2012-01-04&	21&	420&	$K_\text{p}$ &	Fitzgerald\\
HD 61005&	2012-01-04&	21&	420&	$K_\text{p}$ &	Fitzgerald\\
HD 61005&	2012-01-04&	61&	1830&	$K_\text{p}$ &	Fitzgerald\\
HD 114174&	2012-02-02&	92&	1840&	$K_\text{p}$ &	Knutson\\
HD 114174&	2012-02-02&	92&	1840&	$K_\text{p}$ &	Knutson\\
HD 114174&	2012-05-29&	10&	250&	$H$&	Crepp\\
HD 114174&	2012-05-29&	61&	1220&	$K_\text{p}$ &	Crepp\\
HD 114174&	2012-06-24&	51&	1275&	$H$&	Crepp\\
HD 114174&	2012-07-04&	100&	2400&	$J$&	Johnson\\
HIP 107350&	2012-07-22&	30&	600&	$K_\text{s}$&	Macintosh\\
HD 4747&	2012-08-25&	159&	3816&	$K_\text{p}$ &	Johnson\\
HD 19467&	2012-08-26&	20&	600&	$H$&	Crepp\\
HD 19467&	2012-08-26&	20&	480&	$K_\text{p}$ &	Crepp\\
HIP 21547&	2012-09-04&	52&	1560&	$K_\text{s}$ &	Macintosh\\
HD 206860&	2012-10-27&	60&	300&	$K_\text{p}$ &	Morales\\
HIP 6276&	2012-10-27&	40&	1400&	$K_\text{p}$ &	Morales\\
HIP 44526&	2012-12-05&	23&	920&	$K_\text{p}$ &	Hinkley\\
Kappa And&	2013-05-01&	43&	778.3&	$K_\text{p}$ &	Hinkley\\
HIP 116805&	2013-05-30&	38&	1520&	$K_\text{p}$ &	Carpenter\\
HIP 116805&	2013-06-22&	22&	550&	$K_\text{p}$ &	Hinkley\\
HD 182488&	2013-07-03&	28&	840&	$K_\text{s}$ &	Crepp\\
HIP 116805&	2013-08-18&	15&	300&	$K_\text{p}$ &	Johnson\\
HIP 11152&	2013-09-25&	58&	290&	$K_\text{p}$ &	Hinkley\\
HIP 45950&	2014-01-12&	48&	720&	$K_\text{p}$ &	Knutson\\
GJ 504&	2014-05-13&	30&	150&	$Y$&	Hinkley\\
GJ 504&	2014-05-13&	68&	680&	$Y$&	Hinkley\\
HD 114174&	2014-05-21&	10&	100&	$J$&	Knutson\\
HD 114174&	2014-05-21&	10&	100&	$K_\text{p}$ &	Knutson\\
HD 4747&	2014-10-12&	201&	6030&	$K_\text{s}$ &	Crepp\\
HD 203030&	2014-11-09&	60&	1800&	$K_\text{s}$ &	Bowler\\
HD 4747&	2015-01-09&	39&	2340&	$K_\text{s}$ &	Knutson\\
HD 35841&	2015-01-10&	72&	2880&	$K_\text{p}$ &	Hinkley\\
HN Peg&	2015-06-02&	71&	2130&	$K_\text{s}$ &	Knutson\\
HD 203030&	2015-06-03&	105&	3150&	$K_\text{s}$ &	Knutson\\
HIP 60074&	2015-06-06&	172&	344&	$K_\text{p}$ &	Padgett\\
GJ 504&	2015-06-08&	91&	2730&	$Y$ &	Currie\\
Kappa And&	2016-05-26&	29&	580&	$K_\text{s}$ &	Mawet\\
HD 191089&	2016-06-26&	144&	4180.3&	$H$&	Kalas\\
K12&	2016-08-20&	100&	400&	$K_\text{s}$ &	Morales\\
HD 43989&	2017-11-16&	70&	2240&	$K_\text{s}$ &	Choquet\\
Kappa And&	2017-12-10&	10&	300&	$K_\text{p}$&	Currie\\
Kappa And&	2018-01-30&	15&	450&	$K_\text{s}$ &	Bowler\\
Gamma Cep&	2019-07-07&	20&	106&	$K_\text{s}$ &	Bowler
\enddata

\end{deluxetable}

\clearpage
\section{Tuning \texttt{ConStruct} with the HR 8799 Planets} \label{sec:Appendix_C}
\begin{figure}[hbt!]
  \centering
  \centering \begin{tabular}[b]{@{}p{0.42\textwidth}@{}}
    \includegraphics[width=1.0\linewidth]{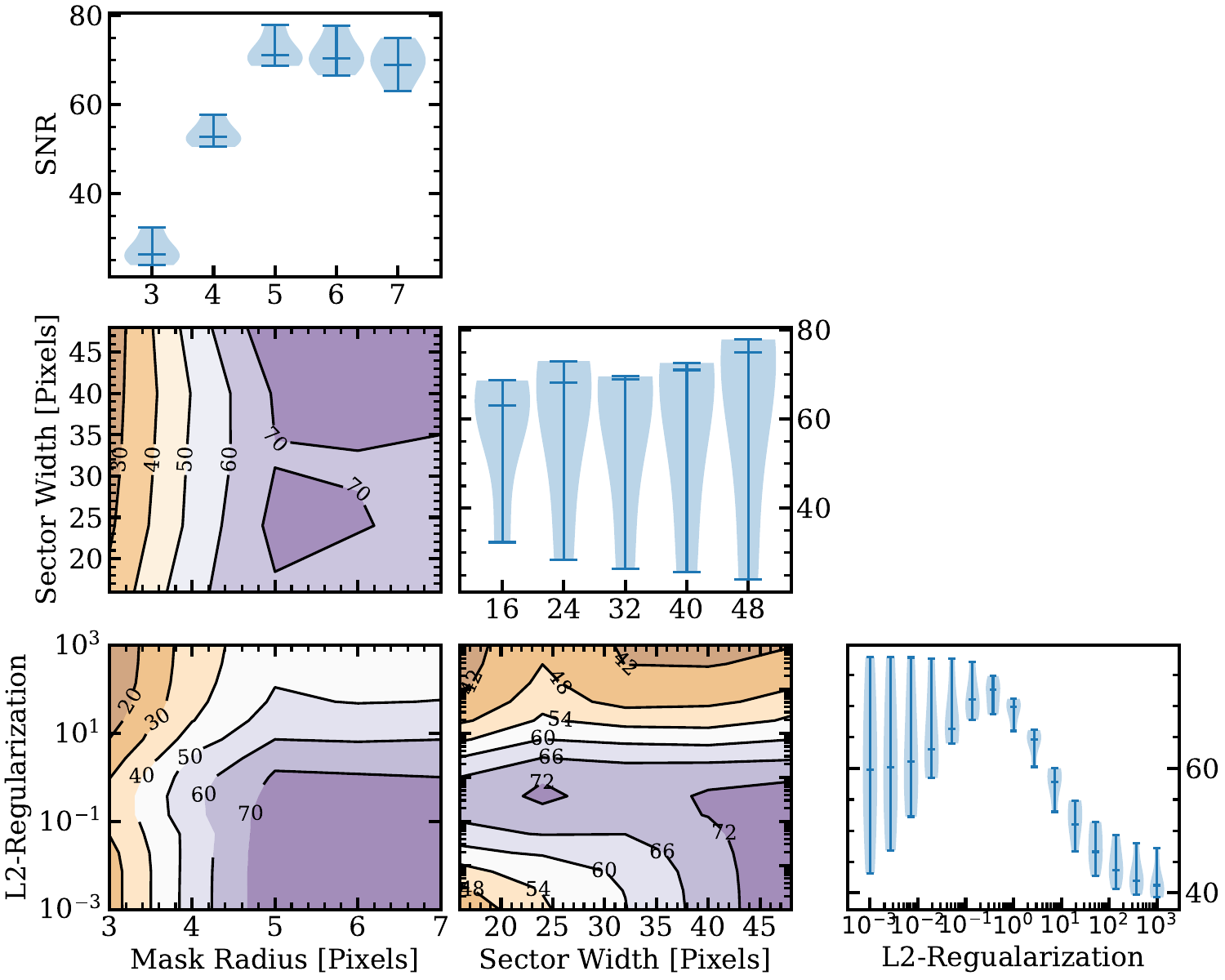} \\
    \centering \small (a) HR 8799 b
  \end{tabular}%
  \quad
  \begin{tabular}[b]{@{}p{0.42\textwidth}@{}}
    \includegraphics[width=1.0\linewidth]{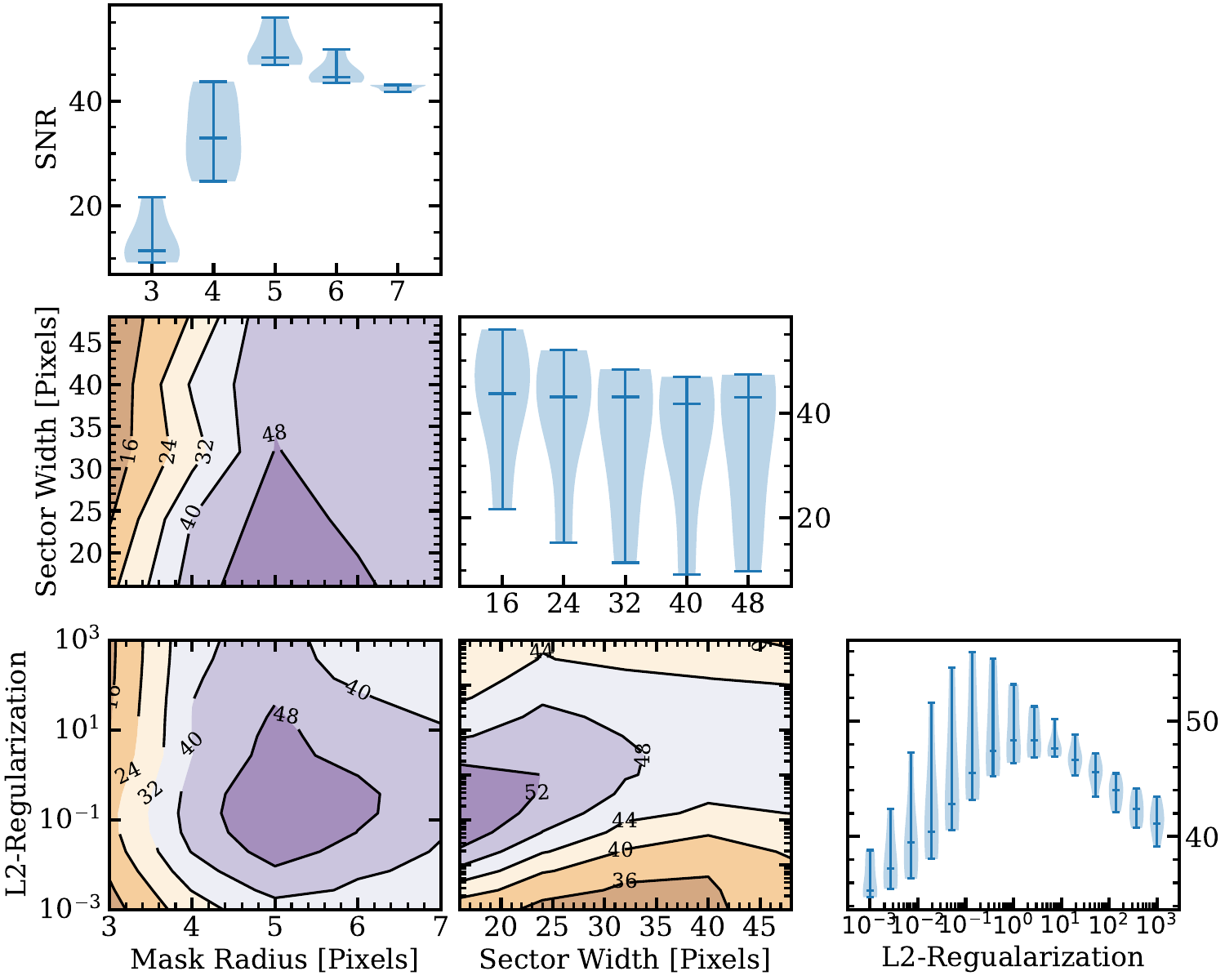} \\
    \centering \small (b) HR 8799 c
  \end{tabular}
  \begin{tabular}[b]{@{}p{0.42\textwidth}@{}}
    \vspace{2mm}\hspace{5mm}
    \includegraphics[width=1.0\linewidth]{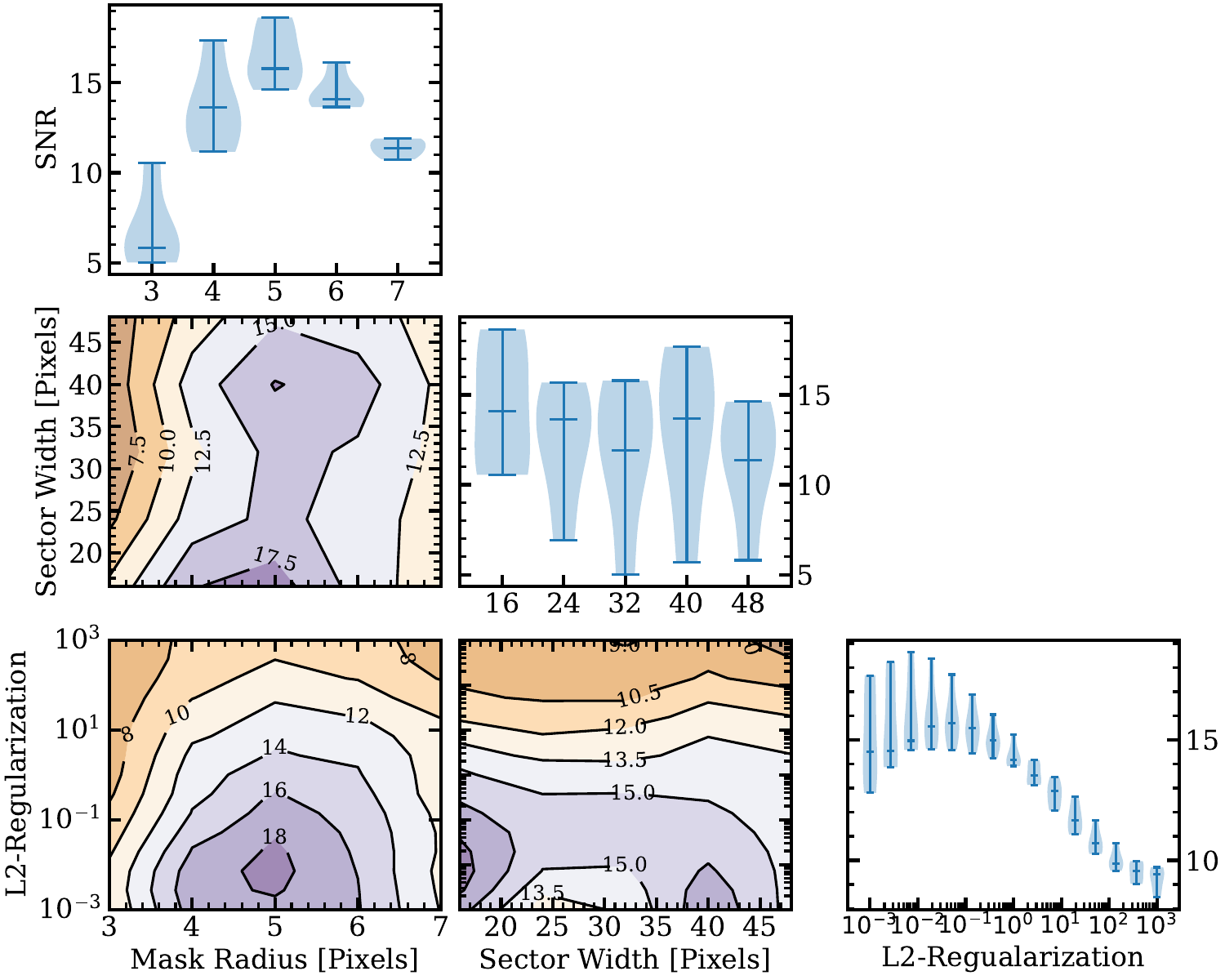} \\
    \centering \small \hspace{10mm} (c) HR 8799 d
  \end{tabular}%
  \quad
  \begin{tabular}[b]{@{}p{0.42\textwidth}@{}}
    \hspace{6mm}\includegraphics[width=1.0\linewidth]{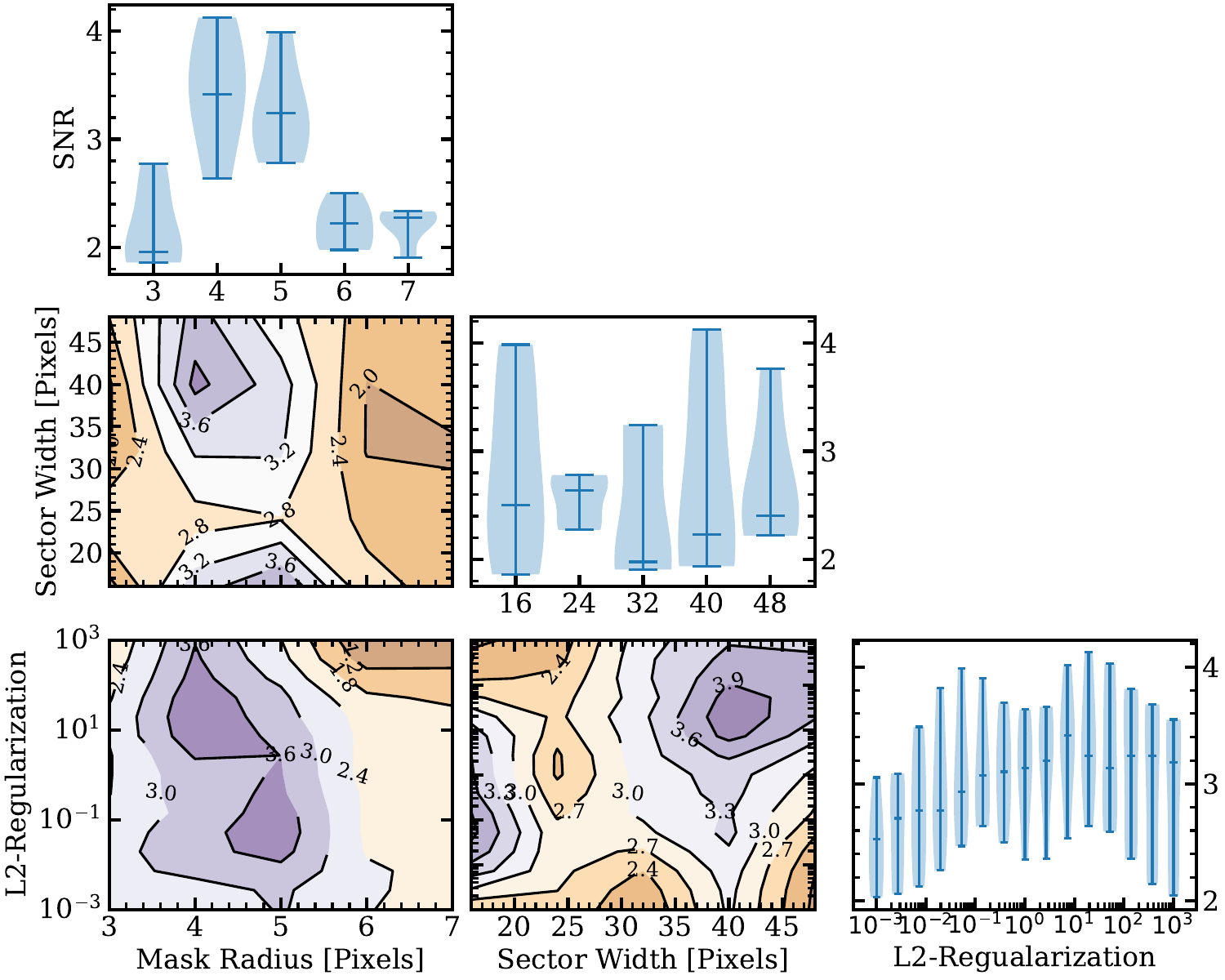} \\
    \centering \small \hspace{10mm} (d) HR 8799 e
  \end{tabular}
  \caption{Marginalized representations for our parametric parameter search for the $2010-07-13$ epoch of HR 8799. Each contour plot shows the S/N values projected onto two of the variables. The projection operation takes the maximum S/N along the axis of the marginalized variable. The violin plots show the spread in the S/N in each variable bin. Panels (a), (b), (c), and (d) correspond to HR 8799 b, c, d, and e, respectively.}
\label{fig:corner_plot_figs_18}
\end{figure}

\begin{figure}[hbt!]
  \centering
  \centering \begin{tabular}[b]{@{}p{0.42\textwidth}@{}}
    \includegraphics[width=1.0\linewidth]{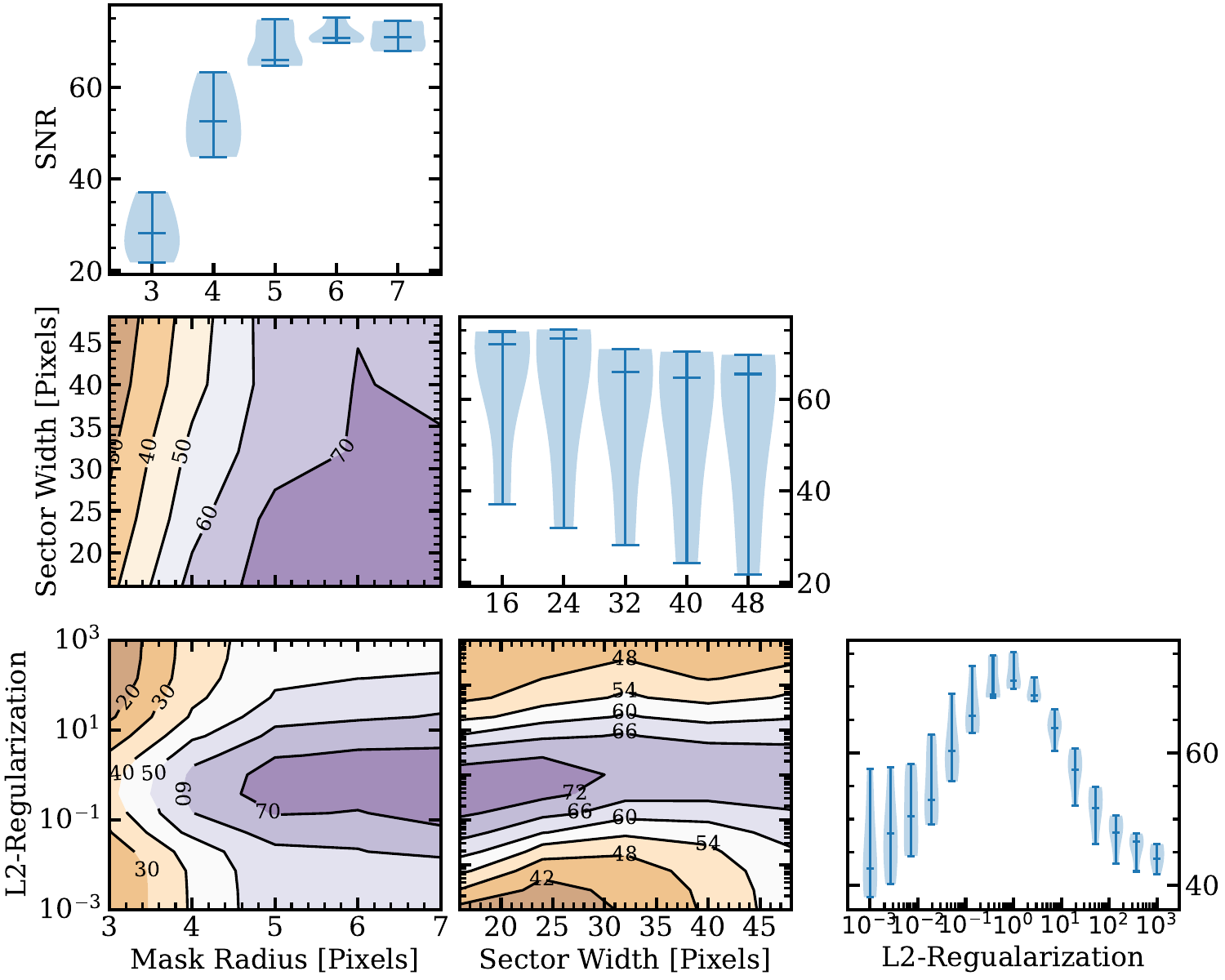} \\
    \centering \small (a) HR 8799 b
  \end{tabular}%
  \quad
  \begin{tabular}[b]{@{}p{0.42\textwidth}@{}}
    \includegraphics[width=1.0\linewidth]{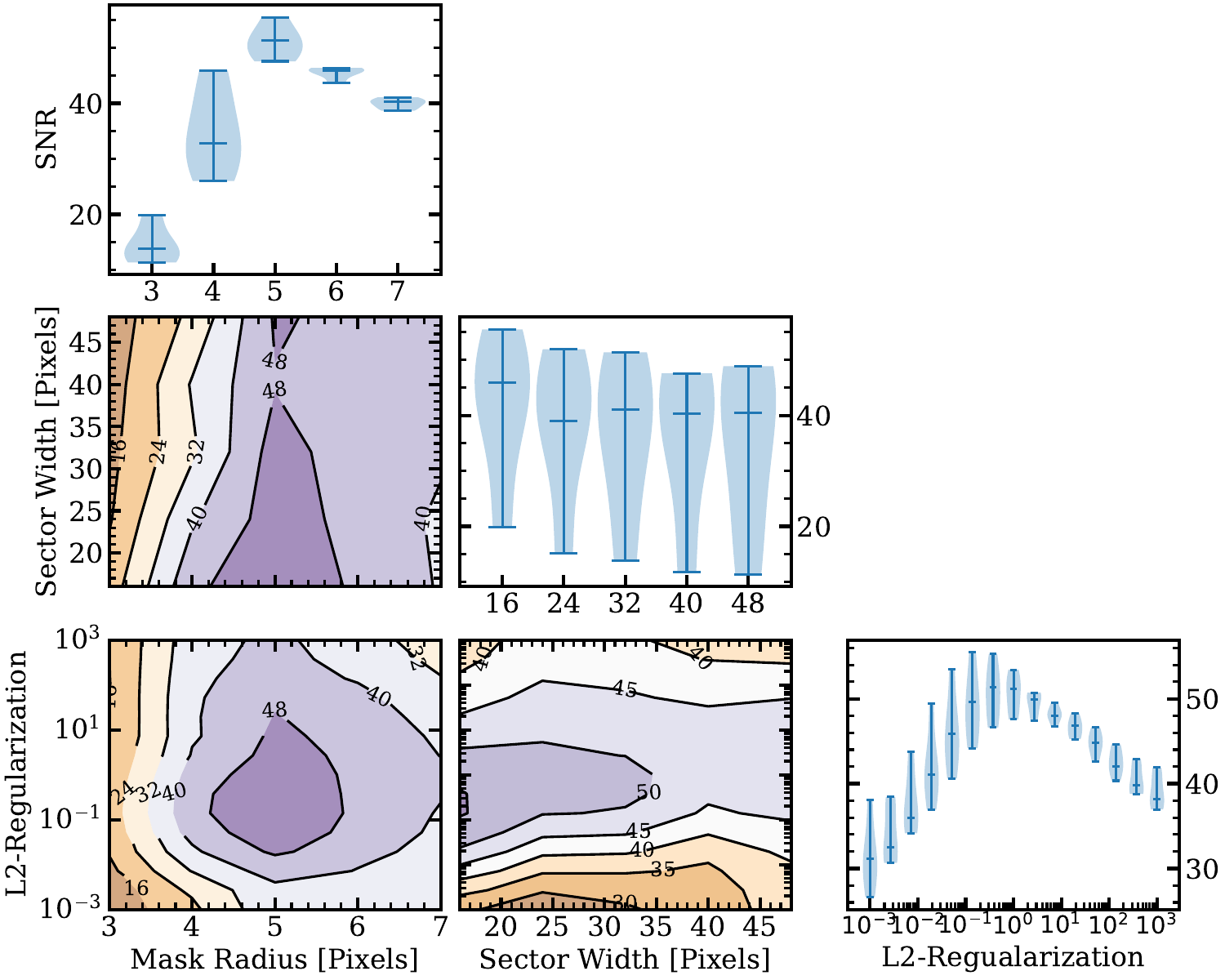} \\
    \centering \small (b) HR 8799 c
  \end{tabular}
  \begin{tabular}[b]{@{}p{0.42\textwidth}@{}}
    \vspace{2mm}\hspace{5mm}
    \includegraphics[width=1.0\linewidth]{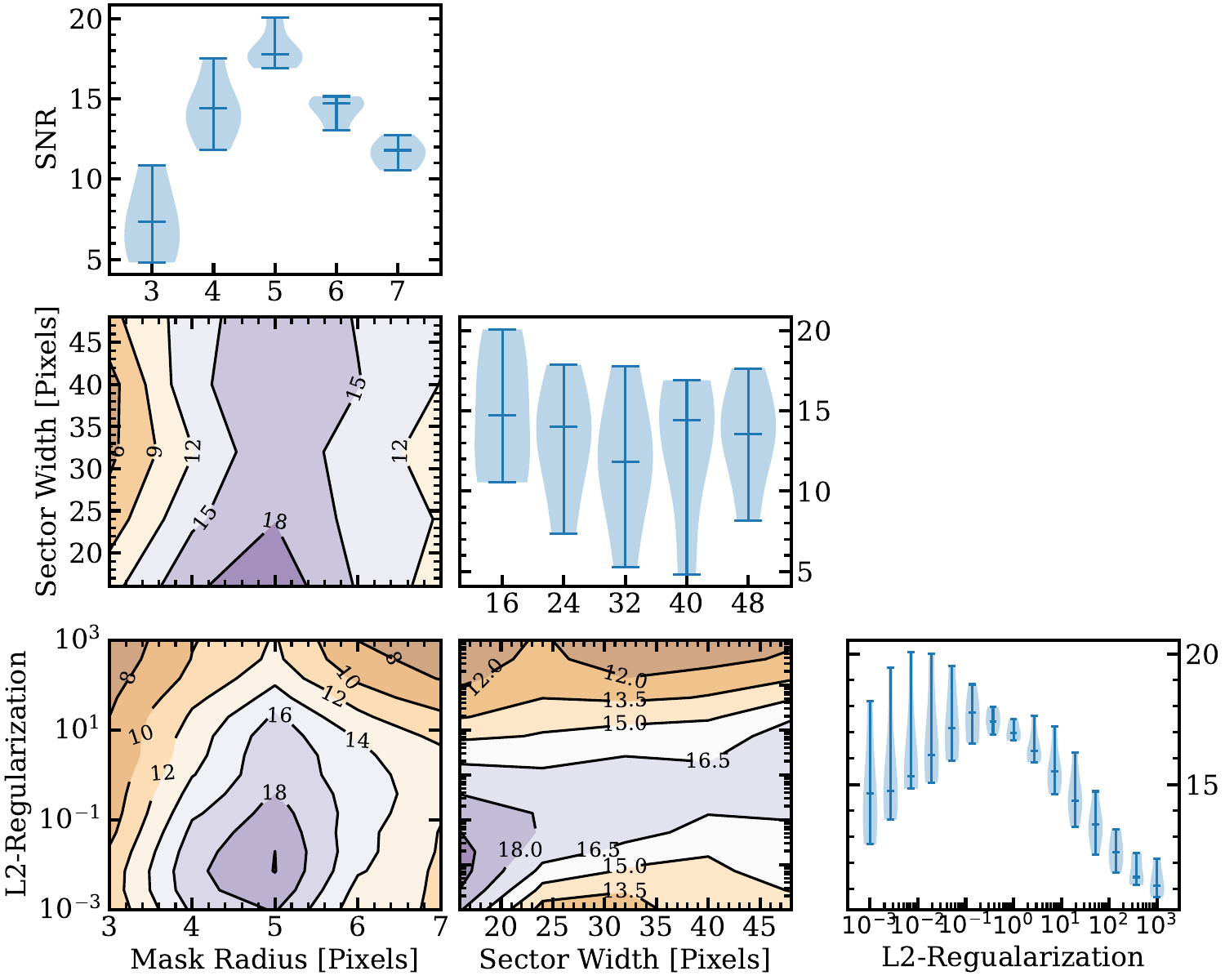} \\
    \centering \small \hspace{10mm} (c) HR 8799 d
  \end{tabular}%
  \quad
  \begin{tabular}[b]{@{}p{0.42\textwidth}@{}}
    \hspace{6mm}\includegraphics[width=1.0\linewidth]{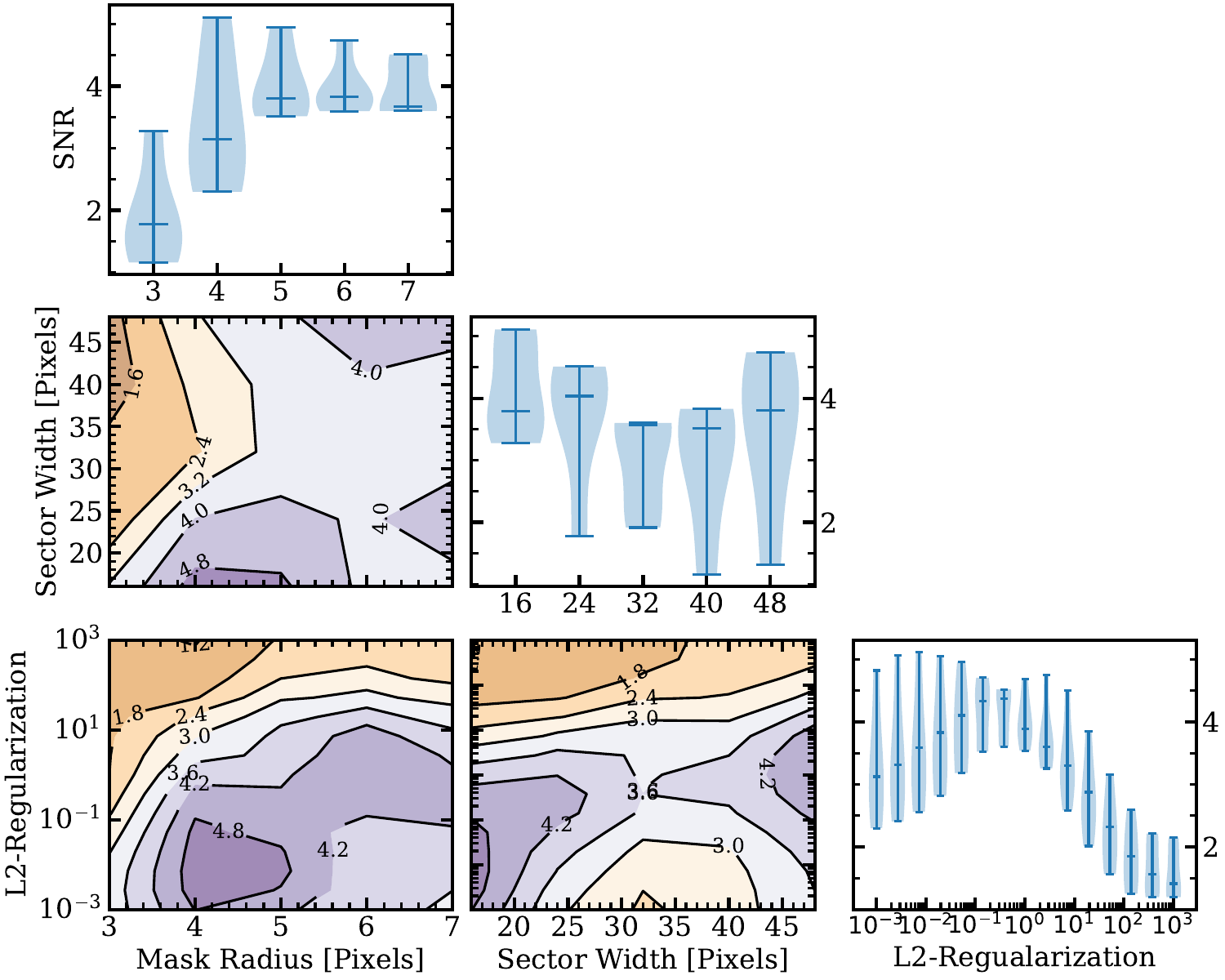} \\
    \centering \small \hspace{10mm} (d) HR 8799 e
  \end{tabular}
  \caption{Marginalized representations for our parametric parameter search for the $2012-10-26$ epoch of HR 8799. Each contour plot shows the S/N values projected onto two of the variables. The projection operation takes the maximum S/N along the axis of the marginalized variable. The violin plots show the spread in the S/N in each variable bin. Panels (a), (b), (c), and (d) correspond to HR 8799 b, c, d, and e, respectively.}
\label{fig:corner_plot_figs_19}
\end{figure}

Here, we show the additional marginalized representations of the 3-D S/N parameter search, used for tuning \texttt{ConStruct}. Figure \ref{fig:corner_plot_figs_18} shows the representation for Sequence 9 in Table \ref{tab:testing_data}. Figure \ref{fig:corner_plot_figs_19} shows the marginalized representation for Sequence 10.

%

%

\clearpage
\section{Tuning PCA reductions with the HR 8799 Planets} \label{sec:Appendix_C2}
Here, we show the performance of PCA post-processing over a grid of parameters tested on the HR 8799 planets. The grid is generated over a range of PCA components and annulus widths. The PCA components are sampled in increments of four components, starting at five, and terminating at the maximum allowable components (i.e., number of frames in the \ADI\hspace{0.1ex} sequence). The grid samples annulus widths in increments of three pixels, starting at 8 pixels and ending at 30. 
%
\begin{figure}[hbt!]
  \centering
  \centering \begin{tabular}[b]{@{}p{0.42\textwidth}@{}}
    \includegraphics[width=1.0\linewidth]{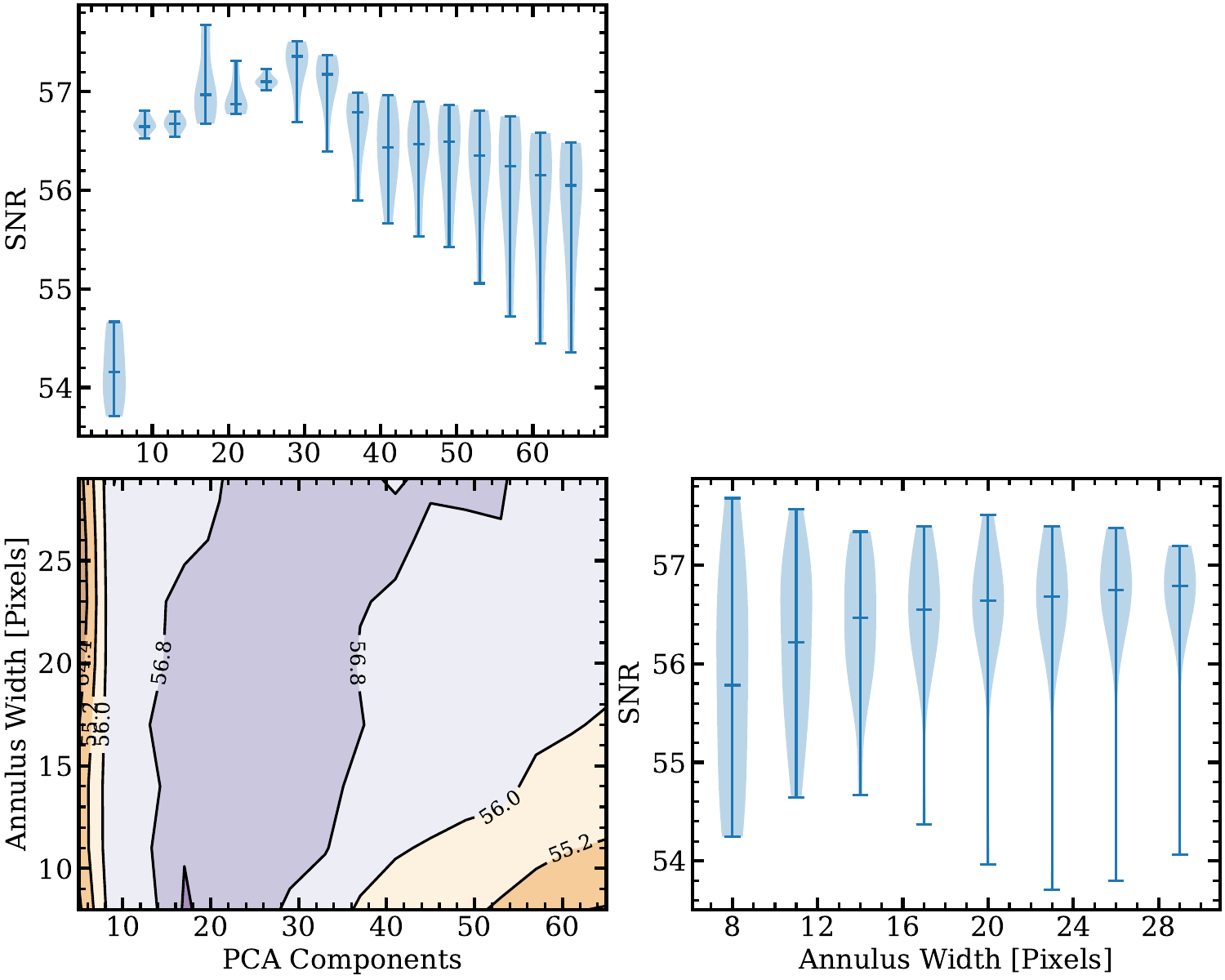} \\
    \centering \small (a) HR 8799 b
  \end{tabular}%
  \quad
  \begin{tabular}[b]{@{}p{0.42\textwidth}@{}}
    \includegraphics[width=1.0\linewidth]{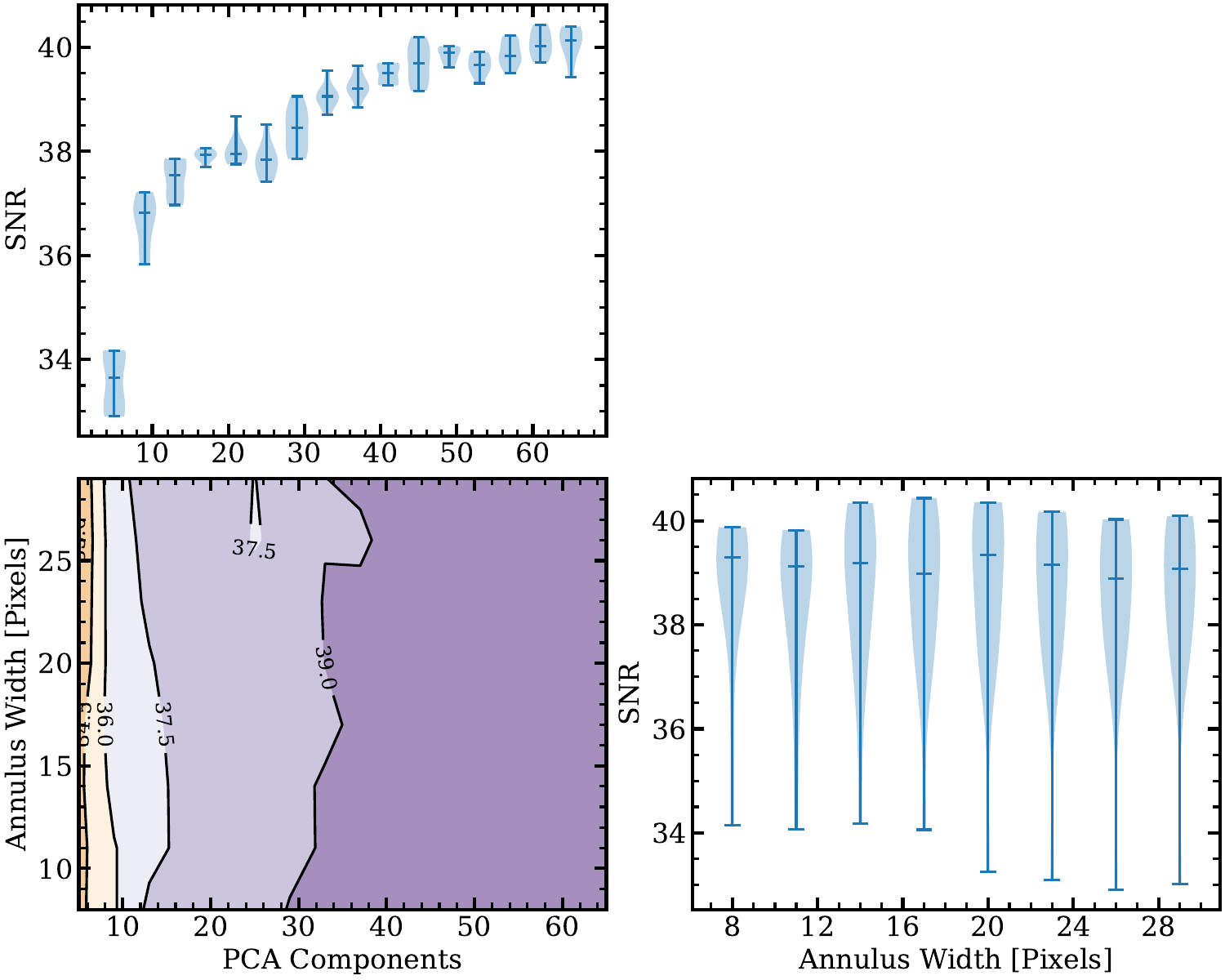} \\
    \centering \small (b) HR 8799 c
  \end{tabular}
  \begin{tabular}[b]{@{}p{0.42\textwidth}@{}}
    \vspace{2mm}\hspace{3mm}
    \includegraphics[width=1.0\linewidth]{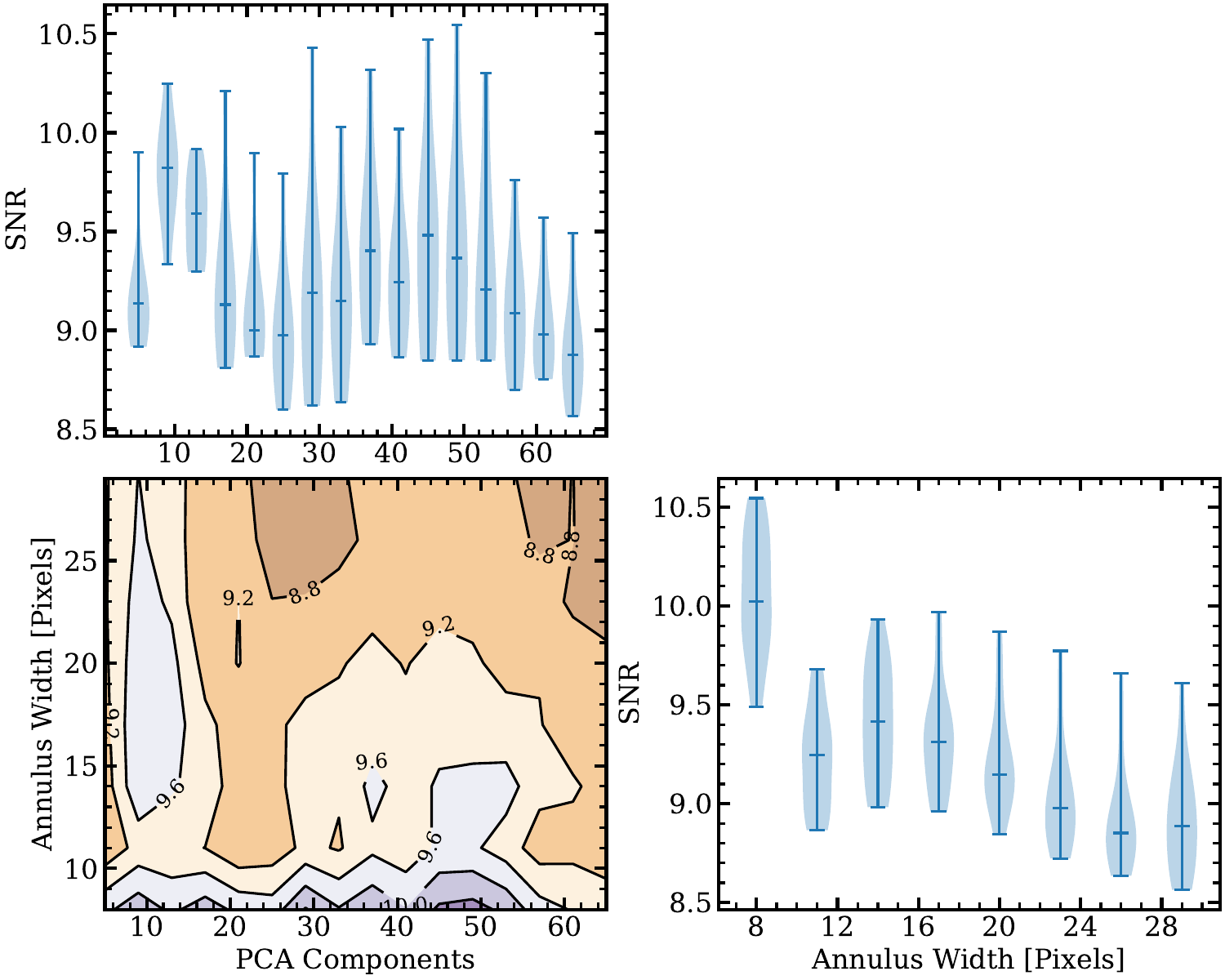} \\
    \centering \small \hspace{10mm} (c) HR 8799 d
  \end{tabular}%
  \quad
  \begin{tabular}[b]{@{}p{0.42\textwidth}@{}}
    \hspace{6mm}\includegraphics[width=1.0\linewidth]{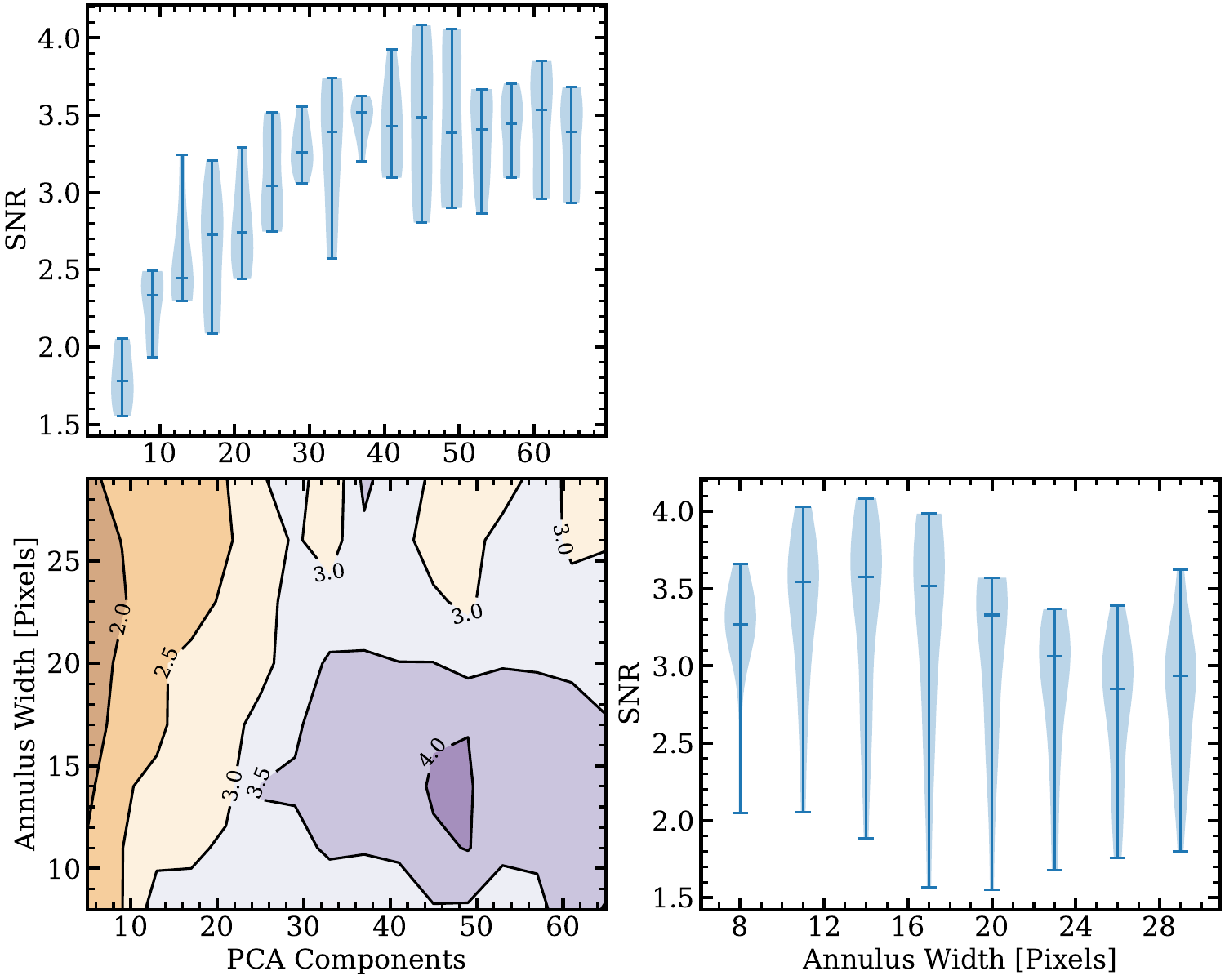} \\
    \centering \small \hspace{10mm} (d) HR 8799 e
  \end{tabular}
  \caption{Parametric PCA post-processing search for the $2010-07-13$ epoch data set of HR 8799. The contour plot shows the recovered S/N over the grid of parameters tested. The violin plots show the spread in the S/N in each variable bin. Panels (a), (b), (c), (d) correspond to HR 8799 b, c, d, and e, respectively.}
\label{fig:corner_plot_figs_18_KLIP}
\end{figure}

%
\begin{figure}[hbt!]
  \centering
  \centering \begin{tabular}[b]{@{}p{0.42\textwidth}@{}}
    \includegraphics[width=1.0\linewidth]{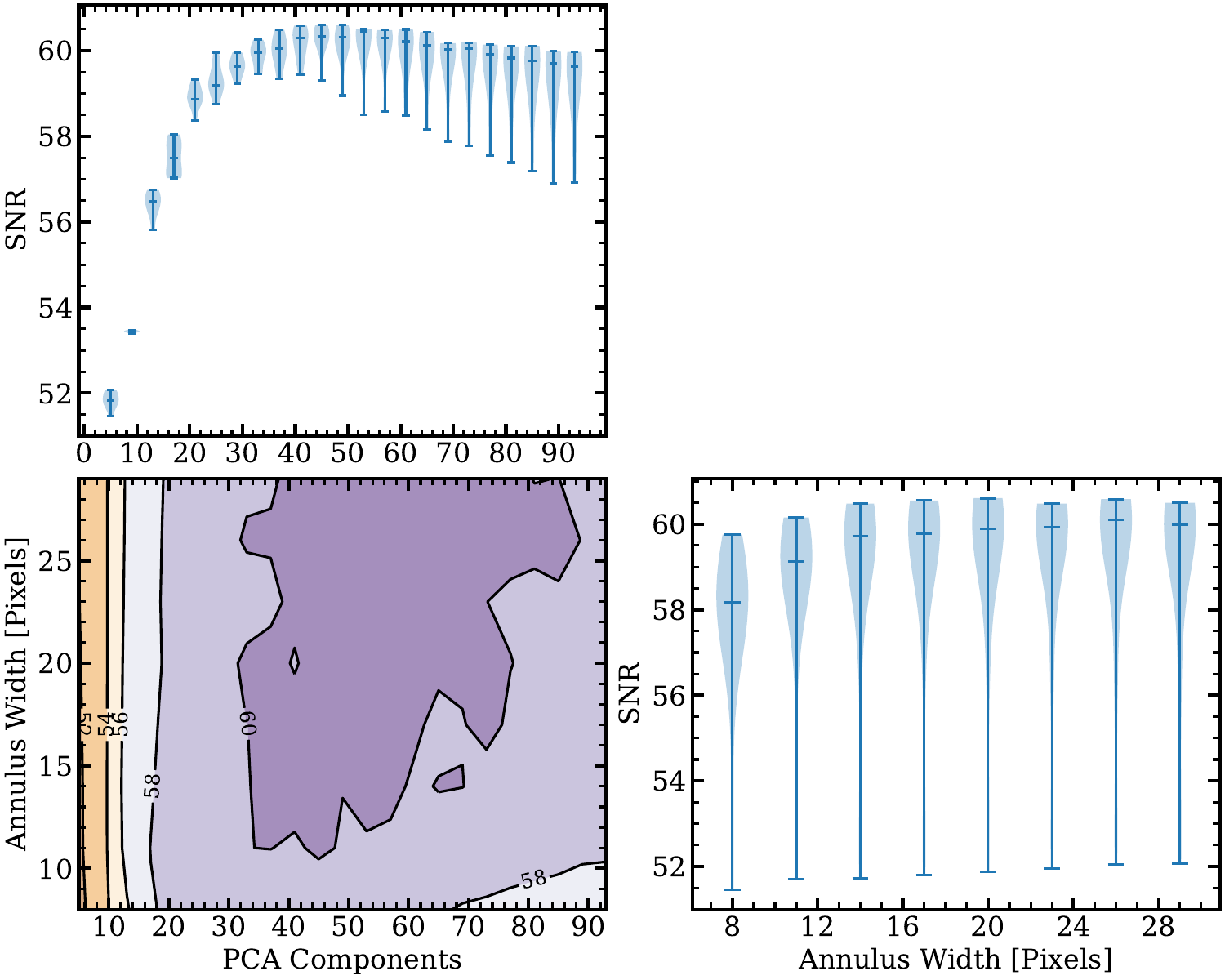} \\
    \centering \small (a) HR 8799 b
  \end{tabular}%
  \quad
  \begin{tabular}[b]{@{}p{0.42\textwidth}@{}}
    \includegraphics[width=1.0\linewidth]{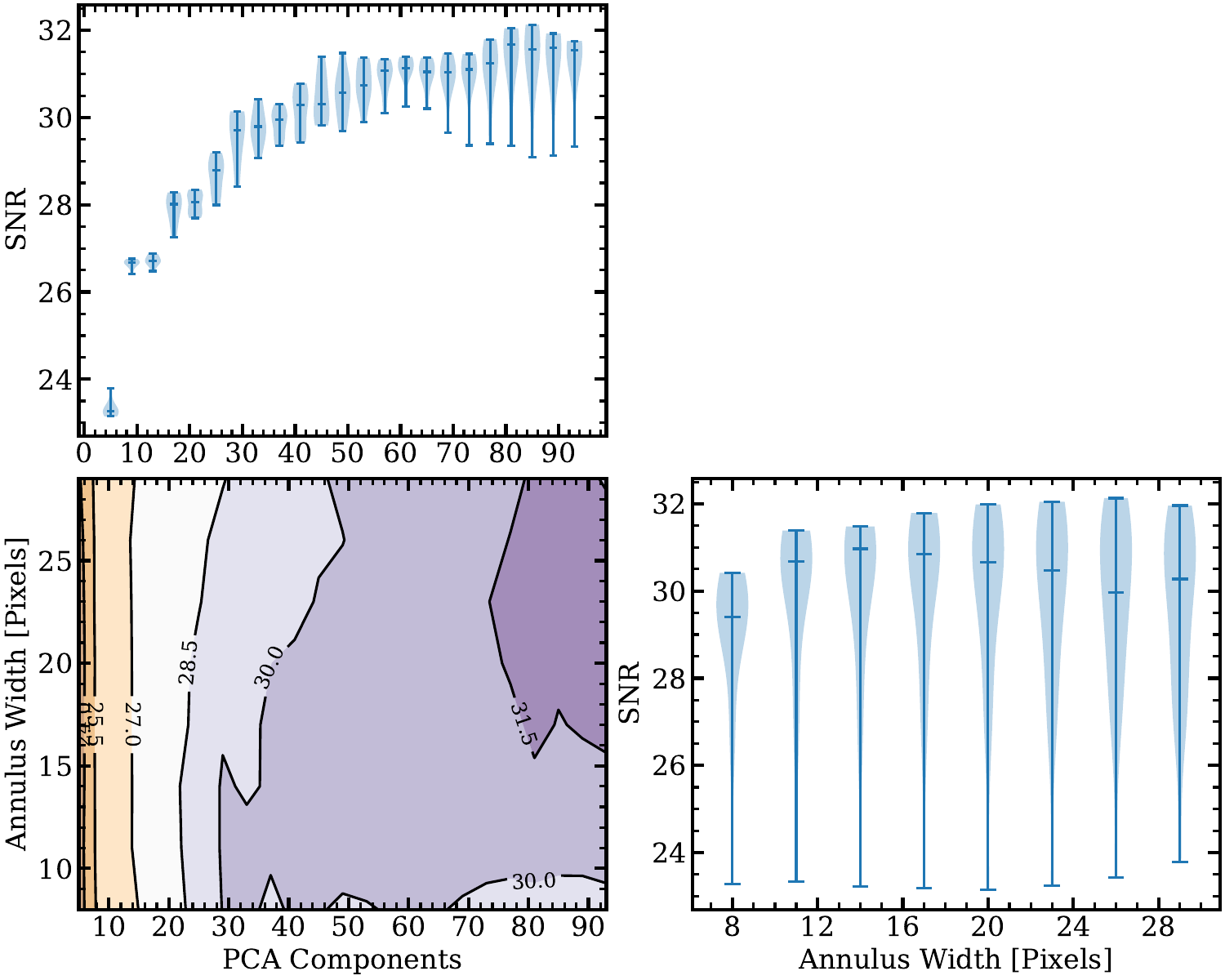} \\
    \centering \small (b) HR 8799 c
  \end{tabular}
  \begin{tabular}[b]{@{}p{0.42\textwidth}@{}}
    \vspace{2mm}\hspace{3mm}
    \includegraphics[width=1.0\linewidth]{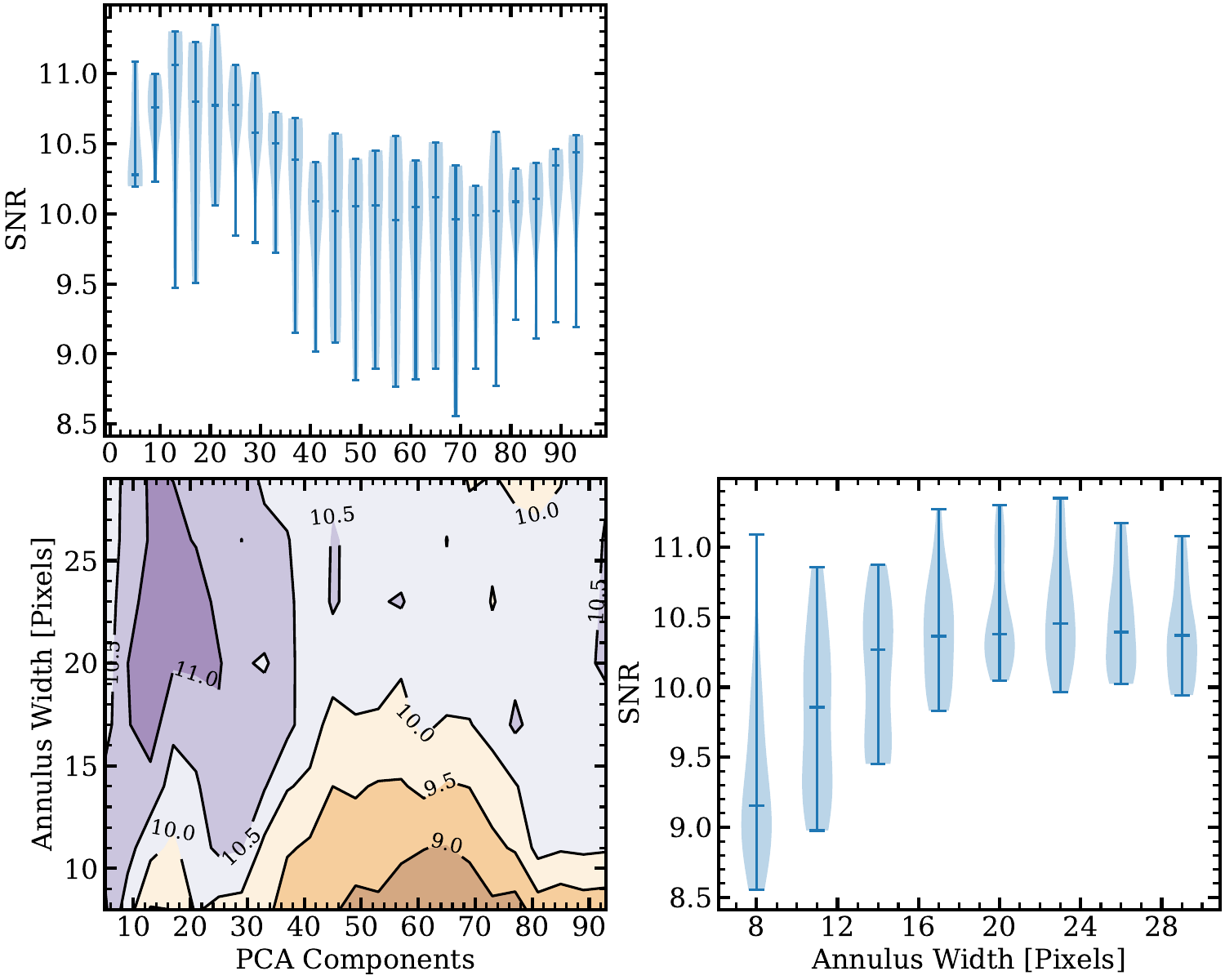} \\
    \centering \small \hspace{10mm} (c) HR 8799 d
  \end{tabular}%
  \quad
  \begin{tabular}[b]{@{}p{0.42\textwidth}@{}}
    \hspace{6mm}\includegraphics[width=1.0\linewidth]{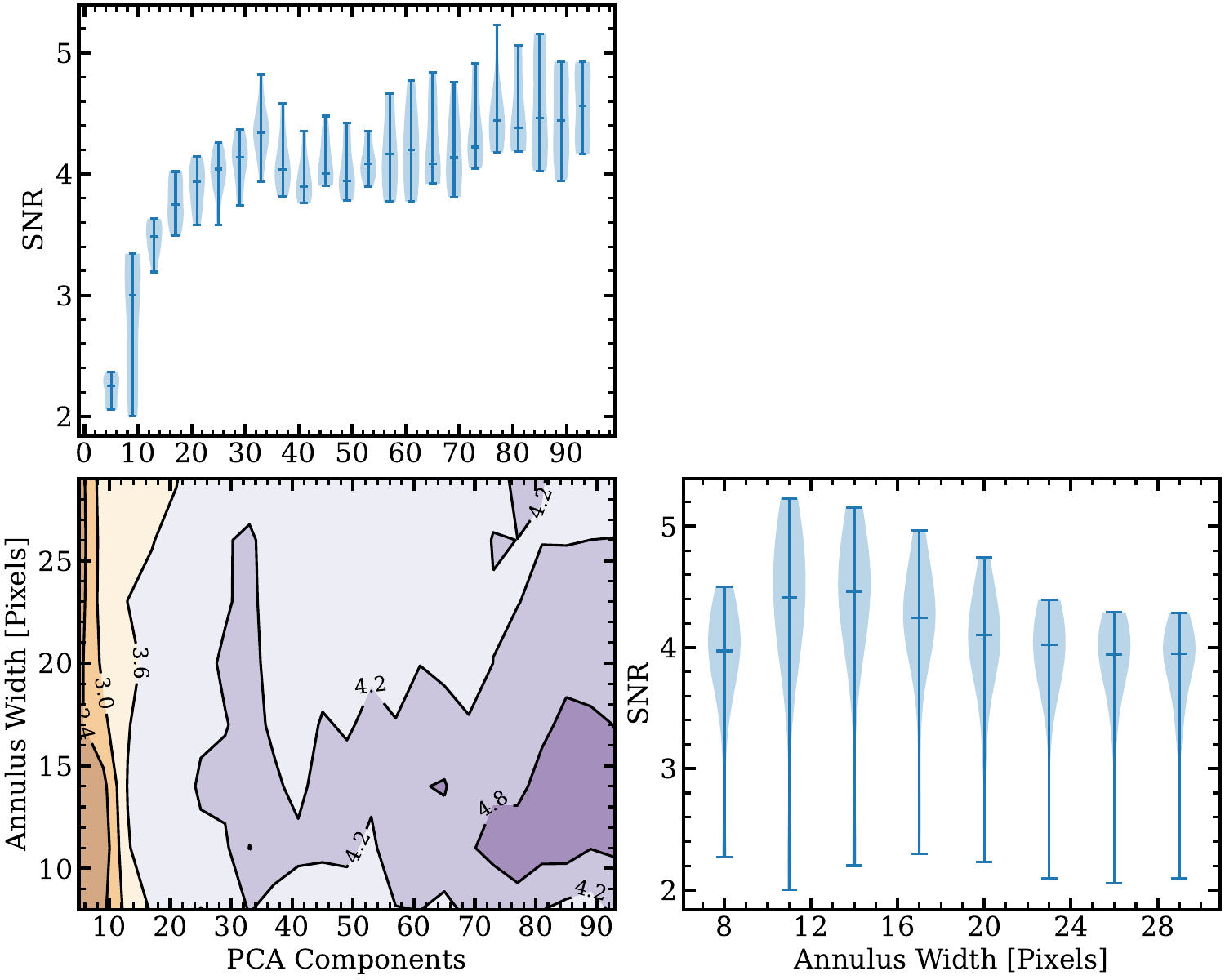} \\
    \centering \small \hspace{10mm} (d) HR 8799 e
  \end{tabular}
  \caption{Parametric PCA post-processing search for the $2012-10-26$ epoch data set of HR 8799. The contour plot shows the recovered S/N over the grid of parameters tested. The violin plots show the spread in the S/N in each variable bin. Panels (a), (b), (c), (d) correspond to HR 8799 b, c, d, and e, respectively.}
\label{fig:corner_plot_figs_19_KLIP}
\end{figure}

%
\begin{figure}[hbt!]
  \centering
  \centering \begin{tabular}[b]{@{}p{0.42\textwidth}@{}}
    \includegraphics[width=1.0\linewidth]{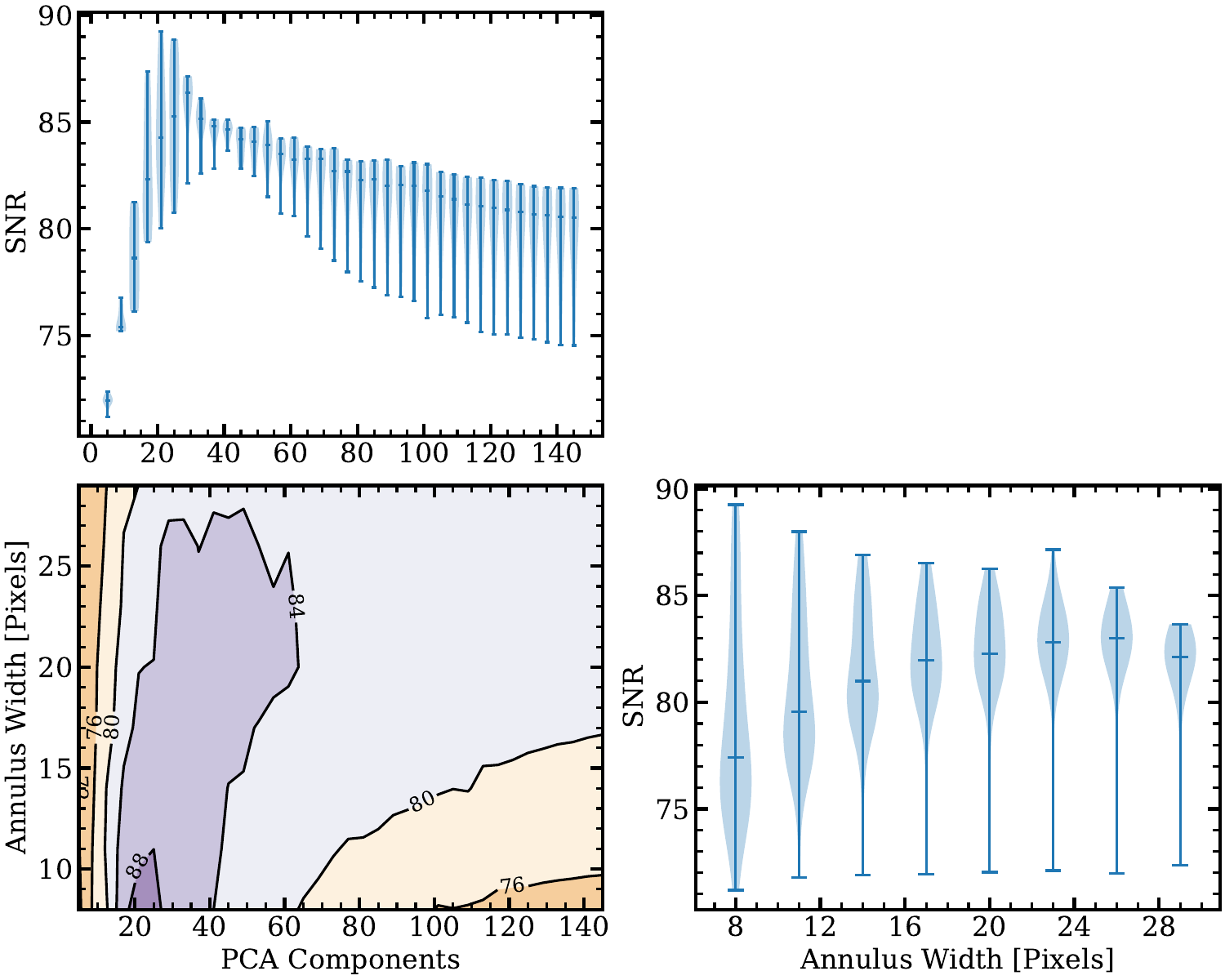} \\
    \centering \small (a) HR 8799 b
  \end{tabular}%
  \quad
  \begin{tabular}[b]{@{}p{0.42\textwidth}@{}}
    \includegraphics[width=1.0\linewidth]{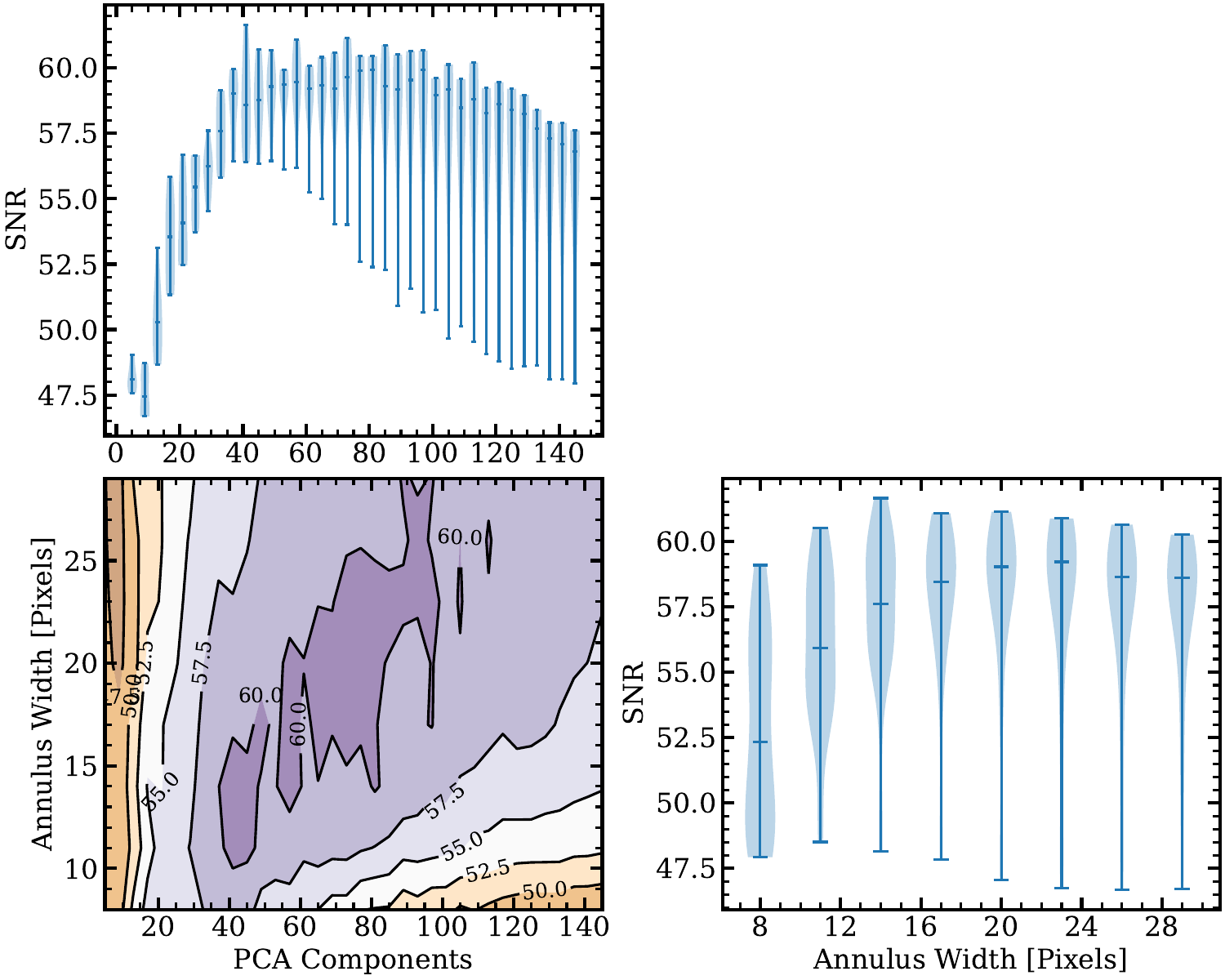} \\
    \centering \small (b) HR 8799 c
  \end{tabular}
  \begin{tabular}[b]{@{}p{0.42\textwidth}@{}}
    \vspace{2mm}\hspace{4mm}
    \includegraphics[width=1.0\linewidth]{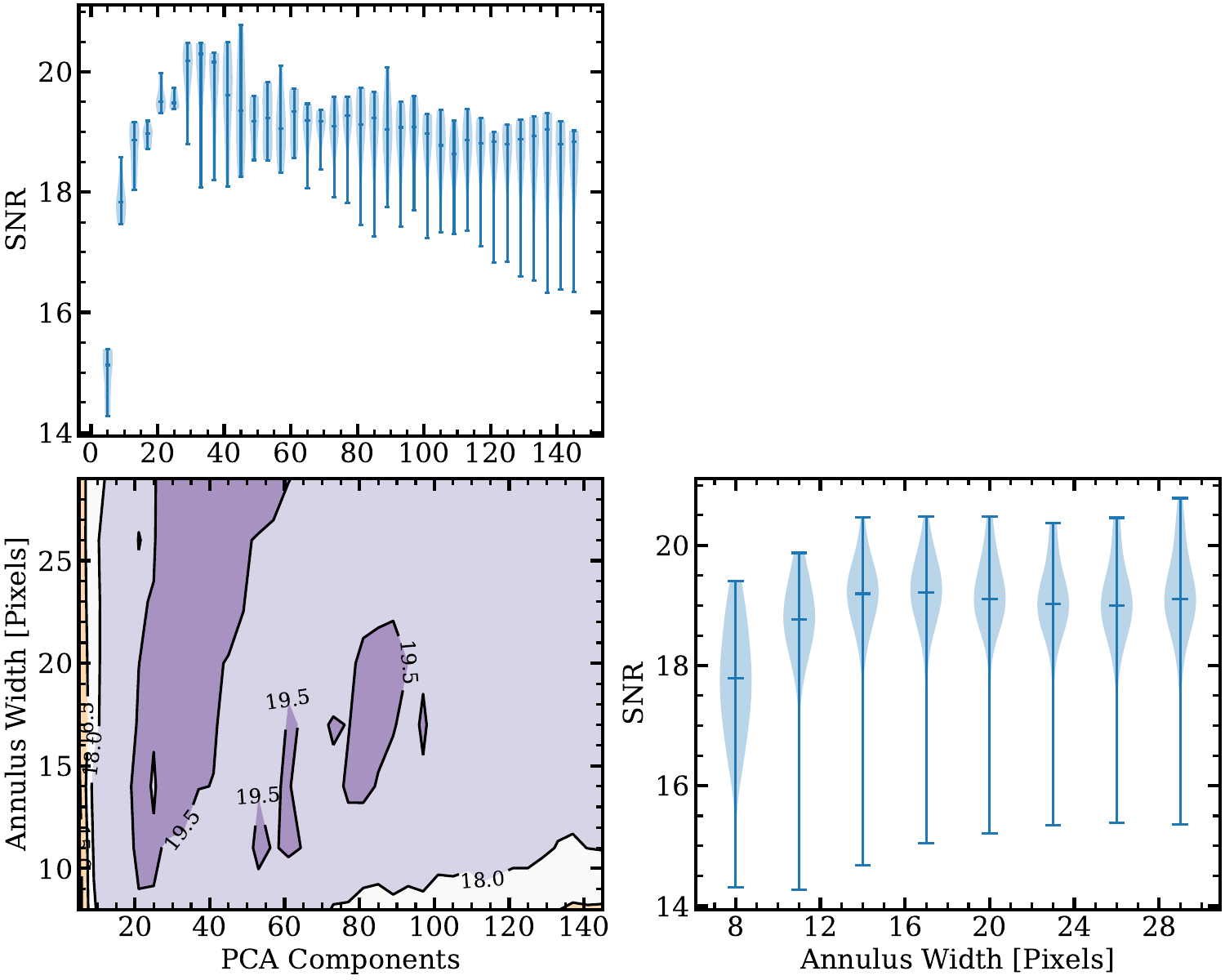} \\
    \centering \small \hspace{10mm} (c) HR 8799 d
  \end{tabular}%
  \quad
  \begin{tabular}[b]{@{}p{0.42\textwidth}@{}}
    \hspace{7mm}\includegraphics[width=1.0\linewidth]{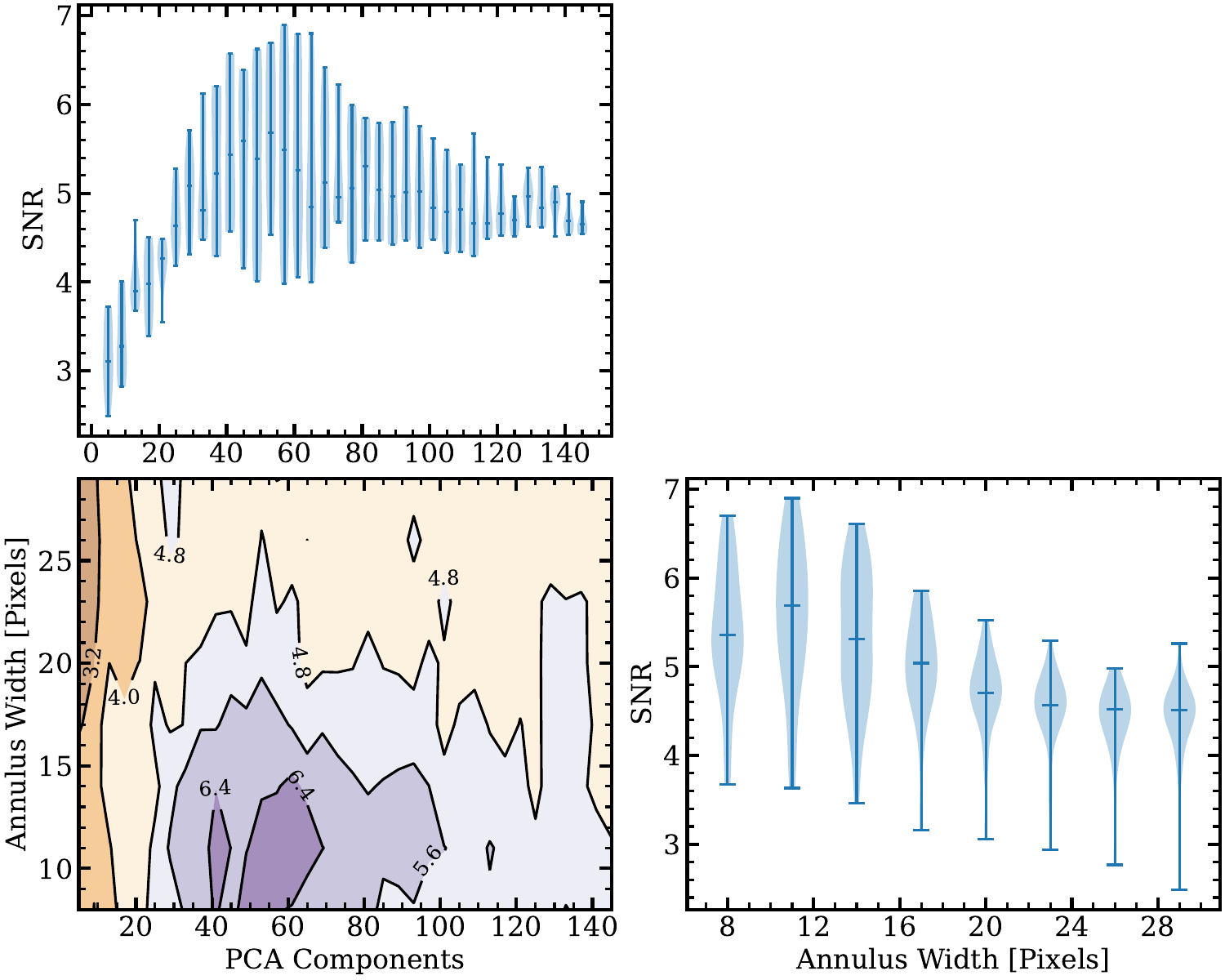} \\
    \centering \small \hspace{10mm} (d) HR 8799 e
  \end{tabular}
  \caption{Parametric PCA post-processing search for the $2011-07-21$ epoch data set of HR 8799. The contour plot shows the recovered S/N over the grid of parameters tested. The violin plots show the spread in the S/N in each variable bin. Panels (a), (b), (c), (d) correspond to HR 8799 b, c, d, and e, respectively.}
\label{fig:corner_plot_figs_20_KLIP}
\end{figure}

\clearpage
\section{Additional Reductions}\label{sec:Appendix_B}
\begin{figure*}[hbt!]
  \centering
  \begin{tabular}[b]{@{}p{1.0\textwidth}@{}}
    \centering\includegraphics[width=.95\linewidth]{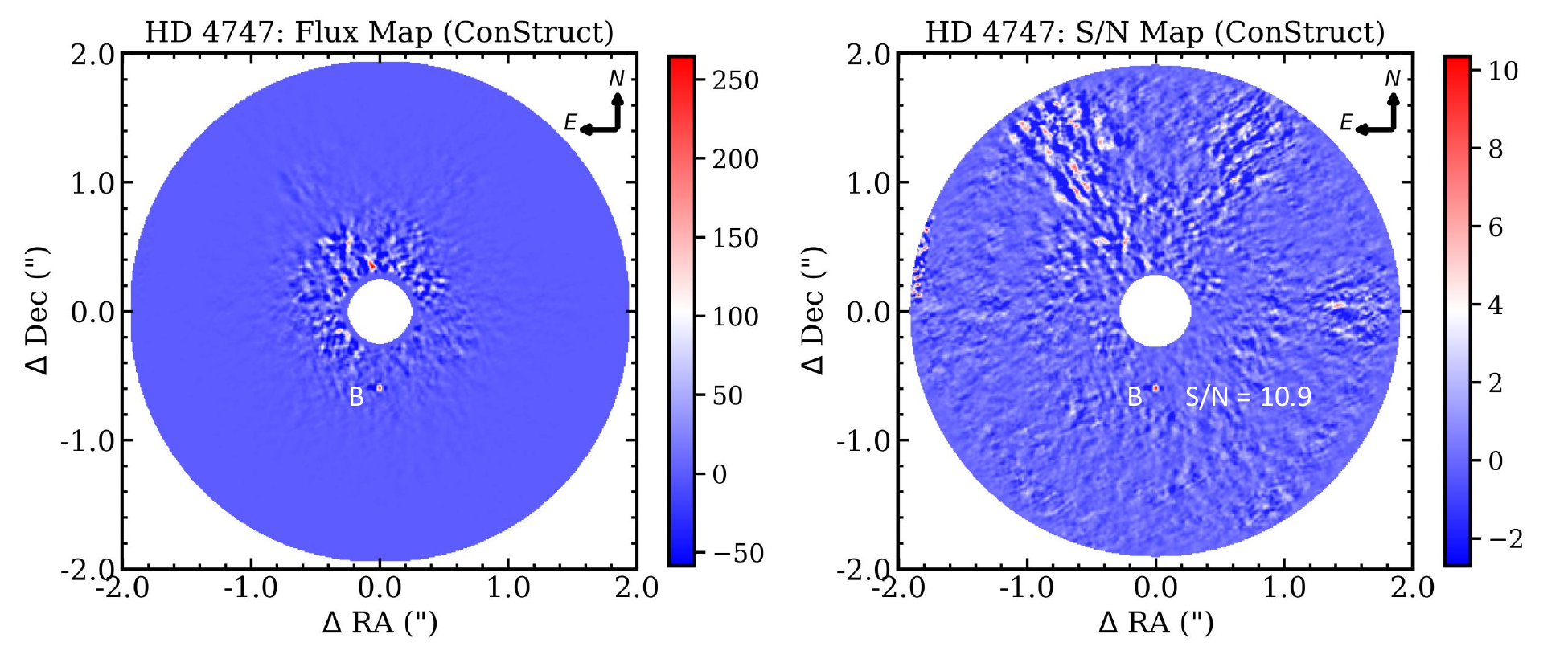} \\
    \centering\small (a) \texttt{ConStruct}
  \end{tabular}%
  \quad
  \begin{tabular}[b]{@{}p{1.0\textwidth}@{}}
    \centering\includegraphics[width=.95\linewidth]{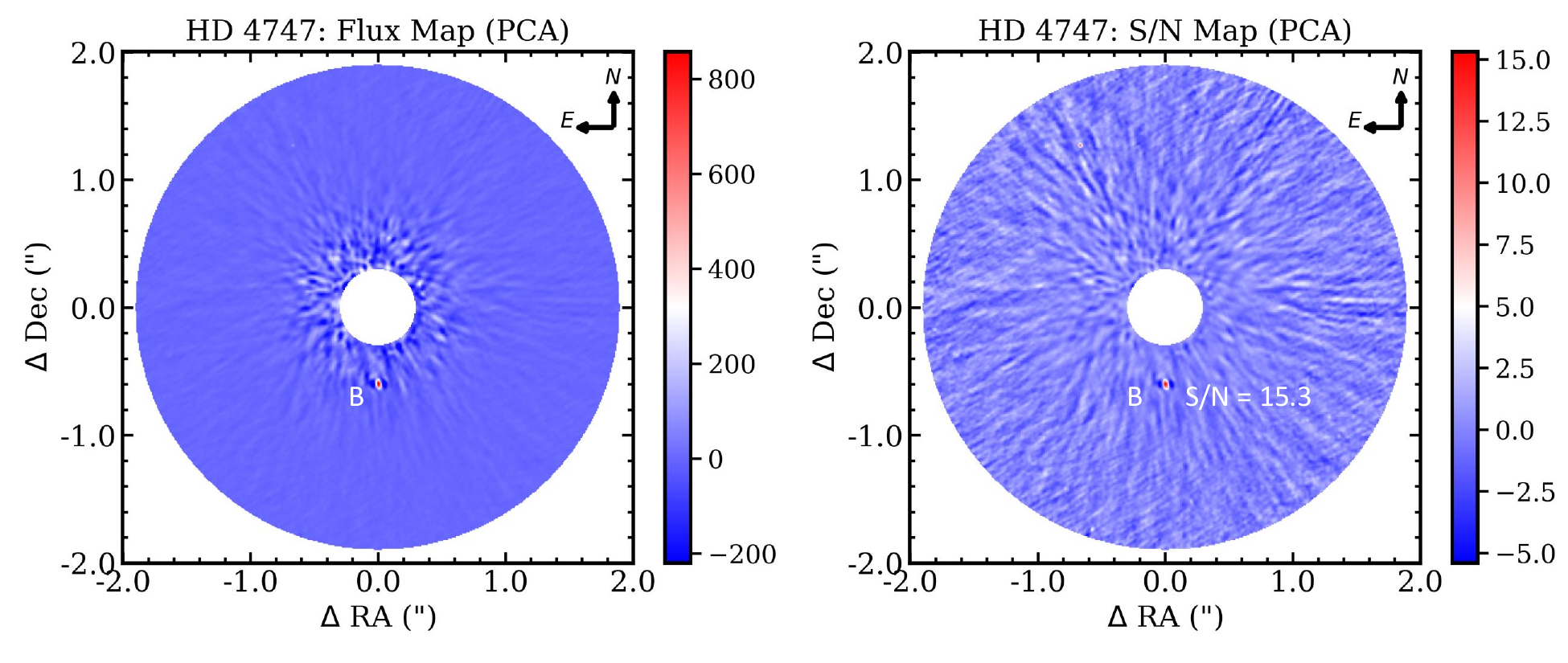} \\
    \centering\small (b) Annular PCA
  \end{tabular}
  \caption{Full-frame reductions with \texttt{ConStruct} and \PCA\hspace{-0.1ex} of HD 4747 \cite{Crepp_2016} for Sequence 1 in Table \ref{tab:testing_data}.}
  \label{fig:reductions_6}
\end{figure*}

\begin{figure*}[hbt!]
  \centering
  \begin{tabular}[b]{@{}p{1.0\textwidth}@{}}
    \centering\includegraphics[width=.95\linewidth]{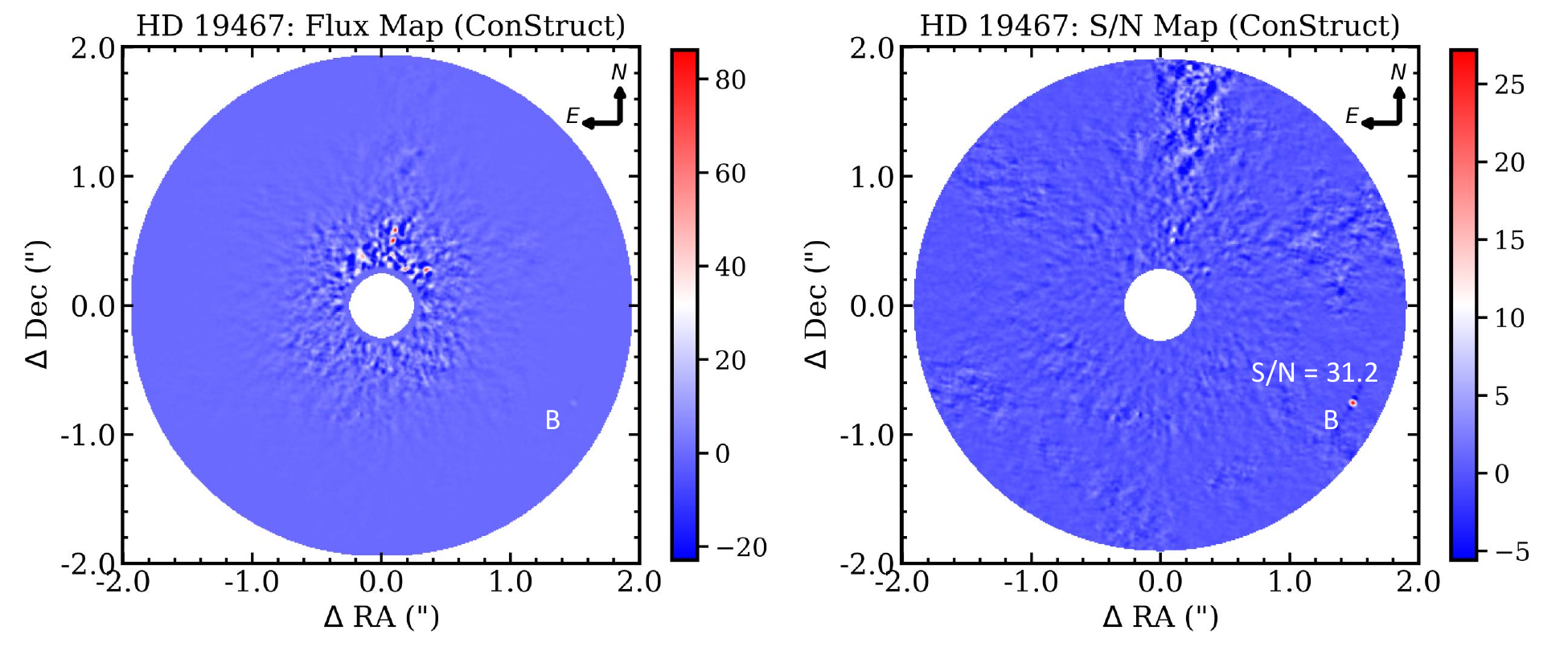} \\
    \centering\small (a) \texttt{ConStruct}
  \end{tabular}%
  \quad
  \begin{tabular}[b]{@{}p{1.0\textwidth}@{}}
    \centering\includegraphics[width=.95\linewidth]{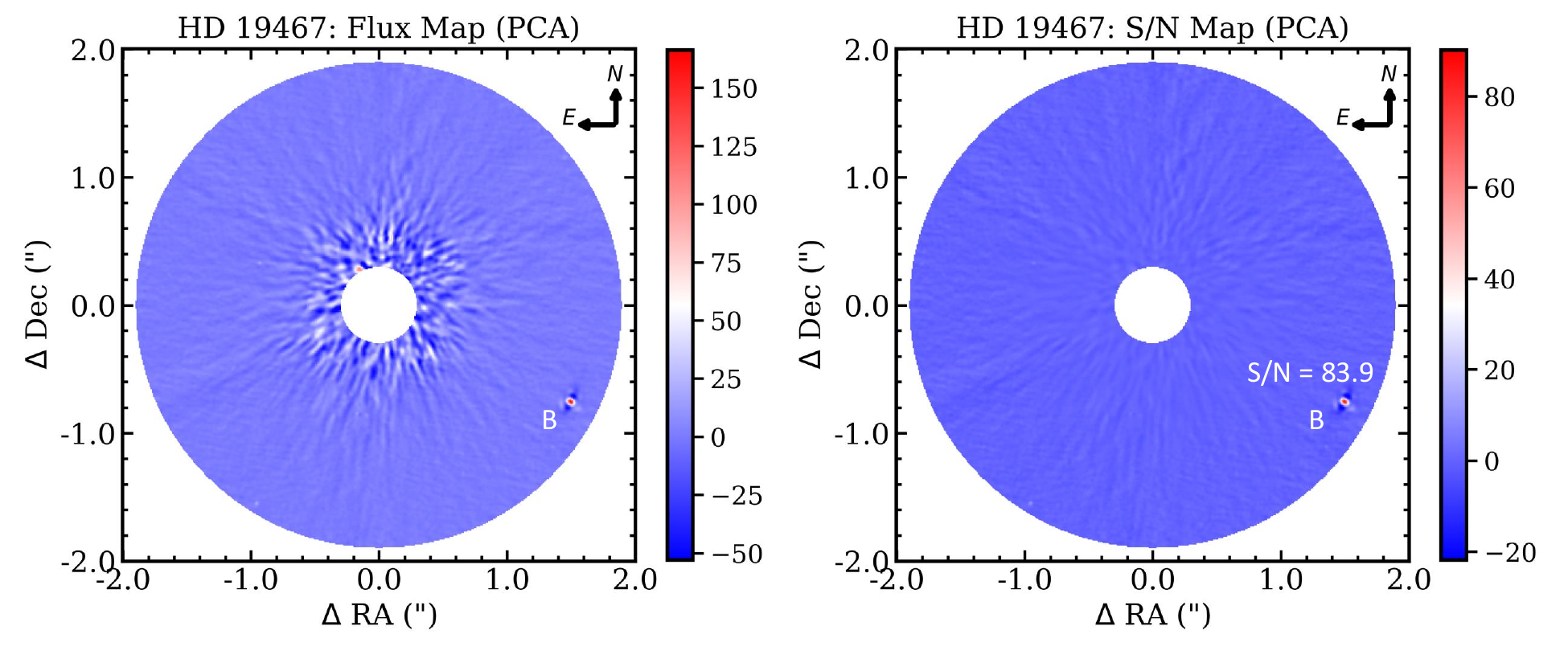} \\
    \centering\small (b) Annular PCA
  \end{tabular}
  \caption{Full-frame reductions with ConStruct and \PCA\hspace{-0.1ex} of HD 19467 \citep{Crepp_2014} for Sequence 2 in Table \ref{tab:testing_data}.}
  \label{fig:reductions_8}
\end{figure*}

\begin{figure*}[hbt!]
  \centering
  \begin{tabular}[b]{@{}p{1.0\textwidth}@{}}
    \centering\includegraphics[width=.95\linewidth]{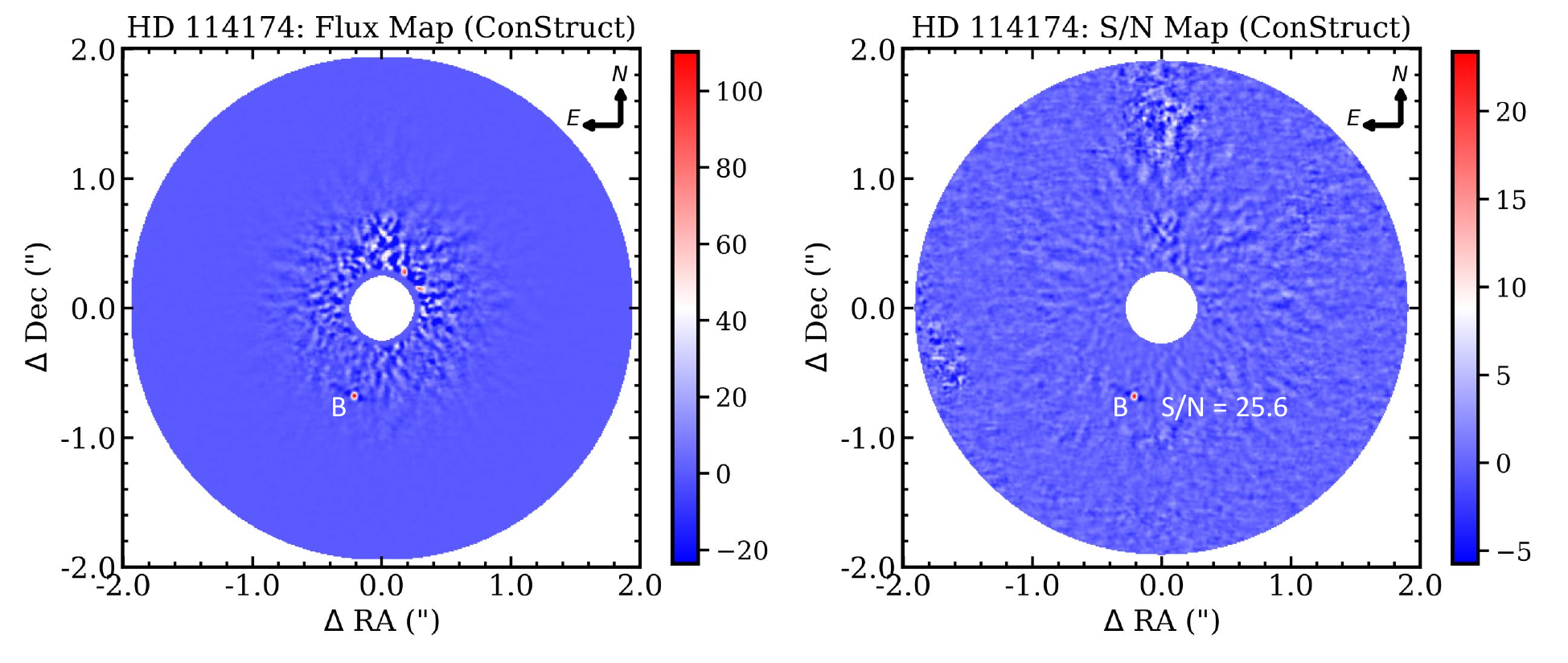} \\
    \centering\small (a) \texttt{ConStruct}
  \end{tabular}%
  \quad
  \begin{tabular}[b]{@{}p{1.0\textwidth}@{}}
    \centering\includegraphics[width=.95\linewidth]{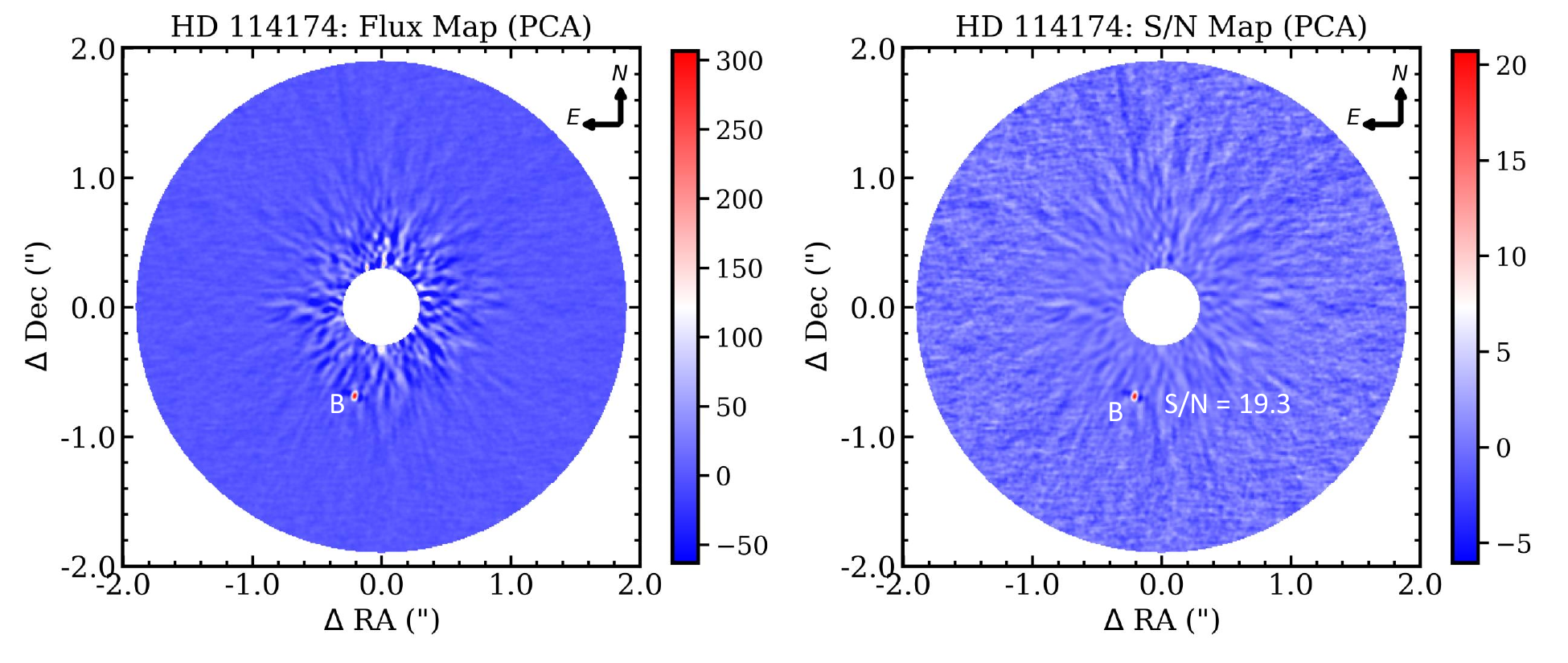} \\
    \centering\small (b) Annular PCA
  \end{tabular}
  \caption{Full-frame reductions with \texttt{ConStruct} and \PCA\hspace{-0.1ex} of HD 114174 \citep{Crepp_2013} for Sequence 4 in Table \ref{tab:testing_data}.}
  \label{fig:reductions_12}
\end{figure*}

\begin{figure*}[hbt!]
  \centering
  \begin{tabular}[b]{@{}p{1.0\textwidth}@{}}
    \centering\includegraphics[width=.95\linewidth]{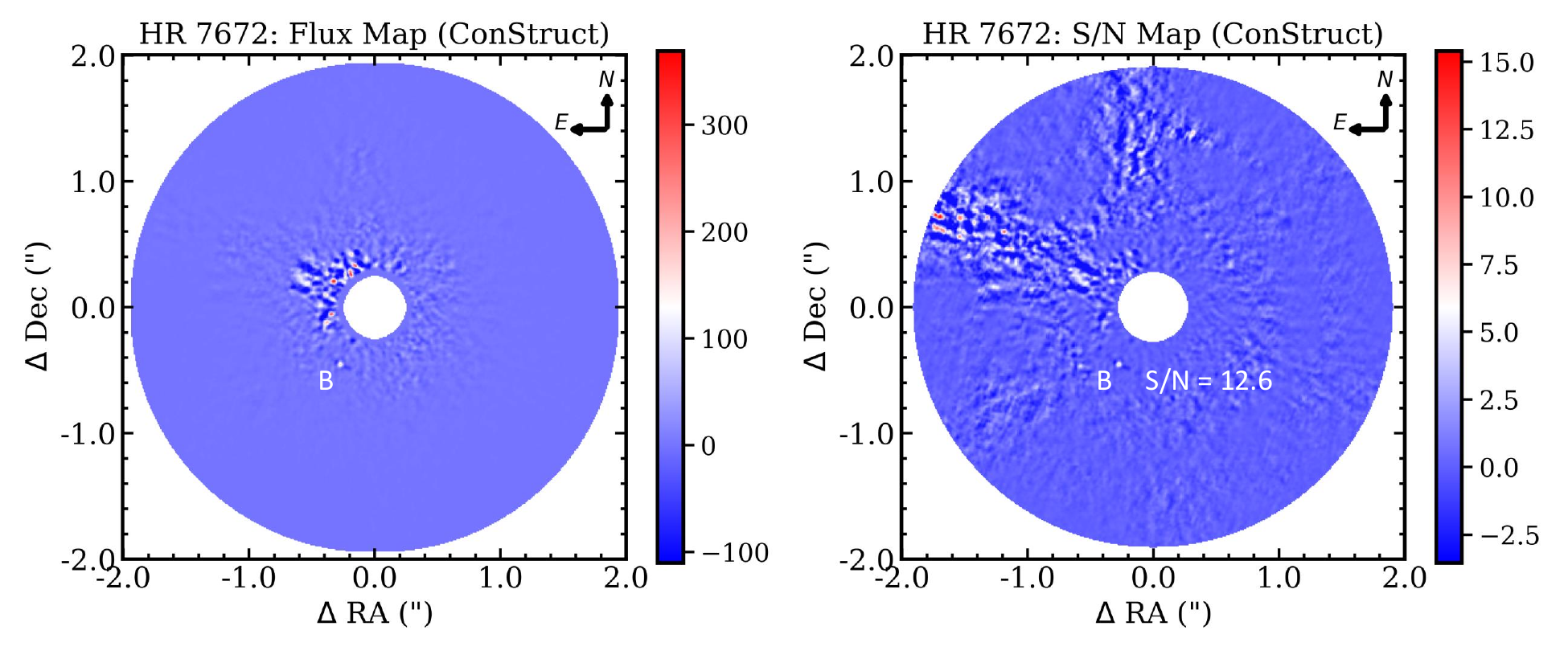} \\
    \centering\small (a) \texttt{ConStruct}
  \end{tabular}%
  \quad
  \begin{tabular}[b]{@{}p{1.0\textwidth}@{}}
    \centering\includegraphics[width=.95\linewidth]{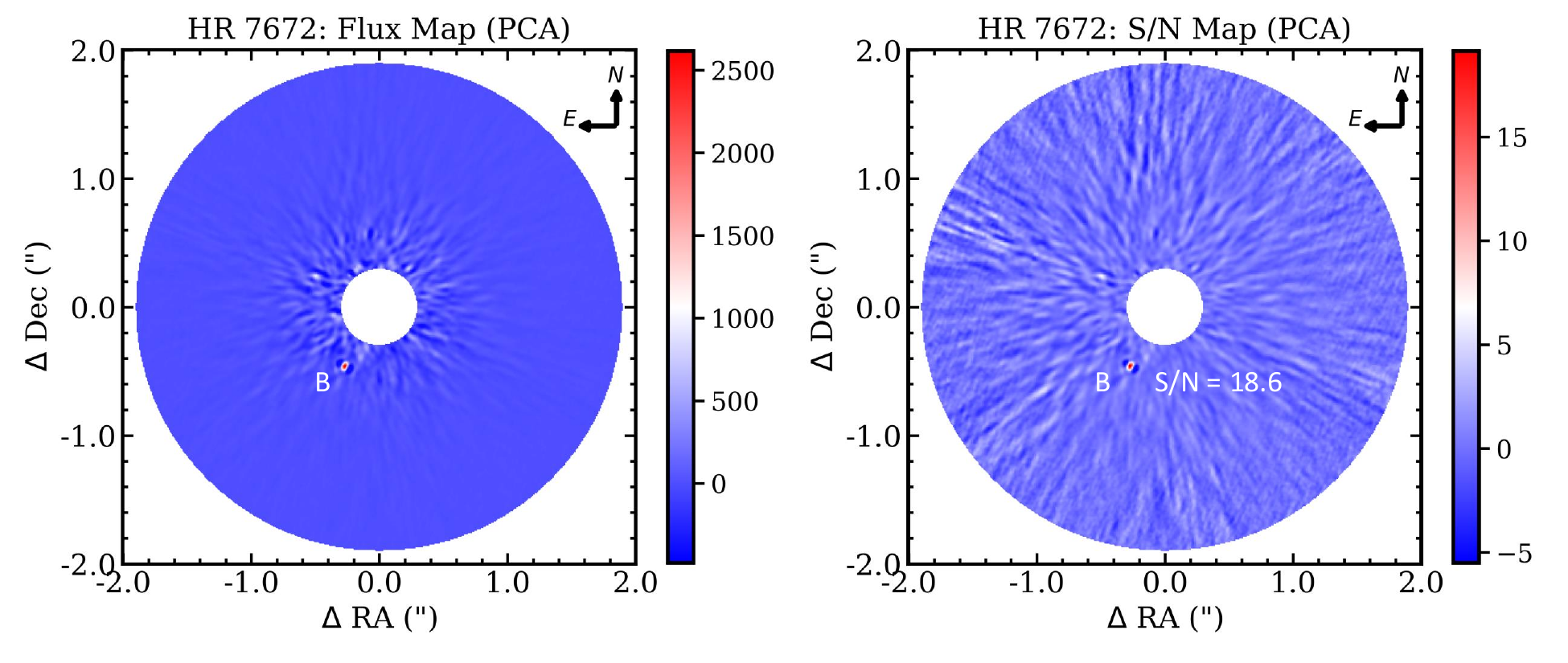} \\
    \centering\small (b) Annular PCA
  \end{tabular}
  \caption{Full-frame reductions with \texttt{ConStruct} and \PCA\hspace{-0.1ex} of HR 7672 \citep{Liu_2002} for Sequence 6 in Table \ref{tab:testing_data}.}
  \label{fig:reductions_14}
\end{figure*}

\begin{figure*}[hbt!]
  \centering
  \begin{tabular}[b]{@{}p{1.0\textwidth}@{}}
    \centering\includegraphics[width=.95\linewidth]{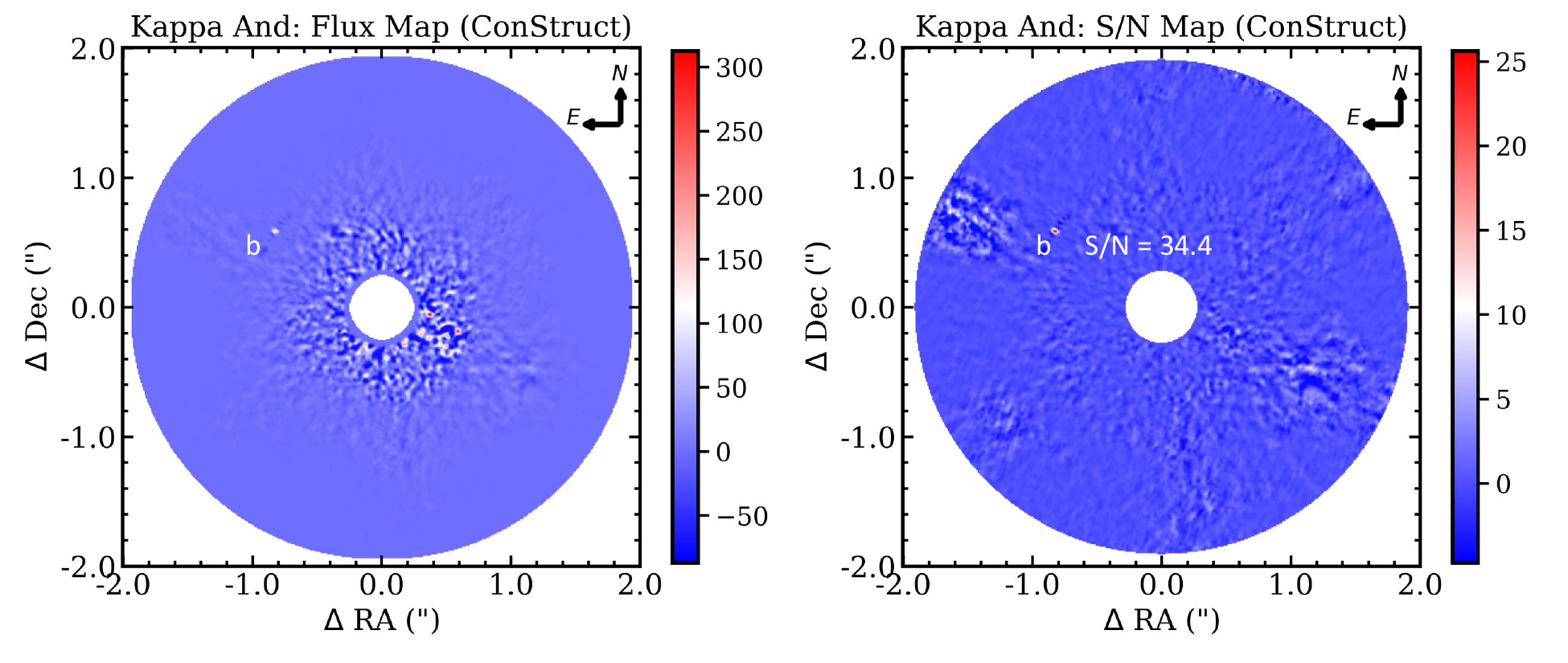} \\
    \centering\small (a) \texttt{ConStruct}
  \end{tabular}%
  \quad
  \begin{tabular}[b]{@{}p{1.0\textwidth}@{}}
    \centering\includegraphics[width=.95\linewidth]{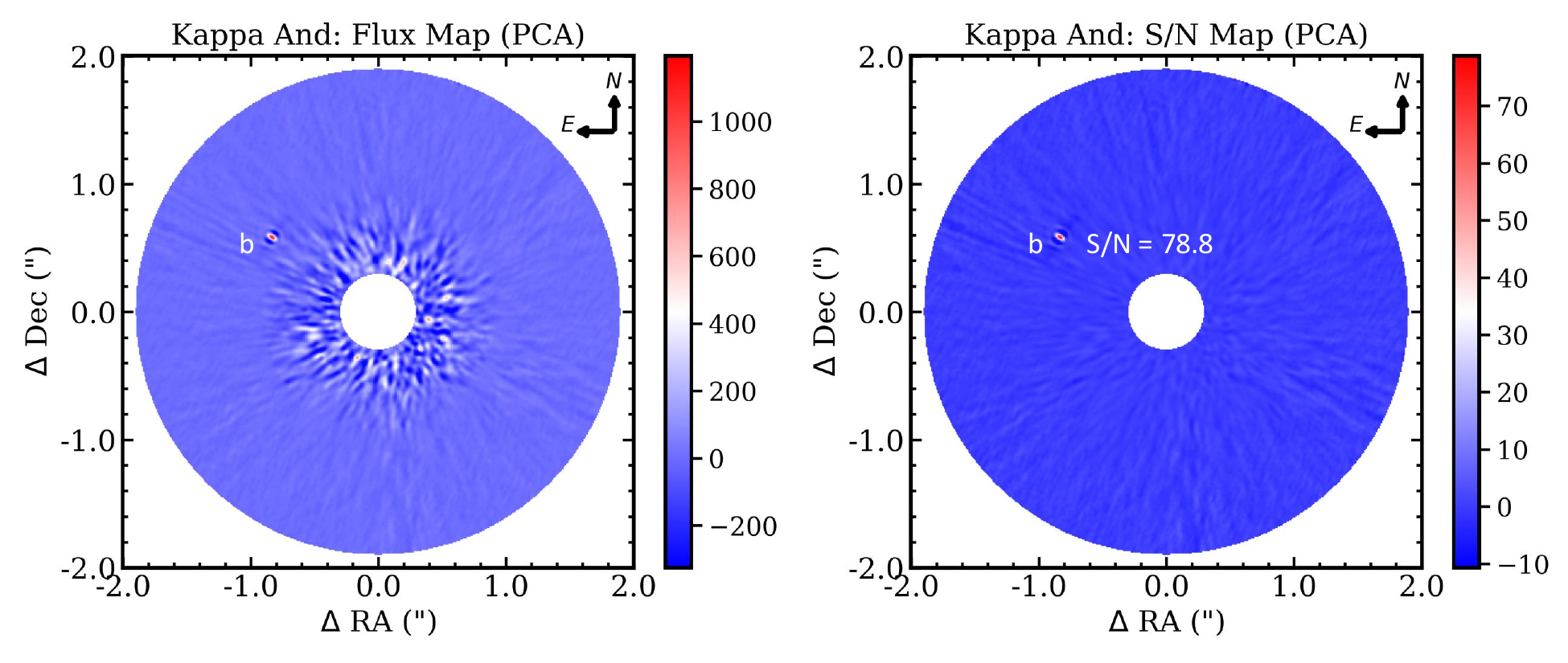} \\
    \centering\small (b) Annular PCA
  \end{tabular}
  \caption{Full-frame reductions with \texttt{ConStruct} and \PCA\hspace{-0.1ex} of Kappa And \citep{Carson_2013} for Sequence 7 in Table \ref{tab:testing_data}.}
  \label{fig:reductions_15}
\end{figure*}

\begin{figure*}[hbt!]
  \centering
  \begin{tabular}[b]{@{}p{1.0\textwidth}@{}}
    \centering\includegraphics[width=.95\linewidth]{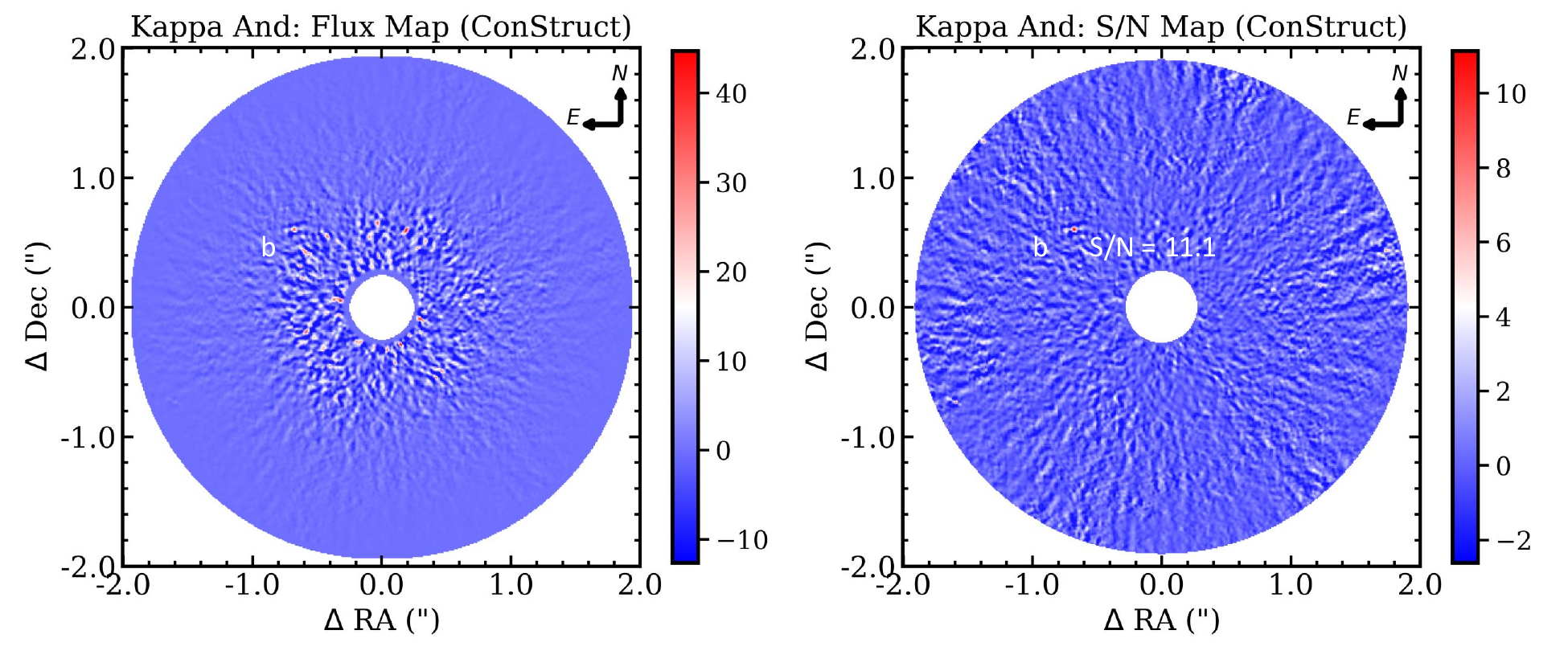} \\
    \centering\small (a) \texttt{ConStruct}
  \end{tabular}%
  \quad
  \begin{tabular}[b]{@{}p{1.0\textwidth}@{}}
    \centering\includegraphics[width=.95\linewidth]{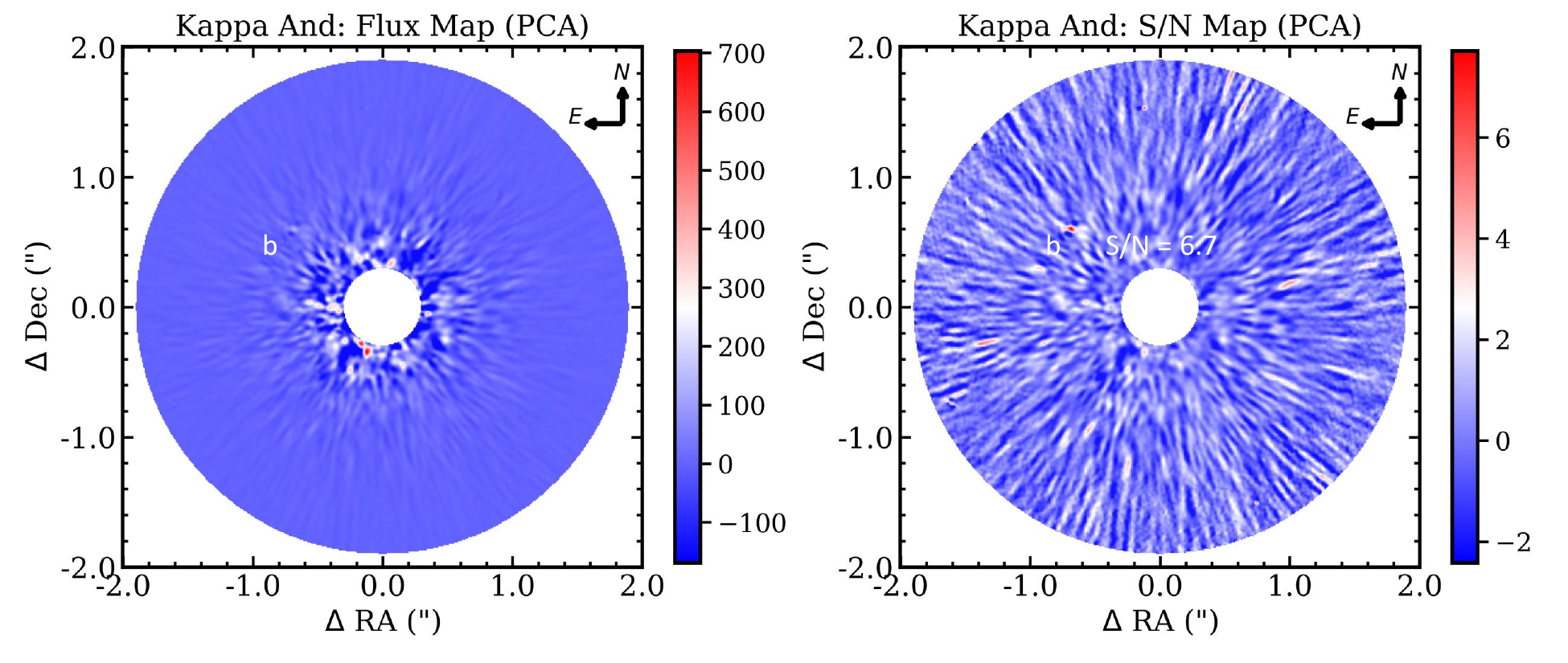} \\
    \centering\small (b) Annular PCA
  \end{tabular}
  \caption{Full-frame reductions with \texttt{ConStruct} and \PCA\hspace{-0.1ex} of Kappa And \citep{Carson_2013} for Sequence 8 in Table \ref{tab:testing_data}.}
  \label{fig:reductions_16}
\end{figure*}

\begin{figure*}[hbt!]
  \centering
  \begin{tabular}[b]{@{}p{1.0\textwidth}@{}}
    \centering\includegraphics[width=.95\linewidth]{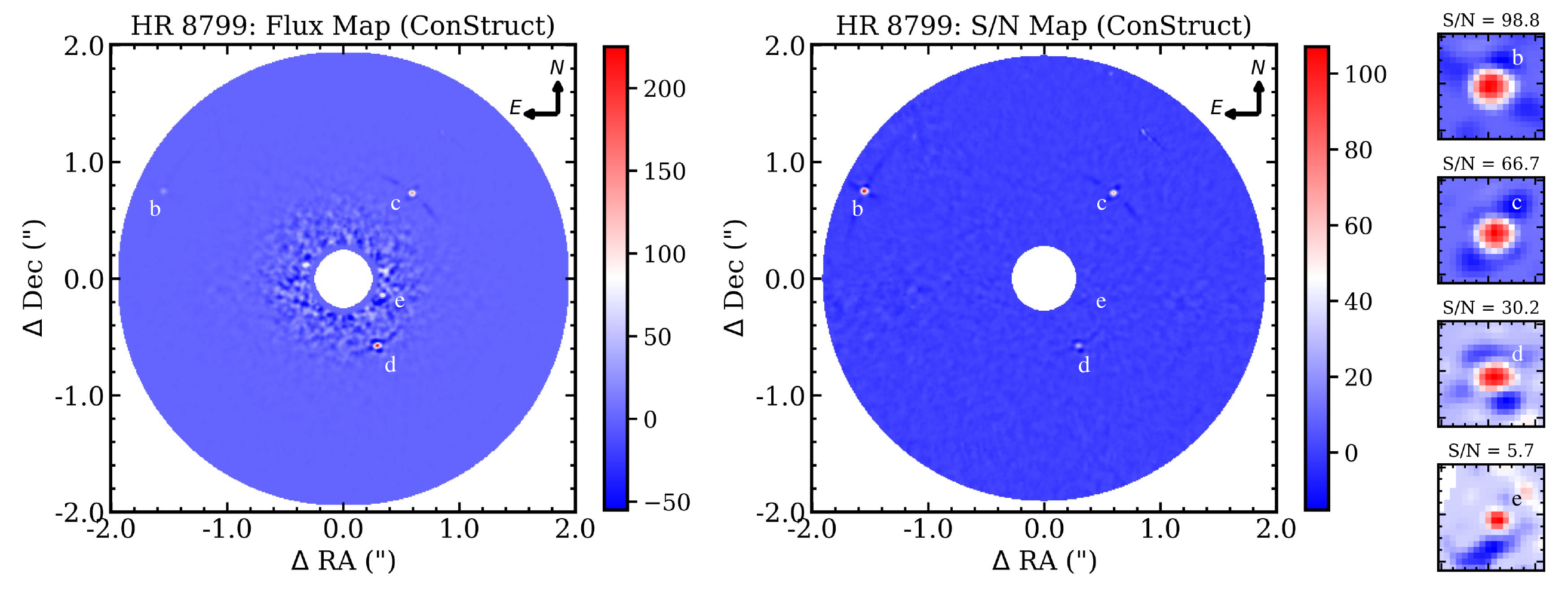} \\
    \centering\small \hspace{-12mm} (a) \texttt{ConStruct}
  \end{tabular}%
  \quad
  \begin{tabular}[b]{@{}p{1.0\textwidth}@{}}
    \centering\includegraphics[width=.95\linewidth]{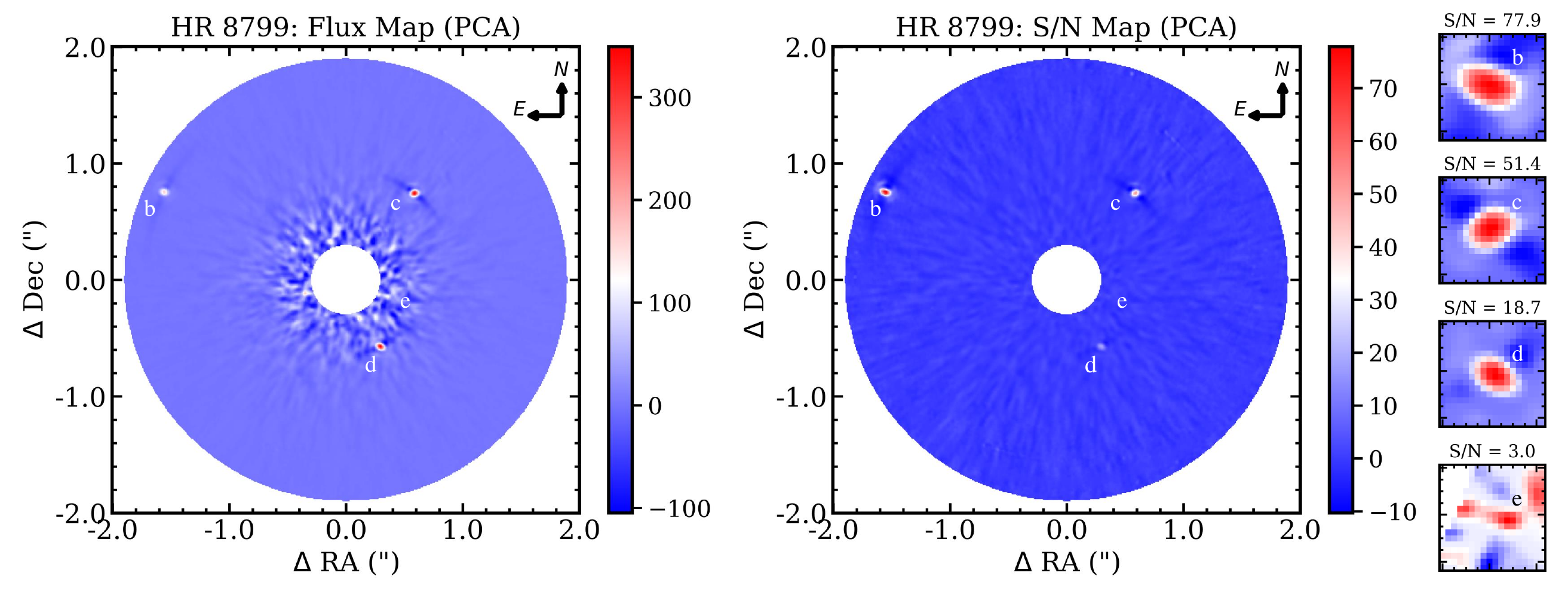} \\
    \centering\small \hspace{-10mm} (b) Annular PCA
  \end{tabular}
  \caption{Full-frame reductions with ConStruct and \PCA\hspace{-0.1ex} of HR 8799 \citep{Marois_2008, Marois_2010} for Sequence 10 in Table \ref{tab:testing_data}.}
  \label{fig:reductions_20}
\end{figure*}

\begin{figure*}[hbt!]
  \centering
  \begin{tabular}[b]{@{}p{1.0\textwidth}@{}}
    \centering\includegraphics[width=.95\linewidth]{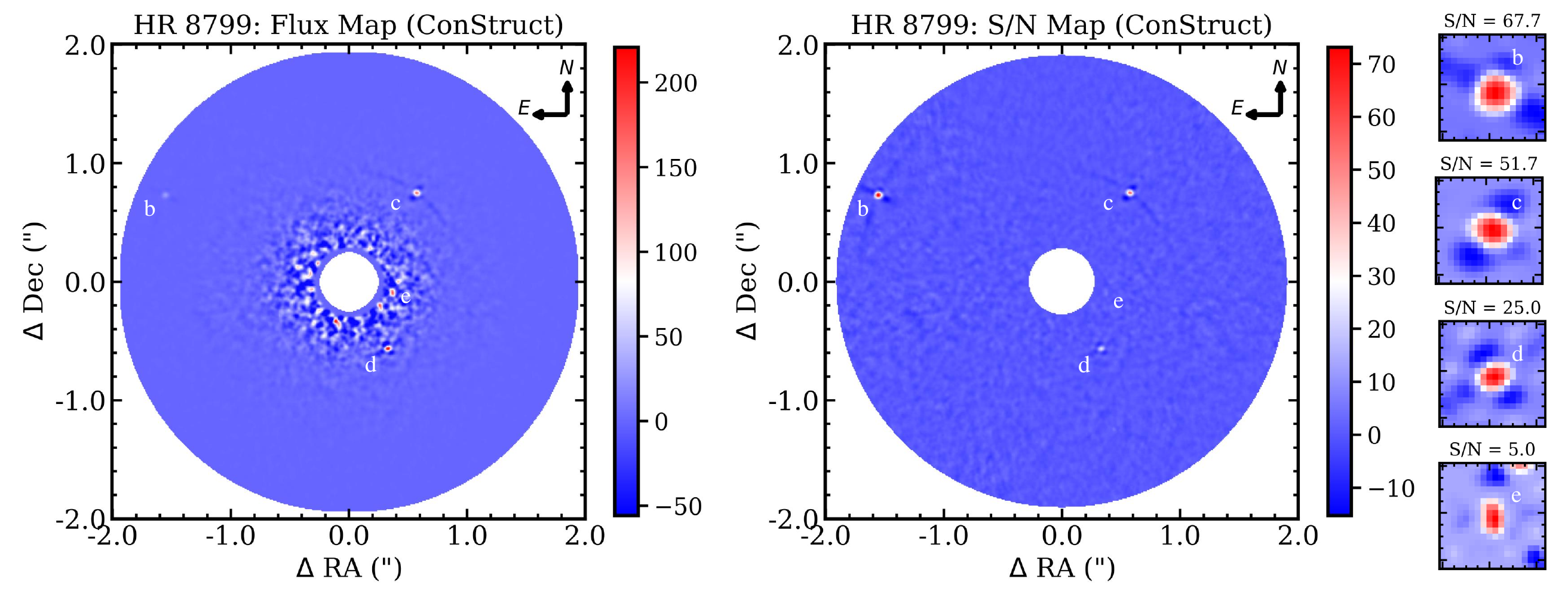} \\
    \centering\small \hspace{-12mm} (a) \texttt{ConStruct}
  \end{tabular}%
  \quad
  \begin{tabular}[b]{@{}p{1.0\textwidth}@{}}
    \centering\includegraphics[width=.95\linewidth]{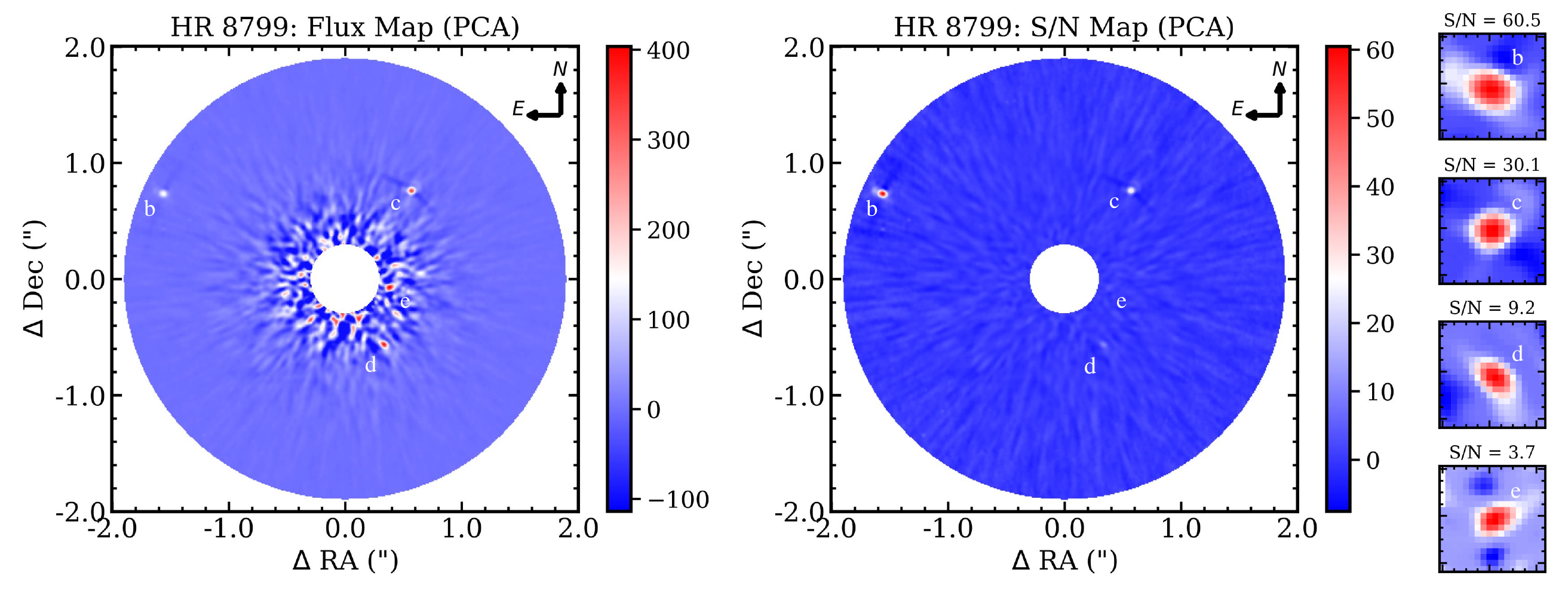} \\
    \centering\small \hspace{-10mm} (b) Annular PCA
  \end{tabular}
  \caption{Full-frame reductions with ConStruct and \PCA\hspace{-0.1ex} of HR 8799 \citep{Marois_2008, Marois_2010} for Sequence 12 in Table \ref{tab:testing_data}.}
  \label{fig:reductions_21}
\end{figure*}

Here, we show additional full-frame reductions of the ADI datasets used in this study processed with both \texttt{ConStruct} and PCA. Figure \ref{fig:reductions_6} shows the reductions for HD 4747 \citep{Crepp_2016}, corresponding to Sequence 1 in Table \ref{tab:testing_data}. Figure \ref{fig:reductions_8} shows HD 19467 \citep{Crepp_2014} for Sequence 2. Figure \ref{fig:reductions_12} is for HD 114174 \citep{Crepp_2013}, corresponding to Sequence 4. Figure \ref{fig:reductions_14} is for HR 7672 \citep{Liu_2002}, corresponding to Sequence 7. Figures \ref{fig:reductions_15} and \ref{fig:reductions_16} show two Kappa And \citep{Carson_2013} reductions corresponding to Sequences 8 and 9, respectively. Figures \ref{fig:reductions_20} and \ref{fig:reductions_21} show two HR 8799 \citep{Marois_2008, Marois_2010} reductions corresponding to Sequences 10 and 11, respectively.

\end{document}